\documentclass[version=last,chapterprefix=true,14pt,BCOR=-12mm,DIV=calc,oneside,a4paper,numbers=noenddot]{scrbook}

\usepackage[left=3cm,right=1.5cm,top=2cm,bottom=2cm]{geometry}           

\usepackage[cp1251]{inputenc}
\usepackage[russian,english,ukrainian]{babel}
\usepackage{amssymb}
\usepackage{graphicx}
\usepackage{amsmath}
\usepackage[makeroom]{cancel}

\usepackage{hyperref}
\usepackage{wrapfig}
\usepackage{array}
\usepackage{subcaption}
\usepackage{comment}

\usepackage{flushend}
\usepackage{cuted}

\usepackage{multicol}
\usepackage{enumitem}

\newcommand\NoIndent[1]{%
	\par\vbox{\parbox[t]{\linewidth}{#1}}%
}

\usepackage{datatool}

\usepackage[stable]{footmisc}

\usepackage[font=small,labelfont=bf]{caption}

\usepackage{color}

\usepackage{amsmath}
\usepackage{amsfonts}

\usepackage{graphicx}

\usepackage[T1]{fontenc}

\usepackage{inputenc}
\usepackage{microtype}

\usepackage{fancyhdr}

\usepackage{array}              
\usepackage{setspace}
\usepackage{multicol}
\usepackage{graphicx}

\usepackage[14pt]{extsizes}

\DeclareGraphicsExtensions{.pdf,.png,.jpg}  
\addtokomafont{disposition}{\rmfamily\mdseries}
\addtokomafont{chapterprefix}{\large}
\setkomafont{dictumtext}{\normalfont\normalcolor\itshape\setlength{\parindent}{1em}\noindent}

\makeatletter
\let\old@makechapterhead\@makechapterhead
\def\fake@makechapterhead#1{%
	\vspace*{50\p@}%
	{\parindent \z@ \raggedright \normalfont
		\ifnum \c@secnumdepth >\m@ne
		\huge\bfseries \strut
		\par\nobreak
		\vskip 20\p@
		\fi
		\interlinepenalty\@M
		\Huge \bfseries #1\par\nobreak
		\vskip 40\p@
}}
\newcommand{\newchapterhead}{\let\@makechapterhead\fake@makechapterhead}
\newcommand{\restorechapterhead}{\let\@makechapterhead\old@makechapterhead}
\makeatother

\let\oldtabular\tabular 
\renewcommand{\tabular}{\fontsize{0.5cm}{0.2cm}\selectfont \oldtabular}

\newcommand\expprefix{}
\let\latexorigthechapter\thechapter
\renewcommand{\thechapter}{\expprefix\latexorigthechapter}

\DeclareTOCStyleEntry[numwidth=2.3em]{default}{chapter}
\DeclareTOCStyleEntry[numwidth=3em, indent=2.3em]{default}{section}

\date{\today}

\begin{document}

\selectlanguage{english}

\begin{titlepage}
	\begin{spacing}{0.5}

		\vspace{10.0cm}
		\begin{center}
			\large \textbf{\uppercase{Lectures on spintronics and magnonics}}
		\end{center}
		\vspace{3.5cm}

		\begin{center}\large
			M. V. Mazanov, V. A. Shklovskij
		\end{center}
		\begin{center}
			 \textit{V. Karazin Kharkiv National University, 61022, Kharkiv, Ukraine}\\
		\end{center}
		
		\vspace{2.5cm}

	\end{spacing}

	\begin{minipage}{0.90\textwidth}
		
		\qquad In this series of lectures, we discuss the basic theoretical concepts of magnonics and spintronics. 
		We first briefly recall the relevant topics from quantum mechanics, electrodynamics of continuous media, and basic theory of magnetism. 
		We then discuss the classical theory of magnetic dynamics: ferromagnetic and antiferromagnetic resonance, dynamic susceptibilities, and spin waves.  
		We open the main discussion with phenomena of spin and exchange spin currents, spin torques, the spin Hall effect, and the spin Hall and Hanle magnetoresistance.  
		Special emphasis is given to the effects of spin transfer torque and spin pumping, where we follow the celebrated derivation utilizing Landauer quantum multi-channel scattering matrix approach.
		Finally, we outline the most important features distinguishing antiferromagnetic dynamics from ferromagnetic one, which make antiferromagnets particularly promising material candidates for spintronics and magnonics. 
		%

	\end{minipage}
	
\clearpage
\newpage	

\

\end{titlepage}
\clearpage
\newpage   
\restoregeometry

\selectlanguage{english}


\renewcommand{\normalsize}{\fontsize{16}{40}\selectfont}
\setlength{\topmargin}{-2.5cm}
\setlength{\oddsidemargin}{-1cm}
\setlength{\evensidemargin}{-1cm}
\setlength{\headheight}{1cm}
\setlength{\headsep}{0.5cm}
\setlength{\topskip}{0.0cm}
\setlength{\textwidth}{18cm} \setlength{\textheight}{25cm}
\setlength{\parindent}{1.27 cm}
\pagestyle{myheadings}

\def\contentsname{CONTENTS}
\def\bibname{REFERENCES}
\def\tablename{Table}
\def\abstractname{Abstract}
\def\conclname{CONCLUSIONS}

\def\chaptername{Lecture}
\def\sectionname{§}
\def\captionname{Fig}
\def\tablename{Table}
\def\abstractname{ABSTRACT}

\addtocounter{page}{0}

\def\chaptername{Lecture}
\def\sectionname{§}
\def\captionname{Fig}
\def\contentsname{CONTENTS}
\def\bibname{REFERENCES}
\def\tablename{Table}
\def\abstractname{ABSTRACT}

\tableofcontents



\chapter*{List of abbreviations and conventions \label{Soglasie}}
\addcontentsline{toc}{chapter}{\protect\numberline{}List of abbreviations and conventions}%

\section*{Abbreviations}
\noindent
\textbf{AFM}~-- Antiferromagnet

\noindent
\textbf{AFMR}~-- Antiferromagnetic Resonance

\noindent
\textbf{ESR}~-- Electron Spin Resonance

\noindent
\textbf{FM}~-- Ferromagnet

\noindent
\textbf{FMR}~-- Ferromagnetic Resonance

\noindent
\textbf{HMR}~-- Hanle magnetoresistance

\noindent
\textbf{ISGE}~-- Inverse Spin Galvanic Effect

\noindent
\textbf{ISHE}~-- Inverse Spin Hall Effect

\noindent
\textbf{NPM}~-- Normal paramagnetic metal

\noindent
\textbf{SHE}~-- Spin Hall effect

\noindent
\textbf{SMR, SHMR}~-- Spin Hall Magnetoresistance

\noindent
\textbf{SOI}~-- Spin-orbit interaction

\noindent
\textbf{SOT}~-- Spin-orbit Torque

\noindent
\textbf{ST}~-- Spin Torque

\noindent
\textbf{STT}~-- Spin Transfer Torque

\noindent
\textbf{SW}~-- Spin wave


\section*{Conventions}
\noindent
The CGS system of units is used throughout the lectures.

\noindent
$ e $ is the modulus elementary charge, $ e>0 $.

\noindent
Bohr magneton $ \mu_B=\frac{e \hbar}{2 m_e c}>0 $.

\noindent
Gyromagnetic factor $ \gamma=-\frac{e}{m_e c}=-\frac{g_e \mu_B}{\hbar}<0  $.

\chapter*{Introduction \label{}}
\addcontentsline{toc}{chapter}{\protect\numberline{}\noindent From the authors}%

\indent

In recent decades, two promising areas have emerged in the physics of magnetism, motivated by rapidly developing manufacturing technology of magnetic nanostructures comparable to electron spin and spin wave coherence lengths. \textbf{Spintronics} (from \textit{spin transport electronics}) studies the control of spin current in metals and semiconductors (similar to electronics, which studies the control of charge current), while \textbf{magnonics} studies the control of spin wave current in magnetic solids.

In spintronics, electron spin is utilised as an additional degree of freedom which allows to increase the efficiency of data storage and transmission and create new types of memory and transmission methods.
In magnonics, spin-wave transport in ferromagnetic insulators such as YIG (Yttrium iron garnet) is considered. In ferromagnetic insulators, high transfer efficiency of the exchange spin current could be achieved due to large spin wave decay times. Moreover, spin waves are already a well-studied object, with methods developed much earlier than the isolation of magnonics as a line of research.

Magnetic materials with movable domain walls displaced by a current or an external magnetic field are also considered as intermediate elements of future ``spintronic circuits''.
Recently, antiferromagnets (AFM) have also been considered promising materials: due to compensation of magnetic sublattices magnetizations, they possess minimal demagnetizing and stray fields, while the characteristic frequencies of spin waves in AFM are much higher than in ferromagnets. In addition, in antiferromagnets, the spin torque which an external current exerts on the domain wall is very nonlocal, and at a current density above some critical value it can even trigger THz self-oscillations of sublattice magnetizations. 



This course, designed for theoretical students, acquaints readers with some of the basic concepts of spintronics and magnonics, starting with the fundamental concepts of magnetism.
The first two lectures concisely set out the quantum-mechanical concept the electron spin, as well as magnetic susceptibility from the electrodynamics of continuous media, along with a brief description of susceptibilities of the two most common types of magnetic substances: para- and ferromagnets.
In the next lecture, the susceptibility of para- and ferromagnets is generalized to the case of an alternating external field.
In the fourth lecture, the classical dispersion law for spin waves in a ferromagnet is derived, and the concept of an exchange spin current in solids, carried by spin waves, is considered.
In the fifth lecture we discuss the spin torques through which the spin density interacts with localized magnetization. 
As an example of bulk spin torques, within the framework of a simple model, we derive the four main types of spin torque in a conducting ferromagnet.
In the sixth lecture, we discuss the phenomenon of spin pumping, in which the dynamics of magnetization in a ferromagnet induces a spin current in a neighboring metallic paramagnet, along with the opposite phenomenon~-- the Spin Transfer Torque (STT). The derivation of expressions for the emerging spin current is given in Landauer formulation, since in modern studies the concept of spin-mixing conductance, which arises in this approach, is widely used as a quantitative estimate of the spin transfer efficiency.
In the seventh lecture, we briefly consider the phenomenology of direct and inverse spin Hall effects, thanks to which the electrical detection of spin current became possible, as well as the Hanle and spin-Hall magnetoresistance phenomena, due to which the nonequilibrium spin density at the edges of the sample can also be detected.
In the last lecture, we outline the main similarities and differences between antiferromagnets and ferromagnets. 
All lectures end with a series of control questions.

In the list of recommended references, we include the textbooks [1–9,15], in which the necessary concepts of the theory of magnetism are presented, as well as the list of seminal articles and reviews containing a presentation of the discussed concepts of spintronics and magnonics. 
The additional reference list includes books, articles and reviews that elaborate and supplement the course material.


\setchapterpreamble{%
	\dictum{%
		
		In this section, we briefly discuss the concepts commonly used to describe the spin of particles and excitations: the \textbf{spin operator}, the \textbf{spin density matrix} and the \textbf{spin polarization vector}. The notion of spin polarization vector will be of great use in discussing the spin torques (Lecture 5) and the Spin Hall Effect (Lecture 7), while the spin operator and the notion of spin density matrix will be used in the discussion of spin pumping (Lecture 6).
		
	}%
	\vspace{24pt}%
}

\chapter{Concepts from quantum mechanics}


\section{Electron spin}\

Electron, as an elementary particle, possessing negative electric charge and responsible for the various properties of condensed matter. 
In addition to the electric charge $ -e<0 $\footnotemark\footnotetext{see. \emph{Conventions}.}, an electron has a mechanical spin momentum and an associated magnetic moment.

In quantum mechanics, spin and magnetic moment correspond to vector operators $\hat{\mathbf{s}}$ and $\hat{\boldsymbol{\mu}}=\gamma \hat{\mathbf{s}}$ \footnotemark[1]. As a momentum operator, $\hat{\mathbf{s}}$ \emph{has no vector eigenvalues}: in any state, it is impossible to know with certainty all three components of its angular momentum simultaneously.
This is reflected in the well-known commutation relations for the components $\hat{s}_x,\hat{s}_y,\hat{s}_z$ of the spin operator $\hat{\mathbf{s}}$ in some orthogonal basis $ \{x,y,z\} $: 
\begin{equation}
\left[\hat{s}_i,\, \hat{s}_j\right]=i\hbar \varepsilon_{ijk} \hat{s}_k.
\end{equation}
The components of the spin operator are the projections of the vector operator $\hat{\mathbf{s}}$ on the coordinate axes $ \mathbf{e}_i $:
\begin{equation}
\hat{s}_i=\hat{\mathbf{s}}\cdot \mathbf{e}_i.
\end{equation}
\textit{Spin projection operator} on an arbitrary axis $ \mathbf{l} $ then reads
\begin{equation}
\hat{s}_l=\hat{\mathbf{s}}\cdot \mathbf{l}.
\end{equation}
The components of angular momentum operator $ \hat{L}_i $ have eigenvalues that are multiples of $ \hbar/2 $. Spin, however, is an intrinsic property of an electron, and its spin components eigenvalues are also inseparable and invariable properties of an electron, like its charge.
From experiment we know that the operator of any electron spin component $ \hat{s}_l $ (including $\hat{s}_x,\hat{s}_y$ and $\hat{s}_z$) has only two eigenvalues: $ +\hbar/2 $ and $ -\hbar/2 $. Consequently, $ \hat{s}_l $ can be represented by a \emph{two-row matrix}, and the spin part of the wave function can be represented by a \emph{column of two components (spinor)}. The canonical matrix representation of the spin operator was introduced by W. Pauli:
\begin{equation}
\hat{\boldsymbol{s}}=\frac{\hbar}{2} \hat{\boldsymbol{\sigma}},
\end{equation}
where \emph{Pauli matrices} are
\begin{equation}
\hat{\sigma}_x=
\begin{pmatrix}
0 & 1 \\
1 & 0
\end{pmatrix}, \qquad
\hat{\sigma}_y=
\begin{pmatrix}
0 & -i \\
i & 0
\end{pmatrix}, \qquad
\hat{\sigma}_z=
\begin{pmatrix}
1 & 0 \\
0 & -1
\end{pmatrix}.
\end{equation}
Components $\hat{s}_i$ act as matrix multiplication by spin column vector, or a \textit{spinor}:
\begin{equation}
\label{every spinor}
\chi=\begin{pmatrix}
a_1 \\
a_2
\end{pmatrix}.
\end{equation}
Norm of spinor reads
\begin{equation}
\chi^\dag \chi = (a_1^*\,\,a_2^*)
\begin{pmatrix}
a_1 \\
a_2
\end{pmatrix}
=
a_1^* a_1 + a_2^* a_2
=
|a_1|^2 + |a_2|^2
.
\end{equation}
Usually spinor is normalized by $ 1 $: $ |a_1|^2 + |a_2|^2=1 $. In doing so, we determine the spinor up to the arbitrary phase factor $ e^{i \phi},\,\forall \phi $.
Normalized eigenspinors of $ \hat{s}_z $ corresponding to states with $s_z=\hbar/2$ and $s_z=-\hbar/2$ read:
\begin{equation}
\label{z_eigenspinors}
\chi^{(z)}_{\uparrow}=\begin{pmatrix}
1 \\
0
\end{pmatrix}, \qquad
\chi^{(z)}_{\downarrow}=\begin{pmatrix}
0 \\
1
\end{pmatrix}.
\end{equation}
Now it is easy to verify directly that in these states $ \hat{s}_x $ and $ \hat{s}_y $ have no definite value. Take, for example, the state $ \chi^{(z)}_{\uparrow} $ and projection $ \hat{s}_x $:
\begin{equation}
\hat{s}_x \chi^{(z)}_{\uparrow} =
\frac{\hbar}{2}\hat{\sigma}_x \chi^{(z)}_{\uparrow} =
\frac{\hbar}{2}
\begin{pmatrix}
0 & 1 \\
1 & 0
\end{pmatrix}
\begin{pmatrix}
1 \\
0
\end{pmatrix}
=
\frac{\hbar}{2}
\begin{pmatrix}
0 \\
1
\end{pmatrix},
\end{equation}
then the average value of $ \hat{s}_x $ in this state is
\begin{equation}
\langle\chi^{(z)}_{\uparrow}|\hat{s}_x|\chi^{(z)}_{\uparrow}\rangle
=
\frac{\hbar}{2}
(1^*\,\,0^*)
\begin{pmatrix}
0 & 1 \\
1 & 0
\end{pmatrix}
\begin{pmatrix}
1 \\
0
\end{pmatrix}
=
\frac{\hbar}{2}
(1\,\,0)
\begin{pmatrix}
0 \\
1
\end{pmatrix}
=
0.
\end{equation}
It is also easy to find the eigenspinors of the operator $ \hat{s}_l $ of the spin projection onto an arbitrary axis $ \mathbf{l}=(\sin{\theta}\cos{\phi},\sin{\theta}\sin{\phi},\cos{\theta}) $ corresponding to states with $s_l=\hbar/2$ and $s_l=-\hbar/2$:
\begin{equation}
\label{theta_phi_eigenspinors}
\chi^{(l)}_{\uparrow}\equiv\chi^{(\theta,\phi)}_{\uparrow}=\begin{pmatrix}
\cos{\frac{\theta}{2}} \\
e^{i \phi} \sin{\frac{\theta}{2}}
\end{pmatrix}, \qquad
\chi^{(l)}_{\downarrow}\equiv\chi^{(\theta,\phi)}_{\downarrow}=\begin{pmatrix}
-e^{-i \phi} \sin{\frac{\theta}{2}} \\
\cos{\frac{\theta}{2}}
\end{pmatrix}.
\end{equation}
In particular, for directions $ \mathbf{e}_z $ and $ \mathbf{e}_y $ we get:
\begin{equation}
\label{x_eigenspinors}
\chi^{(x)}_{\uparrow}\equiv\chi^{(\pi/2,0)}_{\uparrow}=\frac{1}{\sqrt{2}}\begin{pmatrix}
1 \\
1
\end{pmatrix}, \qquad
\chi^{(x)}_{\downarrow}\equiv\chi^{(\pi/2,0)}_{\downarrow}=\frac{1}{\sqrt{2}}\begin{pmatrix}
1 \\
-1
\end{pmatrix},
\end{equation}
\begin{equation}
\label{y_eigenspinors}
\chi^{(y)}_{\uparrow}\equiv\chi^{(\pi/2,0)}_{\uparrow}=\frac{1}{\sqrt{2}}\begin{pmatrix}
1 \\
i
\end{pmatrix}, \qquad
\chi^{(y)}_{\downarrow}\equiv\chi^{(\pi/2,0)}_{\downarrow}=\frac{1}{\sqrt{2}}\begin{pmatrix}
1 \\
-i
\end{pmatrix}.
\end{equation}
It is also useful to notice the relation for spinors for opposite directions $ \mathbf{l} $ and $ -\mathbf{l} $:
\begin{equation}
\label{polezno-0}
\chi^{(-l)}_{\uparrow}=\chi^{(l)}_{\downarrow},
\end{equation}
or
\begin{equation}
\label{polezno-1}
\chi^{(\pi-\theta,\phi+\pi)}_{\uparrow}=\chi^{(\theta,\phi)}_{\downarrow}.
\end{equation}
Together with one of the spin components, for example, $ \hat{s}_z $, the square of the spin $ \hat{\mathbf{s}}^2 $ can also have a definite value. This follows from the commutation relation \footnotemark\footnotetext{see. \emph{Properties of Pauli matrices} at the end of the lecture.}:
\begin{equation}
\label{s_bold squared}
\left[ \hat{\mathbf{s}}^2,\hat{s}_z \right]=
\left(\frac{\hbar}{2}\right)^3 \left[ \hat{\boldsymbol{\sigma}}^2,\hat{\sigma}_z \right]=
\left(\frac{\hbar}{2}\right)^3 \left[ 3 \hat{\sigma}_0,\hat{\sigma}_z \right]=0
.
\end{equation}
Unlike $ \hat{\mathbf{s}} $, $ \hat{\mathbf{s}}^2 = (\hbar/2)^2 \cdot 3 \hat{\sigma}_0 $ has one (doubly degenerate in spin direction) eigenvalue equal to $ 3 (\hbar/2)^2 = (3/4) \hbar^2 $ \footnotemark. 
\footnotetext{The above relations for $ \hat{\mathbf{s}}^2 $ are especially useful if the Hamiltonian includes terms proportional to $ \hat{\mathbf{s}}^2 $.}

Expressions for the electron spin and magnetic moment eigenvalues arise in a surprising way ``in passing'' when deriving the relativistic \emph{Dirac equation} for a free electron (as a result, $s_z$ is obtained) and its nonrelativistic approximation, taking into account the electron interaction with the electromagnetic field $ (\mathbf{A},\phi) $ (this way, $\mu_z$ is found).
The last equation is called the \emph{Pauli equation} for the electron and has the form of the nonrelativistic Schr\"{o}dinger equation with the Hamiltonian
\begin{equation}
\label{PH}
\hat{H}
=
\frac{1}{2m} \left(\mathbf{\hat{p}} - \frac{e}{c} \mathbf{A} \right)^2 + e \phi - \boldsymbol{\hat{\mu}}\mathbf{H},
\end{equation}
where electron magnetic moment operator
$\hat{\boldsymbol{\mu}}=\gamma \hat{\mathbf{s}}=-\mu_B \hat{\boldsymbol{\sigma}}$\footnotemark.

\footnotetext{see \emph{Conventions}.}

\subsection{Electron spin polarization \footnotemark}\

\footnotetext{See also \cite{Kessler}, pp. 1-19.}

The electron spin is essentially of quantum nature: in a given state and in a given basis, only one spin component can have definite value. For example, in the basis $ \{x,y,z\} $ there are two states \eqref{z_eigenspinors} with a certain value of $ \hat{\sigma}_z $, but indefinite values $ \hat{\sigma}_x $ and $ \hat{\sigma}_y $.
There are also other states in which neither $ \hat{\sigma}_x $, nor $ \hat{\sigma}_y $, nor $ \hat{\sigma}_z $ have a definite meaning\footnotemark\footnotetext{Such are all spinors except for \eqref{z_eigenspinors},\eqref{x_eigenspinors},\eqref{y_eigenspinors}.}.
However, for each direction $ (\theta,\phi) $ there are two eigenspinors on the unit sphere \eqref{theta_phi_eigenspinors}. The question arises: is every conceivable spinor \eqref{every spinor} an eigenspinor for some direction?
Indeed, one can show\footnotemark\footnotetext{e.g., note that all conceivable physically different normalized spinors and all directions to the unit sphere are both given by \emph{two} independent parameters.} that any spin state of an electron $ |\chi_n\rangle $ can be associated with a unit \emph{polarization vector}, $ \mathbf{P}_n =(\sin{\theta}\cos{\phi},\sin{\theta}\sin{\phi},\cos{\theta}) $, the spin component $ \hat{\mathbf{s}}\cdot \mathbf{P}_n $ along which has in this state the defined value $ \hbar/2 $.
Note that $ \mathbf{P}_n $ \emph{is just an alternative representation of the spinor} $ |\chi_n\rangle $.
The polarization vector $ \mathbf{P}_n $ corresponds to the mean value of the operator $ \hat{\boldsymbol{\sigma}} $ in the state $ |\chi_n\rangle $:
\begin{equation}
\label{Pn sigma}
\mathbf{P}_n=\langle\chi_n|\hat{\boldsymbol{\sigma}}|\chi_n\rangle.
\end{equation}
This can be verified directly by calculating the mean \eqref{Pn sigma} using the spinor $ \chi^{(\theta,\phi)}_{\uparrow} $ \eqref{theta_phi_eigenspinors}.

\begin{figure}[h!]
	\selectlanguage{english}
	\centering
	\includegraphics
	[width=0.6\textwidth]{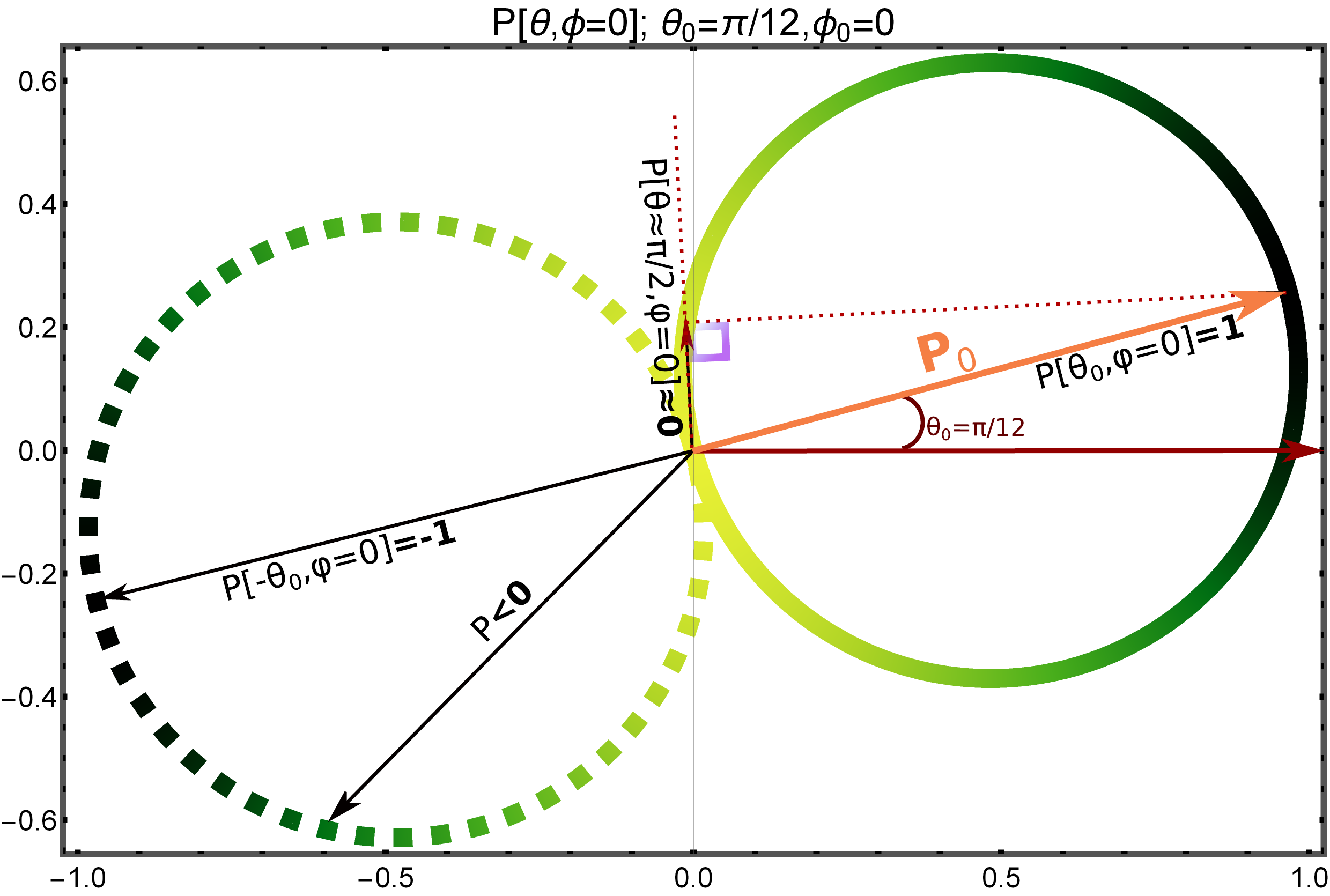}
	\caption{Dependence of $ P_{0,\mathbf{l}}\equiv P_{(\theta,\phi)} $ on the direction of measurement $ (\theta,\phi) $ at $ \phi=0 $ for the state $ \chi^{(\theta_0=\pi/12,\phi_0=0)}_{\uparrow} $. The direction of polarization vector in this state is marked as $ \mathbf{P_0} $.}
	\label{Spin-Graph}
\end{figure}

By analogy, the scalar value, \emph{electron polarization $ P_{n,\mathbf{l}} $ along direction} $ \mathbf{l} $ in the state $ |\chi_n\rangle $, is defined:
\begin{equation}
\label{Pn l}
P_{n,\mathbf{l}}\equiv\langle\chi_n|\hat{\boldsymbol{\sigma}}\cdot \mathbf{l}|\chi_n\rangle
=
\langle\chi_n|\hat{\boldsymbol{\sigma}}|\chi_n\rangle \cdot \mathbf{l}
=
\mathbf{P}_n \cdot \mathbf{l}
,
\end{equation}
which is simply the component of $ \mathbf{P} $ along the direction $ \mathbf{l} $. 
The physical meaning of $ P_{n,\mathbf{l}} $ is the average value of the spin component (in units of $ \hbar/2 $) along the direction $ \mathbf{l} $.
The plot \ref{Spin-Graph} shows the dependence of $ P_{0,\mathbf{l}}\equiv P_{(\theta,\phi)} $ on the direction of measurement $ (\theta,\phi) $ at $ \phi=0 $ for the state $ \chi^{(\theta_0=\pi/12,\phi_0=0)}_{\uparrow} $ and the direction of polarization vector $ \mathbf{P_0} $ in this state.

Note that the (unnormalized) superposition of two states $ |\chi_1\rangle $ and $ |\chi_2\rangle $ with polarization vectors $ \mathbf{P}_1 $ and $ \mathbf{P}_2 $, $ |\chi_{sup}\rangle = |\chi_1\rangle + |\chi_2\rangle $ gives again a state with some polarization vector $ \mathbf{P}_{sup} $, $ |\mathbf{P}_{sup}|=1 $. 
An analogy can be drawn with light: a superposition of two waves with circular polarizations of opposite handedness gives a linearly polarized wave (and not an unpolarized wave).

\section{Statistical ensemble (beam) of electrons}\

In this section, we will consider a statistical ensemble (beam) of noninteracting electrons. The spin part of the wave function of electrons in the beam in a region of space without magnetic field\footnotemark\footnotetext{We consider electrons in stationary states of the spinless Hamiltonian of a free particle $ \hat{H} = \hat{p}^2/2m_e $.} is invariable in time (in what follows, by the wave function we will mean its spin part). We denote all electron states in the beam as $ |\chi_n\rangle $, and the probability of registering an electron in the $ |\chi_n\rangle $ state in an arbitrary beam section as $ p_n $.

As is well known \cite{LL3}, statistical ensembles of particles (or \emph{\textbf{mixed} states}) are fully described by the normalized \textit{density matrix}:
\begin{equation}
\label{rho def}
\hat{\rho}=\sum_{n} p_n |\chi_n\rangle\langle\chi_n|,
\end{equation}
where all spinors $ |\chi_n\rangle $ are normalized, $ \langle\chi_n|\chi_n\rangle=1,\,\forall n $, and the sum of the probabilities is also normalized: $ \sum_{n} p_n = 1 $. It is easy to check that the normalized density matrix has a trace $ 1 $: $ \mathbf{Tr}(\hat{\rho})=1 $.


If the beam consists only ($ p_1=1 $) of electrons in one state $ |\chi_1\rangle = \begin{pmatrix}
a^{(1)}_1 \\
a^{(1)}_2
\end{pmatrix} $, then such a state is called \emph{\textbf{pure}} and is fully described by both the wave function and the density matrix:
\begin{equation}
\hat{\rho}_1=1 \cdot |\chi_1\rangle\langle\chi_1|
=
\begin{pmatrix}
a^{(1)}_1 \\
a^{(1)}_2
\end{pmatrix}
(a_1^{(1)*}\,\,a_2^{(1)*})
=
\begin{pmatrix}
|a^{(1)}_1|^2 & \quad a^{(1)}_1 a^{(1)*}_2 \\
a^{(1)*}_1 a^{(1)}_2 & \quad |a^{(1)}_2|^2
\end{pmatrix}
.
\end{equation}
The average value of any spin operator $ \hat{A} $ in the state $ |\chi_1\rangle $ then reads \cite{LL3}:
\begin{equation}
\langle \hat{A} \rangle
=
\langle\chi_1|\hat{A}|\chi_1\rangle
=
\textbf{Tr}(\hat{\rho}_1 \hat{A}).
\end{equation}

In the general case, the beam state is \textbf{\textit{mixed}}. The density matrix explicitly reads
\begin{equation}
\hat{\rho}
=\sum_{n} p_n |\chi_n\rangle\langle\chi_n|
=
\sum_{n} p_n
\begin{pmatrix}
|a^{(n)}_1|^2 & \quad a^{(n)}_1 a^{(n)*}_2 \\
a^{(n)*}_1 a^{(n)}_2 & \quad |a^{(n)}_2|^2
\end{pmatrix}
.
\end{equation}
The expression for the mean value of the physical quantity $ A $ (now \emph{quantum-mechanical$ + $statistical}) has the form
\begin{equation}
\label{rho_averaging}
\langle \hat{A} \rangle
=
\textbf{Tr}(\hat{\rho} \hat{A}).
\end{equation}
In particular, the \emph{beam polarization vector} $ \mathbf{P} $ is defined as the mean value of the vector spin operator:
\begin{equation}
\mathbf{P}
=
\sum_n p_n
P_n
=
\sum_n p_n
\langle\chi_n|\hat{\boldsymbol{\sigma}}|\chi_n\rangle
=
\langle \hat{\boldsymbol{\sigma}} \rangle
=
\textbf{Tr}(\hat{\rho} \hat{\boldsymbol{\sigma}}).
\end{equation}
In the case of a mixed state, $ |\mathbf{P}|\leq 1 $ (equality is attained for a pure state). This is the difference between the \emph{quantum-mechanical superposition of states} in the one-electron case from the \emph{statistical mixture of states} of electrons in the beam. The density matrix of the quantum-mechanical superposition of states $ |\psi\rangle=|\chi_1\rangle+|\chi_2\rangle $ is not generally equal to the sum of the density matrices of each state:
\begin{equation}
\rho_\psi=|\psi\rangle\langle\psi|=|\chi_1\rangle\langle\chi_1|+|\chi_2\rangle\langle\chi_2|+\boldsymbol{(|\chi_1\rangle\langle\chi_2|+|\chi_2\rangle\langle\chi_1|)},
\end{equation}
where the interference term is highlighted in brackets. In the case of a completely unpolarized electron beam, this term is statistically averaged to zero.
It is easy to check that $ \hat{\rho} $ \emph{is uniquely defined by the beam polarization vector} $ \mathbf{P} $:
\begin{equation}
\label{rho as P sigma}
\hat{\rho}
=
\frac{1}{2}(\hat{\sigma}_0 + \mathbf{P}\cdot \hat{\boldsymbol{\sigma}})
=
\frac{1}{2}
\begin{pmatrix}
1+P_z & P_x-i P_y \\
P_x+i P_y & 1-P_z
\end{pmatrix}
.
\end{equation}
In particular, for a \emph{completely unpolarized beam,} $ \mathbf{P}=0 $,
\begin{equation}
\label{unpolarized}
\hat{\rho}_u
=
\frac{1}{2}
\begin{pmatrix}
1 & 0 \\
0 & 1
\end{pmatrix}
.
\end{equation}
For a beam that is fully or partially ($ |\mathbf{P}|=P\leq 1 $) polarized along the $ z $ axis, $ \mathbf{P}=P \mathbf{e}_z $,
\begin{equation}
\label{puchok ez}
\hat{\rho}
=
\frac{1}{2}
\begin{pmatrix}
1+P & 0 \\
0 & 1-P
\end{pmatrix}
=
P\cdot
\begin{pmatrix}
1 & 0 \\
0 & 0
\end{pmatrix}
+
(1-P)\cdot
\begin{pmatrix}
1/2 & 0 \\
0 & 1/2
\end{pmatrix}
=
P\cdot
\hat{\rho}_{\mathbf{e}_z}
+
(1-P)\cdot
\hat{\rho}_u,
\end{equation}
where $ \hat{\rho}_{\mathbf{e}_z} = \begin{pmatrix}
1 & 0 \\
0 & 0
\end{pmatrix} $ is the density matrix of a beam completely polarized along the $ z $ axis. 

According to the definition of \eqref{rho def}, a beam with a density matrix \eqref{puchok ez}, is \emph{completely indistinguishable} from a beam in which the $ P $-th part of electrons is polarized along $ \mathbf{e}_z $, and the remaining $ (1-P) $-th part is equivalent to a completely unpolarized beam \eqref{unpolarized}. Therefore, \emph{the description of the beam by its density matrix does not describe the beam completely quantum-mechanically}, but only gives a statistical representation of it from the point of view of a classical ``hard'' detector. Interactions with other particles or with an external magnetic field can distinguish two beams in the example above (e.g., they will deflect differently via spin-orbit interactions with impurities). For this reason, in spintronics it is often necessary to retain the idea of a beam as a collection of particles with certain wave functions.

\section{Spin current}\

In condensed matter physics, two types of electron currents are distinguished, corresponding to the charge and spin of an electron: \emph{electric current} and \emph{spin current}. Significant progress in the fabrication of nanomaterials allowed to control the spin currents, despite the relatively short spin coherence times $\tau_{sc}$.

The spin current density operator in the three-dimensional case reads
\begin{equation}
\hat{j}^s_{\alpha \beta} 
= 
(\hbar/2) \hat{v}_{\alpha} \hat{\sigma}_{\beta} 
,
\end{equation}
where $ \hat{\mathbf{v}} $ is the electron velocity operator.
The spin current tensor is calculated as the average \eqref{rho_averaging} for a known density matrix in spin indices, with component $ j^s_{\alpha \beta}  $ being equal to the mean value of the projection of the spin density in the direction $ \mathbf{e}_\beta $, flowing per unit time through a unit area perpendicular to the direction $ \mathbf{e}_\alpha $.

In spintronics, nanostructures are often investigated in which the effects of spin coherence can be very important. This reflects the significant difference between charge transport and spin transport: the electron charge in solid-state physics does not directly correspond to any part of electron wave function,
while the electron spin corresponds to spinor part of a wavefunction, the coherence of which is lost at times $ \tau_{sc} $. Perhaps the most impressive effect where spin coherence lies at the very foundation is the ``magnetoelectronic spin echo'' \cite{Magnetoelectronic Spin Echo}. In this effect, the coherent spin current passes through a metallic magnet, then flows in a non-magnetic region, and then passes through another ferromagnet with magnetization opposite to the first ferromagnet. Finally, the spin current is injected into a Pt spin-Hall bar, where a voltage is registered, indicating that the spin current retained its coherence in the process to a great degree, and has reverted nearly to its original state. 
In the general case, it is necessary to use the approaches of semiclassical coherent transport, such as the method of (matrix in spin indices) Green functions \cite{Nazarov book}.

In some cases, the transport of spin-polarized electrons can be described in the classical kinetic Boltzmann approach (e.g., using the phenomenological diffusion equation for the spin density), \emph{however, in this case, the spin coherence will be completely lost.} In other words, the spin density in this approach can only irreversibly decrease or diffuse.
The continuity equation for the nondissipative approximation of charge transport is
\begin{equation}
\frac{\partial \rho}{\partial t}=-\nabla \mathbf{j}_c,
\end{equation}
where $\rho$ is the volume density of electric current, $\mathbf{j}_c$ is the current density. 
By Gauss theorem,
\begin{equation}
\iiint_V \frac{\partial \rho}{\partial t}\, d^3 \mathbf{r}=-\iint_S \mathbf{j}_c \cdot d \mathbf{S}.
\end{equation}
In classical incoherent approximation, spin current $\mathbf{j}_s$ could be defined by analogy from
\begin{equation}
\iiint_V \frac{\partial \boldsymbol{\mathcal{S}}}{\partial t}\, d^3 \mathbf{r}=-\iint_S \mathbf{j}_s \cdot d \mathbf{S},
\end{equation}
where $ \boldsymbol{\mathcal{S}} $ is the local spin density, and spin current $\mathbf{j}_s$ is a \emph{second-rank tensor}.
In local formulation
\begin{equation}
\label{js_eqn}
\frac{\partial \boldsymbol{\mathcal{S}}}{\partial t}=-\nabla \mathbf{j}_s
.
\end{equation}

\subsection{Spin relaxation}\

In reality, the spin current is not conserved.
Conduction electron spin in metals and semiconductors can relax due to scattering by impurities with potentials that do not commute with the spin density operator, such as \emph{random magnetic field} or \emph{spin-orbit interaction} with randomly distributed non-magnetic impurities \cite{TseBra 2005}.
Spin-dependent scattering plays a very important role in spintronics: it is thanks to it that the spin current can be measured from the induced potential difference at the boundaries of the sample (inverse spin-Hall effect (\textbf{ISHE}), see lecture \ref{SHE_chapter}) \cite{SHE_Sinova}.
Therefore, heavy metals with large spin-orbit interaction (platinum $ Pt $, thallium $ Ta $) are usually used for ``spin current detectors'', and materials with the lowest possible spin relaxation rate are usually chosen as ``spin-conducting'' structures. 
Recently, much attention has been drawn to the method of spin current transfer through ferromagnetic insulators (e.g. YIG) in the form of an \emph{exchange spin current}, the coherence times of which are much larger than those of conduction electron spin current (see section \ref{exchange spc}).

In the classical approach, spin relaxation \emph{in a paramagnetic metal} can be accounted for (in the $ \tau $-approximation of the Boltzmann equation) by adding the relaxation term $ \mathbf{T} $ to the right-hand side of \eqref{js_eqn}:
\begin{eqnarray}
\label{js_eqn+T}
&&\frac{\partial \boldsymbol{\mathcal{S}}}{\partial t}
=
-\nabla \mathbf{j}_s + \mathbf{T}, \\
\label{T}
&&\mathbf{T}
=
-(\boldsymbol{\mathcal{S}}-\boldsymbol{\mathcal{S}}_0)/\tau_s,
\end{eqnarray}
where  $ 1 / \tau_s $ is the spin relaxation rate, $ \delta\boldsymbol{\mathcal{S}} = \boldsymbol{\mathcal{S}}-\boldsymbol{\mathcal{S}}_0 $ is the non-equilibrium spin density (see \cite{Gurevich}, p. 41). 
Anisotropic generalization of relaxation times corresponding to different components of $ \delta\boldsymbol{\mathcal{S}} $ is possible: for example, one can introduce two relaxation times: $ \tau_{s\parallel} $ for component $ \delta\boldsymbol{\mathcal{S}}_\parallel $, parallel to $ \boldsymbol{\mathcal{S}}_0 $, and $ \tau_{s\perp} $ for component $ \delta\boldsymbol{\mathcal{S}}_\perp $ perpendicular to $ \boldsymbol{\mathcal{S}}_0 $ (see also section \ref{Bloch_paramagnet_relaxation}).



\section{Addendum to Lecture 1: Properties of Pauli Matrices}

\begin{eqnarray}
&&
\hat{\sigma}_0=
\begin{pmatrix}
1 & 0 \\
0 & 1
\end{pmatrix}, \nonumber \\
&&
\hat{\sigma}_x=
\begin{pmatrix}
0 & 1 \\
1 & 0
\end{pmatrix}, \quad
\hat{\sigma}_y=
\begin{pmatrix}
0 & -i \\
i & 0
\end{pmatrix}, \quad
\hat{\sigma}_z=
\begin{pmatrix}
1 & 0 \\
0 & -1
\end{pmatrix}
\\
&& 
\hat{\sigma}_\alpha \hat{\sigma}_\beta=i\varepsilon_{\alpha\beta\gamma} \hat{\sigma}_\gamma + \delta_{\alpha\beta}\hat{\sigma}_0 \\
&&
\left[ \hat{\sigma}_\alpha,\hat{\sigma}_\beta \right]=2i \varepsilon_{\alpha\beta\gamma} \hat{\sigma}_\gamma \\
&&
\left\{ \hat{\sigma}_\alpha,\hat{\sigma}_\beta \right\}=2 \delta_{\alpha\beta}\hat{\sigma}_0 \\
&&
\hat{\boldsymbol{\sigma}}^2=\hat{\sigma}_x^2 + \hat{\sigma}_y^2 + \hat{\sigma}_z^2 = 3 \hat{\sigma}_0 \\
&&
\mathbf{Tr}(\hat{\sigma}_i \hat{\sigma}_j) = 2 \delta_{ij} \\
\nonumber
\end{eqnarray}

\bigskip
\bigskip
\begin{center}
	\uppercase{Control questions}
\end{center}
\noindent 1. Check that \eqref{theta_phi_eigenspinors} are indeed the eigenspinors of the spin projection operator $ \hat{s}_l $ on an arbitrary axis $ \mathbf{l} $.

\noindent 2. Check that the representation \eqref{rho as P sigma} is valid: i.e. show that $ \langle \hat{\boldsymbol{\sigma}} \rangle
=
\textbf{Tr}(\hat{\rho} \hat{\boldsymbol{\sigma}}) = \mathbf{P} $.

\noindent 3. Which subsystem the angular momentum of spin current can be transferred to: in a ferromagnet; in a paramagnet?

\chapter{Concepts from the electrodynamics of continuous media}

\section{Micro- and macro-fields. Polarization and magnetization\footnotemark}\

\footnotetext{In this section, we follow the book \cite{Turov}.}

This section summarizes the basic equations of electrodynamics of continuous media \cite{Turov, LL9}. We will return to them in the derivation of dispersion law for spin waves.

The \emph{microscopic charge density} and \emph{microscopic current density} associated with the system of particles read
\begin{eqnarray}
\rho_\mu(\mathbf{r})=\sum_{n} e_n \delta(\mathbf{r}-\mathbf{r}_n), \\
\label{jd}
\mathbf{j}_\mu(\mathbf{r})=\sum_{n} e_n \mathbf{v}_n \delta(\mathbf{r}-\mathbf{r}_n),
\end{eqnarray}
where $e_n$ and $\mathbf{v}_n$ are the charge and velocity of the $ n $th particle, respectively.
\emph{Maxwell-Lorentz equations} in vacuum relate the \textit{microscopic fields} $\mathbf{E}_\mu$ and $\mathbf{B}_\mu$ with $\rho_\mu$ and $\mathbf{j}_\mu$:
\begin{eqnarray}
\label{meq1}
\nabla \times \mathbf{E}_\mu &=& -\frac{1}{c} \frac{\partial \mathbf{B}_\mu}{\partial t}, \\
\label{meq2}
\nabla \times \mathbf{B}_\mu &=& \frac{1}{c} \frac{\partial \mathbf{E}_\mu}{\partial t} + \frac{4 \pi}{c} \mathbf{j}_\mu, \\
\label{meq3}
\nabla \cdot \mathbf{E}_\mu &=& 4\pi \rho_\mu, \\
\label{meq4}
\nabla \cdot \mathbf{B}_\mu &=& 0.
\end{eqnarray}
The Maxwell-Lorentz equations should be supplemented with an expression for the Lorentz force acting on point charges $e_n$
\begin{equation}
\mathbf{f}_n=e_n \left( \mathbf{E}_{\mu} + \frac{1}{c} \mathbf{v}_n \times \mathbf{B}_\mu \right),
\end{equation}
and the corresponding equations of motion
\begin{equation}
m_n \frac{d\mathbf{v}_n}{dt} = \mathbf{f}_n.
\end{equation}

The transition from macroscopic Maxwell-Lorentz equations to \emph{equations of macroscopic electrodynamics}, and from the equations of motion for the charges to \emph{constitutive relations} is carried out by averaging over physically infinitesimal volumes $ \Delta V $ of medium and over physically infinitesimal times $ \Delta t $, which both depend on the specific microscopic model of the substance. As a result of averaging, fast spatial and temporal field oscillations, which are not recorded by macroscopic instruments and are not of physical interest in most cases, are smoothed out.
The averaged form of equations (\ref{meq1}-\ref{meq4}) is obtained by substituting $\rho_\mu$ and $\mathbf{j}_\mu$ by the averaged $\left< \rho_\mu \right>$ and $\left< \mathbf{j}_\mu \right>$ and fields $\mathbf{E}_\mu$ and $\mathbf{B}_\mu$ by macro-fields $\mathbf{E}$ and $\mathbf{B}$.

Polarization $\mathbf{P}$ (dipole moment per unit volume) of substance is defined as
\begin{equation}
\int \mathbf{P} d V = \int \mathbf{r} \left< \rho_\mu \right> dV,
\end{equation}
and the magnetization $\mathbf{M}$ (magnetic moment per unit volume) of substance~-- as
\begin{equation}
\int \mathbf{M} d V = \int \mathbf{r} \times \left< \mathbf{j}_\mu \right> dV.
\end{equation}
When the condition of electroneutrality is satisfied
\begin{equation}
\int \left< \rho_\mu \right> dV = 0,
\end{equation}
one can express $\left< \rho_\mu \right>$ in terms of polarization $\mathbf{P}$:
\begin{equation}
\left< \rho_\mu \right> = -\nabla \cdot \mathbf{P},
\end{equation}
and when equality $\frac{\partial \mathbf{P}}{\partial t}=0$ is satisfied, one can express $\left< \mathbf{j}_\mu \right>$ in terms of the magnetization $\mathbf{M}$:
\begin{equation}
\left< \mathbf{j}_\mu \right> = c \, \nabla \times \mathbf{M}.
\end{equation}

If particle spin magnetic moments $\boldsymbol{\mu}_{n}$ contribute to magnetization, an additional \textit{spin current} with density
\begin{equation}
\label{spin current}
\mathbf{j}_s(\mathbf{r})=c \nabla \times \sum_{n} \boldsymbol{\mu}_n \delta(\mathbf{r}-\mathbf{r}_n)
\end{equation}
could be introduced into the microscopic current density \eqref{jd}.

Finally, assuming that the system may additionally contain external charges with volume density $ \rho_0 $ and the flow of external currents with density $ \mathbf{J}_0 $, we obtain the classical Maxwell macroscopic equations:
\begin{eqnarray}
\label{m1}
\nabla \times \mathbf{E} &=& -\frac{1}{c} \frac{\partial \mathbf{B}}{\partial t}, \\
\label{m2}
\nabla \times \mathbf{H} &=& \frac{1}{c} \frac{\partial \mathbf{D}}{\partial t} + \frac{4 \pi}{c} \mathbf{J}_0, \\
\label{m3}
\nabla \cdot \mathbf{D} &=& 4\pi \rho_0, \\
\label{m4}
\nabla \cdot \mathbf{B} &=& 0,
\end{eqnarray}
where new quantities are introduced: \emph{electric induction} vector $\mathbf{D} = \mathbf{E} + 4\pi \mathbf{P}$ and \emph{magnetic field strength} vector $\mathbf{H} = \mathbf{B} - 4\pi \mathbf{M}$.

The constitutive relations connect the vectors $ \mathbf{P} $ and $ \mathbf{M} $ with the fields $ \mathbf{E} $ and $ \mathbf{H} $.
For weak fields, these relations are linear:
\begin{eqnarray}
P_i=\alpha_{ij} E_j, \\
M_i=\chi_{ij} H_j,
\end{eqnarray}
where the quantities $\alpha_{ij}$ and $\chi_{ij}$ are called, respectively, tensors of \emph{dielectric} and \emph{magnetic susceptibility}.

More generally, the vectors $\mathbf{P},\,\mathbf{M}$ are nonlinear functions of the fields $\mathbf{E},\,\mathbf{H}$ and their time and spatial derivatives. 
Terms which are linear in time and spatial derivatives correspond to the \emph{temporal} and \emph {spatial dispersion} of the medium. When considering the effects of spatial dispersion, the scale of the field inhomogeneity (wavelength, penetration depth) is compared with a certain characteristic correlation length in the substance~-- the size of the neighborhood in which the particles ``feel'' the action of the field in the center of the neighborhood \cite{Turov} (interatomic distance, mean free path, coherence length in a superconductor). Similarly, when considering time dispersion, the characteristic frequencies of the field are compared either with the characteristic natural frequencies of waves in a substance, or with some characteristic wave relaxation rates in the substance.

\section{Static susceptibilities} \

The magnetic susceptibility of a medium describes its linear response to an external magnetic field.
This section briefly outlines classical approaches to the static susceptibilities of independent atoms, paramagnets, and ferromagnets. In the following sections, the susceptibilities will be generalized to the case of an alternating high-frequency external magnetic field.

\subsection{Free atom\footnotemark}\

\footnotetext{In this section, we follow §112 in \cite{LL3}.}

Consider an atom in a uniform magnetic field $ \mathbf{H} $ with a vector potential
\begin{equation}
\mathbf{A}=\frac{1}{2} \left[\mathbf{H}\times\mathbf{r}\right].
\end{equation}
Let us assume that the intra-atomic magnetic fields corresponding to spin-orbital and spin-spin interactions are small. In the Pauli Hamiltonian for the atomic electron \eqref{PH}, $ \mathbf{A}_a $ is the vector potential of the magnetic field at point $\mathbf{r}_a $. The sum of Hamiltonians \eqref{PH} (i.e., the atomic Hamiltonian) then takes the form:
\begin{equation}
\label{PH++}
\hat{H}=\hat{H}_0+\mu_B \left( \hat{\mathbf{L}} + 2 \hat{\mathbf{S}} \right)\mathbf{H} + \frac{e^2}{8mc^2} \sum_a \left[\mathbf{H}\times\mathbf{r}_a\right]^2,
\end{equation}
where $\hat{H}_0$ is the atomic Hamiltonian at $ \mathbf{H} = 0 $, and the operators of the total angular momentum and total spin of the atom, $ \hat{\mathbf{L}} $ and $ \hat{\mathbf{S}} $, are normalized to $ \hbar $. Operator
\begin{equation}
\label{hat_mu_at}
\hat{\boldsymbol{\mu}}_{\text{at}}=-\mu_B \left( \hat{\mathbf{L}} + 2 \hat{\mathbf{S}} \right)
\end{equation}
then corresponds to the magnetic moment of the atom.

Let the field $ H $ be weak enough so that $ \mu_B H $ is much smaller than the energy intervals of atomic fine structure. Then the second and third terms on the right-hand side of \eqref{PH++} can be regarded as a perturbation (with the second term dominating). In the vector model of atom (see, e.g., \cite{Shpolsky}, \S72), it can be shown that the linear Zeeman splitting of energy levels then has the form
\begin{equation}
\Delta E = \mu_0 g M_J H,\qquad M_J=-J,\,-J+1,\,...,J,
\end{equation}
where the Land\'{e} factor $g$ reads
\begin{equation}
g=1+\frac{J(J+1)-L(L+1)+S(S+1)}{2J(J+1)},
\end{equation}
and $ M_J $ is the magnetic quantum number.
Since $-\partial\Delta E/\partial H$ is the average value of the magnetic moment of the atom, an atom in a state with a certain value of $ M_J $ then has an average magnetic moment
\begin{equation}
\label{mu_z}
\mu_z=-\mu_B g M_J
\end{equation}
along the magnetic field direction.
Moreover, within the vector model of the atom one can also show that the mean square of atomic magnetic moment reads
\begin{equation}
\label{mu_sq}
\mu^2=\mu^2_B g^2 J (J+1).
\end{equation}

Consider the case when the Zeeman splitting associated with the second term on the right-hand side of \eqref{PH++} vanishes in the first-order approximation. Furthermore, let the atom have neither spin nor orbital angular momentum: $S=L=0$ (then $\mu_{\text{at}}=0$). In this case, the contribution from the second term on the right-hand side \eqref{PH++} in any order of the perturbation theory vanishes, and the third term gives
\begin{equation}
\Delta E = \frac{e^2}{8mc^2} \sum_a \overline{\left[\mathbf{H}\times\mathbf{r}_a\right]^2},
\end{equation}
where the overline denotes the mean value in the state with $S=L=0$. Since $\left[\mathbf{H}\times\mathbf{r}_a\right]^2=H^2 r^2_a \sin^2\theta$, where $\theta$ is te angle between $\mathbf{H}$ and $\mathbf{r}_a$, and since $\overline{\sin^2\theta}=2/3$ for spherically symmetric wave functions, we have
\begin{equation}
\Delta E = \frac{e^2}{12mc^2} H^2 \sum_a \overline{\mathbf{r}_a^2}.
\end{equation}
The magnetic moment of the atom is $-\partial\Delta E/\partial H$. Writing it out as $ \chi H $, the magnetic susceptibility of an atom reads
\begin{equation}
\label{chi}
\chi = -\frac{e^2}{6mc^2} H^2 \sum_a \overline{\mathbf{r}_a^2}
\end{equation}
Since in this case $\chi<0$, then the atom with $S=L=0$ is \emph {diamagnetic}. Note that \eqref{chi} does not contain the Planck constant $ \hbar $, which indicates the possibility of classical derivation from classical laws of electromagnetic induction.

There is one more remarkable case when there is no level shift linear in the field: $J=0,\,S=L\neq0$. In this case, the contribution from the second term on the right-hand side of \eqref{PH++}, as a second-order perturbation, prevails over the first-order diamagnetic contribution from the third term.
Since the second-order correction to the normal state of the atom is always negative, the atom in a state $J=0,\,S=L\neq0$ \emph{is paramagnetic}. This result is known as \emph{Van Vleck's polarization paramagnetism} of free atoms, and is caused by the deformation of electron cloud in magnetic field \cite{VanFl}.



\subsection{Langevin paramagnetism}\

Consider a system of identical weakly interacting atoms in a weak magnetic field $\mathbf{H}_0$ directed along $z$ axis, so that the interaction energy of each atom with magnetic field reads\footnote[1]{\label{rrr} see \eqref{mu_z}, \eqref{mu_sq}; accordingly, linearity of Zeeman effect is assumed.}
\begin{equation}
\label{asympt}
\boldsymbol{\mu}\cdot\mathbf{H}_0=\mu_z H_0 \ll k T,
\end{equation}
where $T$ is the temperature. 
In addition, we assume that the atoms are in the $J\neq 0$ state, so that the effects of atomic diamagnetism and polarization paramagnetism make only a small contribution to the magnetization. It can be shown (for the derivation, we refer the reader to the book~\cite{Smart}) that the temperature dependence of the sample magnetization $ M $ in the entire temperature range is described by a function of the form:
\begin{equation}
\label{MBrill}
M=n g \mu_0 J B_J(x),
\end{equation}
where $n$ i the volume concentration of atoms in a sample, parameter $x=g \mu_0 J H_0/k T$, and $B_J(x)$ is the Brillouin function:
\begin{equation}
\label{BrillFunc}
B_J(x)=\frac{2J+1}{2J}\coth\left(\frac{2J+1}{2J} x\right)
-\frac{1}{2J}\coth\left(\frac{x}{2J}\right).
\end{equation}
Considering the asymptotics \eqref{asympt} corresponding to small $x\ll 1$, we have \[ B_J(x)\simeq \frac{J+1}{3J}x, \]
and for the volume susceptibility $\chi_0=M/H_0$ we obtain the Curie law:
\begin{equation}
\label{chi_lang}
\chi_0=\frac{n\mu^2}{3k T},
\end{equation}
where the previous expression \eqref{mu_sq} is kept for $\mu^2$.

Exactly the same result was obtained by P. Langevin in the framework of the classical approach, which, of course, did not take into account spatial quantization. The classical case can be obtained by passing to the limit $J\rightarrow\infty,\, \mu_0\rightarrow0,\,\lim_{J\rightarrow\infty,\,\mu_0\rightarrow0}=\mu$, where $\mu$ is now the ``classical'' magnetic moment of the atom.

In sufficiently strong magnetic fields ($\mu_z H_0 \gg T$),
full orientation of the moments is possible. The saturation of the magnetization is described by the exact formula \eqref{MBrill}\footnotemark.
\footnotetext{
	Further refinements of the theory are associated with taking into account \cite{Smart}: (1) the contribution of excited levels to the susceptibility; (2) the interaction of paramagnetic atoms with diamagnetic ones in the case of a complex lattice; (3) exchange and magnetic-dipole interactions of paramagnetic atoms. Although these effects can be small corrections to the susceptibility in an external static field, they can play a significant role in the case of an alternating field. For example, in the theoretical consideration of EPR resonance lines, it turns out that corrections for the so-called exchange narrowing of the resonance lines should be introduced (see, e.g., \cite{Kittel}, p. 624).
}

\subsection{Weiss ferromagnetism \label{FM}}\

Ferromagnets are substances with \emph{spontaneous} (i.e. nonzero in the absence of external magnetic field) magnetic moment.
Gyromagnetic experiments have shown that the magnetic moment of ferromagnets is almost exclusively due to the orientation of the \emph{spin} moments of atoms. 
Consequently, some ``molecular'' forces of an electromagnetic nature, which are responsible for the spontaneous orientation of atomic spins, must act in ferromagnets.

According to the phenomenological theory of P. E. Weiss, called the \emph{molecular field approximation} (1907), ferromagnetism can be explained by phenomenologically introducing some \emph{effective magnetic field}
\begin{equation}
\label{eff_field}
\mathbf{H}_{eff}=\lambda_w \mathbf{M},
\end{equation}
which describes the effect of forces which atoms exert on the magnetic moment of a certain atom that they surround, and is proportional to the magnetization of the substance $ \mathbf{M} $; $ \lambda_w $ is some molecular field constant.

The effective magnetic field is associated with the minimum energy of \emph{exchange interaction} of atoms of a ferromagnet for a state with parallel arrangement of atomic spins. Exchange interaction is, in turn, an effective interaction describing the dependence of electrostatic energy of a system of atomic electrons on the total spin of the system (see, e.g., \cite {LL3}, \S62).
In microscopic theory, the expression for $ \mathbf{H}_{eff} $ can be obtained from the Heisenberg exchange Hamiltonian of the form
\begin{equation}
\mathcal{H}_{ex}=-\sum_{i>j,j} J_{ij} \mathbf{S}^z_i\cdot \mathbf{S}^z_j ,
\end{equation}
where summation is carries out by all electron pairs, $\mathbf{S}^z_i$ are $z$-projections of atomic spins, and $J_{ij} > 0$ are the exchange integrals (see e.g. \cite{HS}, chapter 5, \S5). In the self-consistent field approximation, the expression for $\mathcal{H}_{ex}$ takes the form
\begin{equation}
\mathcal{H}_{ex}=-\sum_{k} \mathbf{S}^z_k \left(\sum_{i\neq k} J_{ik} \left\langle\mathbf{S}^z_i\right\rangle\right),
\end{equation}
which has the structure of total interaction energy of atomic magnetic moments $g\mu_0\mathbf{S}^z_k$ (cf. \eqref{mu_z} with $M_j=S_z$) in an effective field
\begin{equation}
\label{Heff_micro}
\mathbf{H}_{eff}=\frac{1}{g\mu_B}\sum_{j\neq i} J_{ij} \left\langle\mathbf{S}^z_j\right\rangle,
\end{equation}
where $\left\langle\mathbf{S}^z_j\right\rangle$ is the average $z$-projection of the spin of a $\mathbf{S}^z_j$ $j$-th atom. 
In the equilibrium case, the values $\left\langle\mathbf{S}^z_j\right\rangle$ are constant, and then the expression \eqref{Heff_micro} turns into \eqref{eff_field}, where magnetization reads $\mathbf{M}=\mathbf{M}_z=g\mu_B\left\langle\mathbf{S}^z_j\right\rangle/\Omega_0$ ($\Omega_0$ is the crystal unit cell volume), and the phenomenological parameter has the form $\lambda_w=\left(\Omega_0/(g\mu_B)^2\right)\sum_{j\neq i} J_{ij}$.
The short-range nature of the exchange interaction of atoms corresponds to a rapid decrease of exchange integrals $J_{ij}$ with an increase in the distance between the atoms $ i $ and $ j $; accordingly, the main contribution to the sum \eqref{Heff_micro} will be made by the interactions of the nearest atoms\footnotemark\footnotetext{A similar property is inherent in the coefficients of the dynamic force matrix in the theory of crystal lattice vibrations~\cite{Kosevich}.}.

\begin{figure}[h!]
	\centering
	\includegraphics [width=0.35\textwidth]{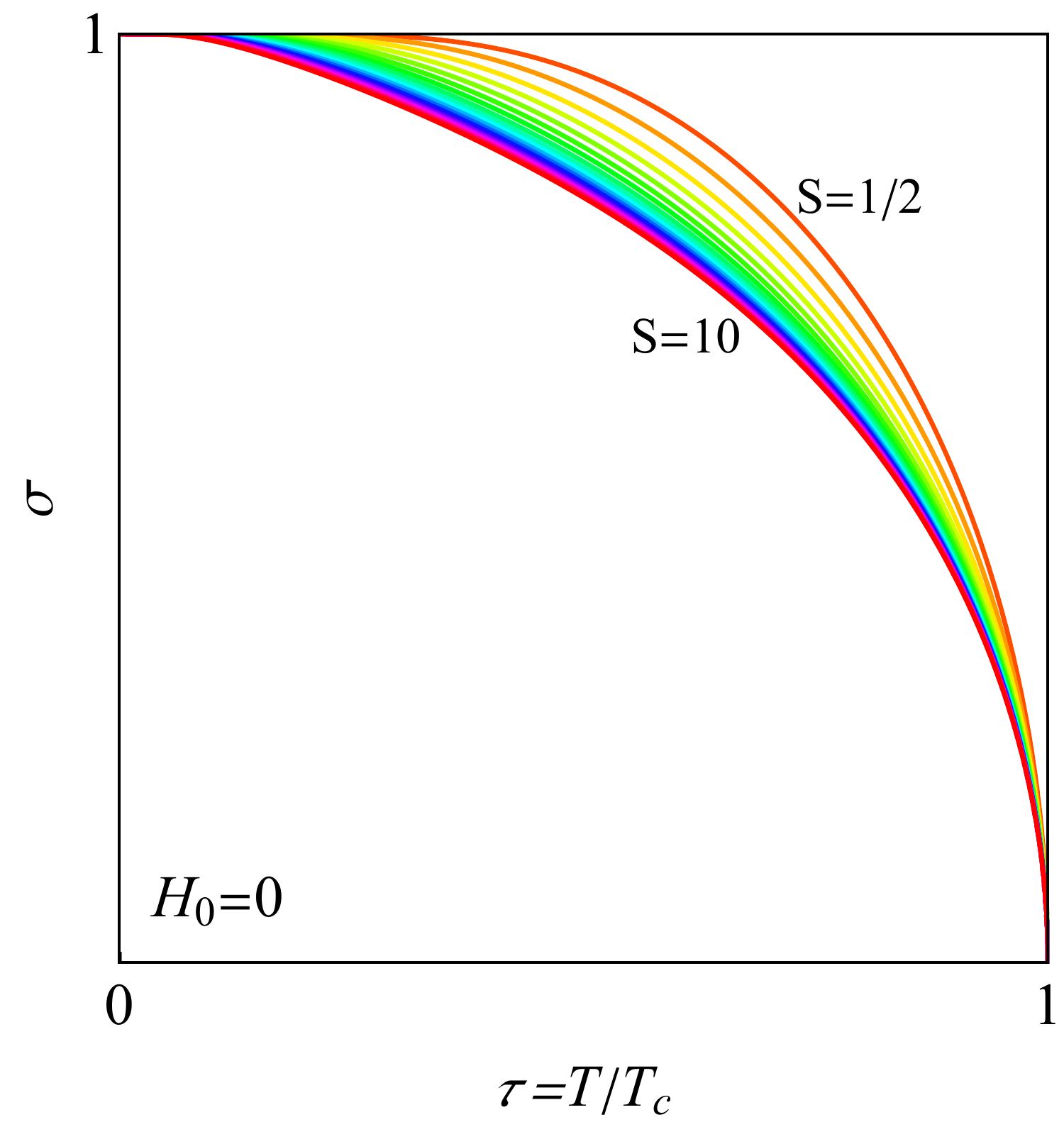}
	\includegraphics [width=0.35\textwidth]{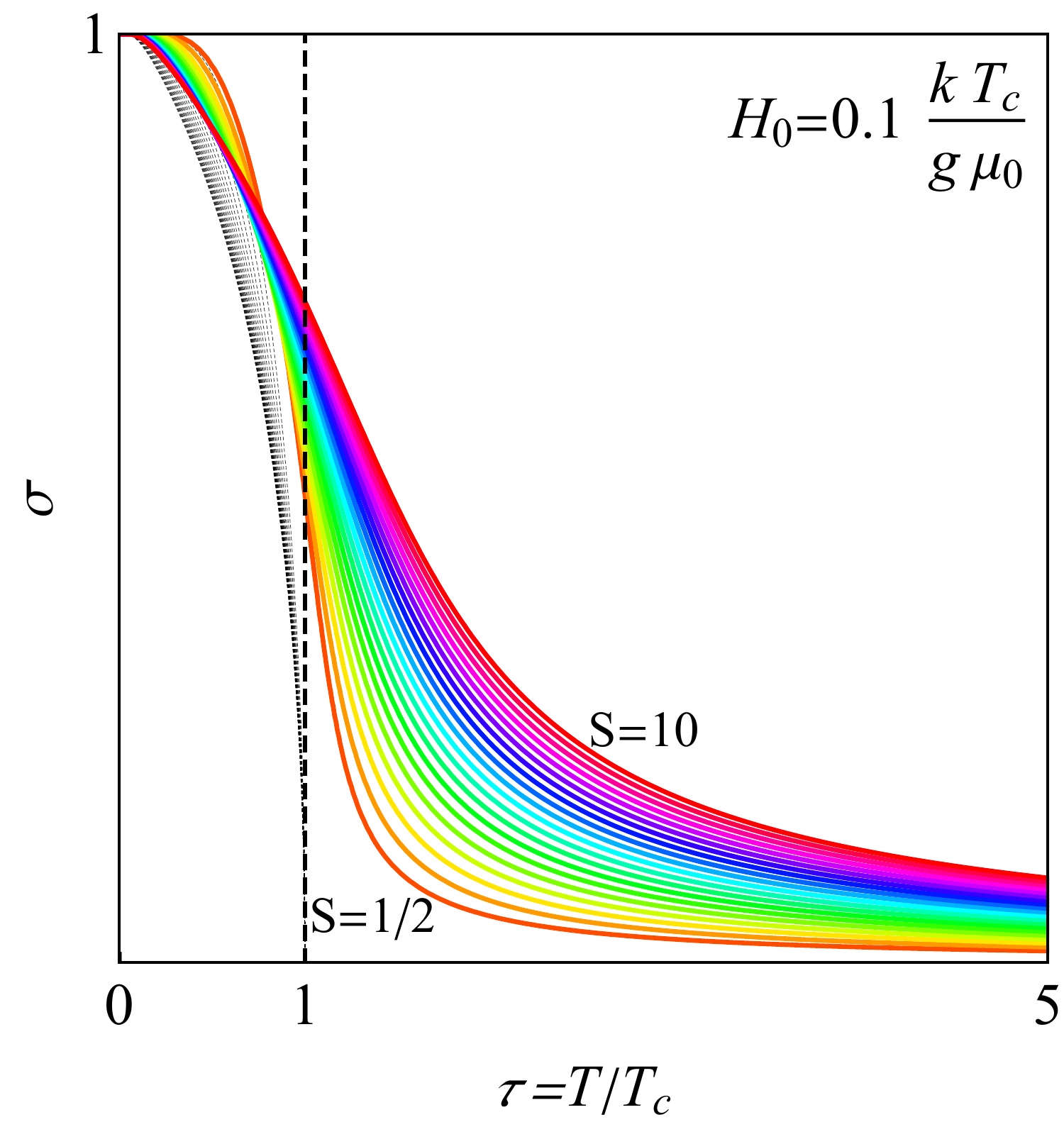}
	\caption{Dependencies of the relative magnetization $\sigma=M/Ng\mu_0 S$ of a ferromagnet on the dimensionless temperature $\tau=T/T_c$ in the Weiss molecular field theory. The curves are plotted for different atomic spins $S=1/2,1,3/2,...,10$. \emph{Left panel}: in the absence of an external field, $ H_0 = 0 $. \emph{Right panel}: in an external field $H_0=0.1\,k T_c/ g \mu_0$ (coloured solid lines), and at $ H_0 = 0 $ (gray dashed lines).}\label{CL}
\end{figure}

To calculate the dependence of the ferromagnet magnetization on the temperature and applied external field, one can use the Langevin theory of paramagnetism with the field $\mathbf{H}_0+\mathbf{H}_{eff}$, taking into account the Weiss molecular field. The main results~\cite{Smart} are as follows:

1) In the absence of external field ($ H_0 = 0 $): a) at temperatures $ T $ lower than \emph{Curie temperature} $ T_c $, the ferromagnet has spontaneous magnetization, which decreases with increasing temperature due to thermal suppression of the effects of exchange interaction; b) at $ T > T_c $ there is no spontaneous magnetization. The corresponding dependencies are shown on the left panel in Fig. \ref{CL}.

2) In an external field ($ H_0 \neq 0 $): the spontaneous magnetization is not completely destroyed at $ T = T_c $, but for usual values of atomic spin ($ S < 5 $) it rapidly decreases at $ T \gtrsim T_c $, so that at high temperatures $ k T \gg g \mu_B H_0 S $ the Curie-Weiss law is valid for the volume susceptibility $ \chi = M / H_0 $:
\begin{equation}
\chi=\frac{n\mu^2}{3k} \frac{1}{T-T_c},
\end{equation}
where $\mu=g\mu_B \sqrt{S(S+1)}$. The indicated dependencies are shown in the right panel in Fig. \ref{CL}.
The Curie temperature contains only the fundamental parameters that characterize the magnet.
The result for the simplest model of nearest-neighbor interaction reads \cite{Smart}:
\begin{equation}
T_c=\frac{2 z \mathcal{J} S(S+1)}{3k},
\end{equation}
where $z$ is the the number of nearest neighbors of a magnetic atom (coordination number), and $ \mathcal{J} $ is the value of the nearest-neighbor exchange integral.
Comparison of the obtained law with a similar result for paramagnetism \eqref{chi_lang} implies that \textit{at high temperatures a ferromagnet becomes a paramagnet}.

\bigskip
\bigskip
\begin{center}
	\uppercase{Control questions}
\end{center}
\noindent 1. Why can the operator \eqref{hat_mu_at} be regarded as atomic magnetic moment operator? (Consider the analogy with Pauli equation for a single \textit{electron} \eqref{PH})

\noindent 2. What is the physical nature of the exchange field \eqref{eff_field} in ferromagnets? How do the microscopic parameters of a system of atoms determine the strength of this field?

\noindent 3. What state does a ferromagnet transition into above its Curie temperature?

\setchapterpreamble{%
	\dictum{%
		
		This lecture discusses the high-frequency susceptibilities of para- and ferromagnets in an external alternating magnetic field, as well as their temporal and spatial dispersion.
	}%
	\vspace{24pt}%
}

\chapter{Magnetic media in alternating external magnetic field}\

\section{Electron spin resonance (ESR) and its temporal dispersion\footnotemark}\
\footnotetext{This section is based on the derivation scheme proposed in \S11.1 of the book \cite{Turov}.}

In paramagnetic media, a phenomenon of resonant absorption of magnetic field energy occurs at the frequency of Larmor precession of individual magnetic moments that make up the medium. This phenomenon is usually called the electron spin resonance, ESR (or, sometimes, the electron paramagnetic resonance, EPR). In this section, the high-frequency susceptibility of a paramagnet will be derived; its poles will give the resonant frequencies.

The atomic magnetic dipoles constituting a paramagnet have magnetic moments $ \boldsymbol{\mu}_i $ associated with the total mechanical moments of atoms $ \mathbf{I}_i $ by means of the \emph{gyromagnetic constant} $ \gamma $: $ \boldsymbol{\mu}_i = \gamma \mathbf{I}_i $.
In the macroscopic approach, magnetization is the magnetic moment of a unit volume of a magnet, $\mathbf{M}=\sum_{i}^{}\boldsymbol{\mu}_i=\gamma\mathbf{J}$, where $\mathbf{J}=\sum_{i}^{}\mathbf{I}_i$ is the total excess atomic mechanical moment of a unit volume.

Consider a system of magnetic dipoles in an external uniform magnetic field $\mathbf{H}=\mathbf{H}_0+\mathbf{h}(t)$, where $\mathbf{H}_0$ is a constant field directed along $ z $ axis, and $ \mathbf{h}(t) $ is a weak ($h\ll H_0$) variable field with an arbitrary direction. We will seek a solution to the equations of magnetization dynamics as the sum of unperturbed stationary part in the field $ \mathbf{H}_0 $ and a small correction associated with the perturbation $ \mathbf{h}(t) $.
Let the susceptibility in the constant field $ \mathbf{H}_0 $ be
$\chi_0$ \eqref{chi_lang}:
\begin{equation}
\label{M0}
\mathbf{M}_0=\chi_0\mathbf{H}_0.
\end{equation}
In an additional weak external alternating field $\mathbf{h}(t)=\mathbf{h}_{0}e^{-i\omega t}$ magnetization $\mathbf{M}(t)$ reads
\begin{equation}
\label{Mt}
\mathbf{M}(t)=\mathbf{M}_0+\mathbf{m}(t),
\end{equation}
and our task is to find a small correction $\mathbf{m}(t)$, which is the response to perturbation $ \mathbf{h}(t) $.

Based on the equation of motion for the mechanical moment of the magnetic dipole (the rate of change of the mechanical moment is equal to the acting torque \cite{LL1}), we heuristically obtain the \textit{equation of magnetization dynamics} 
\begin{equation}
\label{M-time-evol}
\frac{\partial\mathbf{M}(t)}{\partial t}=\gamma[\mathbf{M}\times\mathbf{H}(t)],
\end{equation}
as the sum of equations for individual dipoles (using $\mathbf{I}=\boldsymbol{\mu}/\gamma$).
This \emph{classical} derivation (according to Kittel) corresponds to the simplest model without dipole-dipole interaction, anisotropy field and relaxation processes. In more general models, the equation is preserved if the external magnetic field $\mathbf{H}(t)$ is replaced by an effective field $\mathbf{H}^*=-\delta E / \delta \mathbf{M}$ (functional derivative of magnetic energy density by magnetic moment per unit volume; for the derivation see e.g.~\cite{LL9, KosKov}).

\begin{figure}
	\begin{center}
		\includegraphics[width=0.2\textwidth]{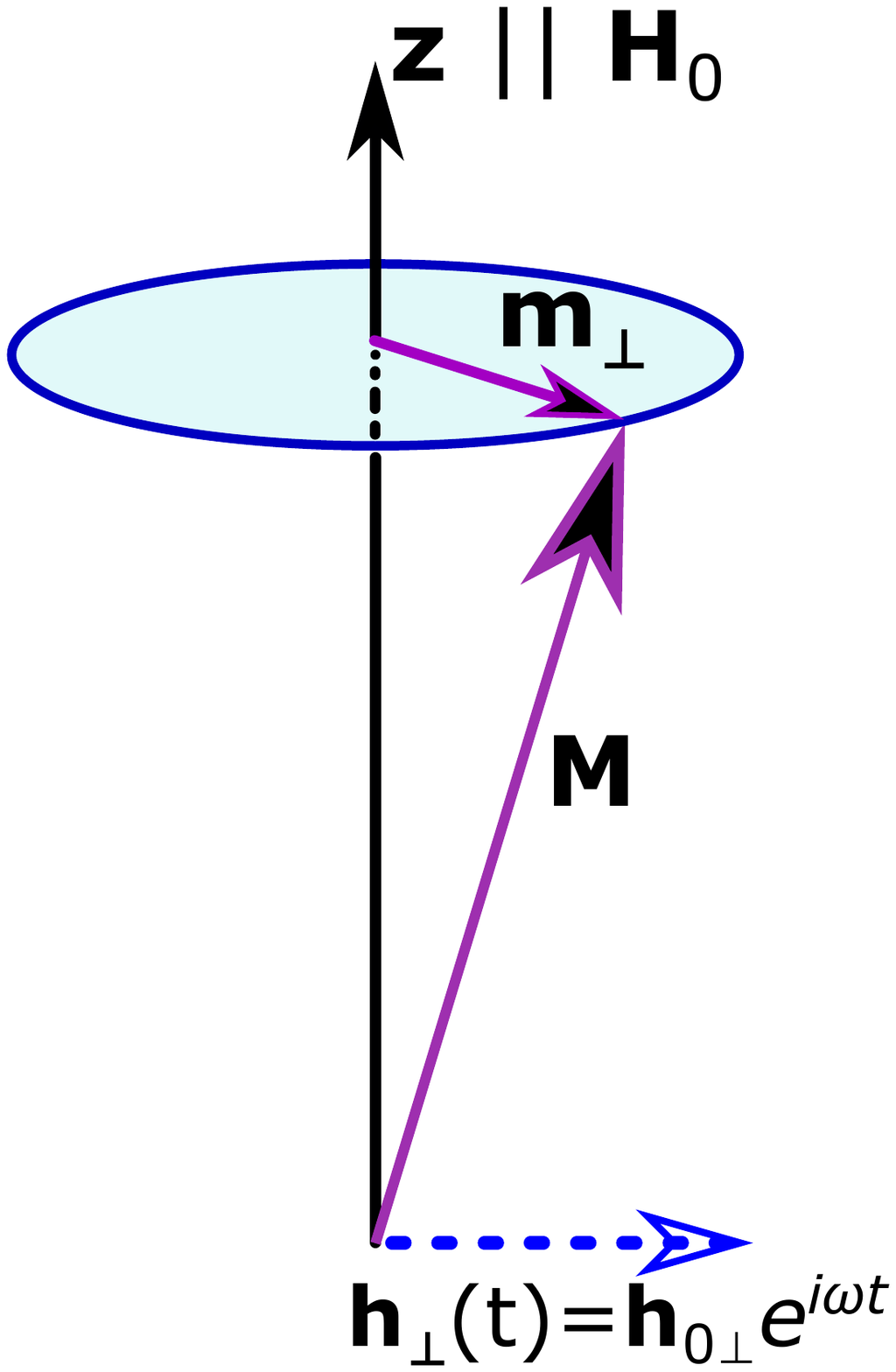}
	\end{center}
	\vspace{-15pt}
	\caption{Magnetization precession in an external alternating magnetic field $\mathbf{H}_0 + \mathbf{h}_{\perp}(t)$. ESR resonance occurs when the frequency $ \omega $ of external field $ \mathbf{h}_{\perp}(t) $ and the frequency $ \omega_0 $ of precession of individual magnetic moments coincide.}\label{f0}
\end{figure}

Let us return to paramagnets, and linearize the equation \eqref{M-time-evol} by represen-ting $\mathbf{M}(t)$ according to \eqref{Mt}, leaving only first-order corrections ($\propto h\ll H_0$ and $\propto m\ll M_0$), and taking into account the stationarity of $\mathbf{M}_0 \parallel \mathbf{H}_0$ according to \eqref{M0}:
\begin{equation}
\label{M-time-evol-lin}
\frac{d \mathbf{m}(t)}{d t}=\gamma[\mathbf{m}\times\mathbf{H}_0]+\gamma[\mathbf{M}_0\times\mathbf{h}].
\end{equation}
First, we consider the natural oscillations of the magnetic moment $\mathbf{m}(t)$ in the absence of alternating field. For the longitudinal ($\mathbf{z}$-) component $\mathbf{m}_{\parallel}(t)$ we have $dm_z/dt=0$, therefore, $m_z=const$. In a weak alternating field, the longitudinal component $\mathbf{M}_{\parallel}(t)$ of magnetization vector thus will differ from the equilibrium value $\mathbf{M}_0$ \eqref{M0} by a second-order value, while in the first order of perturbation theory $ m_z = 0 $. For the transverse component $\mathbf{m}_{\perp}(t)$ we have
\begin{equation}
\label{mperp}
\frac{d\mathbf{m}_{\perp}}{dt}=\gamma[\mathbf{m}_{\perp}\times\mathbf{H}_0].
\end{equation}
Taking the scalar product of \eqref{mperp} with $\mathbf{m}_{\perp}$, we obtain:
\begin{equation}
\mathbf{m}_{\perp}\cdot\frac{d\mathbf{m}_{\perp}(t)}{dt}=\frac{1}{2}\frac{d(\mathbf{m}_{\perp}(t))^2}{dt}=
\gamma\mathbf{m}_{\perp}\cdot[\mathbf{m}_{\perp}\times\mathbf{H}_0]=0,
\end{equation}
whence follows $(\mathbf{m}_{\perp})^2=const$.
Substituting into a time-harmonic solution of the form $\mathbf{m}_{\perp}(t)=\mathbf{m}_{0}e^{-i\omega t}$, we get
\begin{equation}
\label{mp_rec}
-i\omega\mathbf{m}_{\perp}=\gamma[\mathbf{m}_{\perp}\times\mathbf{H}_0],
\end{equation}
and then substitute this ``recurrent'' expression for $\mathbf{m}_{\perp}$ to its right-hand side:
\begin{equation}
\label{mp_rec2}
-\omega^2\mathbf{m}_{\perp}-\gamma^2\left[[\mathbf{m}_{\perp}\times\mathbf{H}_0]\times\mathbf{H}_0\right]=0.
\end{equation}
By expanding the double vector product in this equation and using $(\mathbf{m}_{\perp}\cdot\mathbf{H}_0)=0$, we get
\begin{equation}
\label{}
-\omega^2\mathbf{m}_{\perp}+\omega^2_0\mathbf{m}_{\perp}=0 \quad \Rightarrow \quad \omega=\omega_0,
\end{equation}
where $\omega_0=\gamma H_0$ is the \emph{Larmor precession frequency}. 



We also obtain an explicit form of the solution of the equation \eqref{mperp}. To this end, we write it out in projections on $ x $ and $ y $ axes:
\begin{eqnarray}
\label{mperp_x}
\frac{dm_x(t)}{dt}&=&\gamma m_y [\mathbf{e}_y\times\mathbf{H}_0]_x=\gamma m_y H_0, \\
\label{mperp_y}
\frac{dm_y(t)}{dt}&=&\gamma m_x [\mathbf{e}_x\times\mathbf{H}_0]_y=-\gamma m_x H_0.
\end{eqnarray}
Multiplying the equation \eqref{mperp_y} by $i$ and adding to \eqref{mperp_x}, we get
\begin{equation}
\label{before_comp}
\frac{d(m_x(t)+im_y(t))}{dt}=\gamma H_0(m_y-im_x)=-i\gamma H_0(m_x+im_y).
\end{equation}
We introduce a complex quantity $m^{\mathbb{C}}_{\perp}=m_x+im_y$, which replaces the vector $\mathbf{m}_{\perp}$. Using \eqref{before_comp}, we obtain
\begin{equation}
\label{m_c}
\frac{dm^{\mathbb{C}}_{\perp}(t)}{dt}=-i\gamma H_0m^{\mathbb{C}}_{\perp}.
\end{equation}
The equation \eqref{m_c} has a time-harmonic solution
\begin{equation}
\label{}
m^{\mathbb{C}}_{\perp}(t)=m^{\mathbb{C}}_{\perp 0} e^{-i\omega_0 t},
\end{equation}
where $\omega_0=\gamma H_0$ is the precession frequency. Indeed, since absolute values $\mathbf{m}_{\perp}$ and $\mathbf{m}_{\parallel}$ are time-independent, and the phase of the introduced complex function $m^{\mathbb{C}}_{\perp}=m_x+im_y$ rotates with the frequency $\omega_0$, then the vector $ \mathbf{m} $ precesses around the $ z $ axis at the same frequency $ \omega_0 $ (see Fig. \ref{f0}).

Now let $\mathbf{h}(t)=\mathbf{h}_0 e^{-i\omega t}\neq0$. The equation \eqref{M-time-evol-lin} projected on $ z $ axis, in view of $\mathbf{M}_0\parallel z$, gives again $ m_z = const $. For the transverse component $\mathbf{m}_{\perp}(t)$ we have
\begin{equation}
\label{mperp+h}
\frac{d\mathbf{m}_{\perp}(t)}{dt}=\gamma[\mathbf{m}_{\perp}\times\mathbf{H}_0]-\gamma[\mathbf{h}_{\perp}\times\mathbf{M}_{0}].
\end{equation}
We search a solution to \eqref{mperp+h} as a time-harmonic function oscillating with the frequency of external perturbing field, $\mathbf{m}_{\perp}(t)=\mathbf{m}_{\perp 0}e^{-i\omega t}$. Substituting this solution, we obtain
\begin{equation}
\label{mp+h_rec}
-i\omega\mathbf{m}_{\perp}(t)=\gamma[\mathbf{m}_{\perp}\times\mathbf{H}_0]-\gamma[\mathbf{h}_{\perp}\times\mathbf{M}_{0}].
\end{equation}
As before, by expressing $\mathbf{m}_{\perp}(t)$ from this equation and substituting it into the first term on the right-hand side we get
\begin{eqnarray}
-i\omega\mathbf{m}_{\perp}(t)=i\frac{\gamma^2}{\omega}
\big{(}
\left[[\mathbf{m}_{\perp}\times\mathbf{H}_0]\times\mathbf{H}_0\right]- \nonumber \\
\label{mp+h_rec2}
\left[[\mathbf{h}_{\perp}\times\mathbf{M}_0]\times\mathbf{H}_0\right]
\big{)}
-\gamma[\mathbf{h}_{\perp}\times\mathbf{M}_{0}].
\end{eqnarray}
Expanding double cross products and using
$(\mathbf{m}_{\perp}\cdot\mathbf{H}_0)=0$ and $(\mathbf{h}_{\perp}\cdot\mathbf{H}_0)=0$, we get
\begin{equation}
\label{}
\omega^2\mathbf{m}_{\perp}=-\gamma^2\left(-\mathbf{m}_{\perp}H^2_0+\mathbf{h}_{\perp}M_0 H_0\right)-i\omega\gamma[\mathbf{h}_{\perp}\times\mathbf{M}_{0}],
\end{equation}
or
\begin{equation}
\label{res+h}
(\omega^2-\omega^2_0)\mathbf{m}_{\perp}=-\chi_0\omega^2_0\mathbf{h}_{\perp}-i\omega\gamma\chi_0[\mathbf{h}_{\perp}\times\mathbf{M}_{0}].
\end{equation}
This ratio can be rewritten in a convenient form, remembering that $m_z=0$:
\begin{eqnarray}
\label{res+h_conven}
\mathbf{m}&=&\mathbf{m}_{\perp}=\chi\mathbf{h}_{\perp}-i\left[\mathbf{G}\times\mathbf{h}_{\perp}\right],\\
\chi&=&\chi(\omega)=\chi_0\omega^2_0/(\omega^2_0-\omega^2), \\
\label{gyr_vec}
\mathbf{G}&=&\chi_0\omega\gamma/(\omega^2_0-\omega^2)\mathbf{H}_0.
\end{eqnarray}
Thus, we find the frequency dispersion of the magnetization $\mathbf{m}(t)$. At a frequency $\omega_0$ equal to the precession frequency of individual magnetic moments $ \mu_i $ in a constant uniform field $ \mathbf{H}_0 $, a resonance occurs if the perturbing high-frequency field $ \mathbf{h}(t) $ has a component $\mathbf{h}_{\perp}(t)$ transverse to $ \mathbf{H}_0 $ (see Fig. \ref{f0}).
The vector $\mathbf{G} \parallel \mathbf{H}_0$ characterizes the \textit{magnetic gyrotropy} of the medium induced by the static component of external field. Indeed, by rewriting the relation \eqref{res+h_conven} in coordinates we get
\begin{equation}
\label{m_xyz}
m_x=\chi h_x+ih_yG, \qquad m_y=\chi h_y-ih_xG, \qquad m_z=0,
\end{equation}
By introducing the tensor of magnetic susceptibility
\begin{equation}
\label{chi_matrix}
\chi_{ij}=
\begin{pmatrix} 
\chi & iG & 0\\
-iG & \chi & 0\\
0 & 0 & 0\\
\end{pmatrix},
\end{equation}
the relations \eqref{m_xyz} can be written in tensor form: $m_i=\chi_{ij}h_j$. From \eqref{chi_matrix} and \eqref{gyr_vec} follows $\chi_{ij}(\mathbf{H}_0)=\chi_{ji}(-\mathbf{H}_0)$, which is indeed the gyrotropy condition.
By introducing the cyclic components of $\mathbf{m}$ and $\mathbf{h}$,
\begin{equation}
\label{m_pm_h_pm}
m_\pm=m_x\pm im_y, \qquad h_\pm=h_x\pm ih_y,
\end{equation}
and rewriting \eqref{m_xyz} in these components, we finally obtain
\begin{eqnarray}
\label{chi_pm}
m_\pm&=&\chi_\pm h_\pm, \\
\label{chi_res}
\chi_\pm&=&\chi\pm G=\chi_0\omega_0/(\omega_0\mp\omega).
\end{eqnarray}
The introduced susceptibilities $\chi_\pm$ correspond to waves with right and left polarizations, while \textit{the resonance occurs only for waves of one polarization}. Indeed, if $\gamma>0$, then $\omega_{res}=\omega_0$, and $\chi_+$ will be resonant; if $\gamma<0$, then $\omega_{res}=-\omega_0$, and $\chi_-$ will be resonant. 
A relationship between $\chi_\pm$ and $(\chi,\,G)$ could also be established:
\begin{equation}
\label{}
\chi=1/2(\chi_+ + \chi_-), \qquad G=1/2(\chi_+ - \chi_-).
\end{equation}
Mathematically, the values $\chi_\pm$ obtained in \eqref{chi_res} have a singularity at $ \omega_{res} $, which is resolved by including damping in \eqref{M-time-evol}: the functions $ \chi_\pm $ then become complex, with real parts responsible for the dispersion of the refractive index (anomalous near resonance), and imaginary parts responsible for the energy absorption (maximum at resonance).

\subsection{Damping in the dynamic equation for paramagnets 
\label{Bloch_paramagnet_relaxation}}

The simplest corrections to the equation of magnetization dynamics \eqref{M-time-evol}, which account for the damping of oscillations in a paramagnet, were first introduced by F. Bloch~\cite{Kittel}, and have different forms for $\mathbf{z}$- and $\mathbf{x,y}$-components of the equation due to symmetry breaking in an external constant magnetic field:
\begin{eqnarray}
\label{Bloch_Eqns}
\frac{\partial M_x}{\partial t}&=&\gamma[\mathbf{M}\times\mathbf{H}(t)]_x - \frac{M_x}{T_2}, \\
\frac{\partial M_y}{\partial t}&=&\gamma[\mathbf{M}\times\mathbf{H}(t)]_y - \frac{M_y}{T_2},  \\
\frac{\partial M_z}{\partial t}&=&\gamma[\mathbf{M}\times\mathbf{H}(t)]_z + \frac{M_0-M_z}{T_1}.
\end{eqnarray}
Here, the relaxation corrections stem from $ \tau $-approximation of Boltzmann kinetic equation: the rate of approach of magnetization components $(M_x, \,M_y, \,M_z)$ to their equilibrium values $(0,\,0,\,M_0)$ due to relaxation is proportional to the deviations $(M_x,\,M_y,\,M_z-M_0)$ taken with the opposite sign. 
The relaxation time $ T_1 $ for the $ \mathbf{z} $ component is called the \emph{spin-lattice relaxation time}. The relaxation processes in this case are mediated by the transfer of magnetic energy to the phonon system. 
The relaxation time $ T_2 $ for the $ \mathbf{x,y} $ component is called the \emph{transverse relaxation time}. The corresponding processes are not associated with energy fluxes outgoing from the magnetic system, and can be explained by \emph{dephasing} of magnetic dipoles precession due to the difference in microscopic fields acting on different dipoles.

\subsection{ESR and NMR: characteristic frequencies}\

Below we give an estimate for the paramagnetic resonance frequency $ \omega_0 = \gamma H_0 $ for electronic and nuclear magnetic systems, respectively:

1) Electron spin resonance (ESR): $\gamma\sim e/mc\sim 10^7$, so that $\omega_0 \sim 10^7 H_0$. Then at $H_0\simeq 10^3\div 10^4 \,\text{Oe}$ we have $\omega_0 \sim 10^{10}\div10^{11} \,\text{Hz}$, which corresponds to decimeter and centimeter microwaves.

2) Nuclear magnetic resonance (NMR): $\gamma\sim e/M_n c\sim 10^4$, where $M_n$ in the nucleon mass (assuming $M_n\simeq 10^3 m$). Then $\omega_0 \sim 10^4 H_0$, which for $H_0\simeq 10^3\div 10^4 \,\text{Oe}$ corresponds to radio waves with frequencies $\omega_0 \sim 10^{7}\div10^{8} \,\text{Hz}$.
Thus, the resonant frequencies for electrons and nuclei are in completely different frequency ranges, which allows to observe NMR signals even in substances in which nuclear magnetization makes a negligible (compared to electronic) contribution to the atomic magnetic moment.

\newpage
\section{Ferromagnetic resonance \label{FMR}}\

In the problem of paramagnetic resonance, the external magnetic field was considered as specified and equal to the internal one. In the case of ferromagnetic resonance (FMR), the internal field depends on the shape of the sample (because of intrinsic magnetic dipole forces), while the specified field is the external field at a sufficient distance from the sample. Therefore, to find the law of magnetization precession, it is first necessary to find the connection between the internal field and the external one using Maxwell's equations \eqref{m1}-\eqref{m4} and boundary conditions on the sample surface.
Usually, in FMR experiments, a non-conducting single-crystal ferromagnet is used, which has the shape of a rectangular sample or a plate and has dimensions that are small in comparison with the length of the electromagnetic wave. 
In this case, Maxwell equations can be approximated by \textit{magnetostatic equations} \cite{LL8}, in which the terms responsible for the effect of wave retardation in the sample are discarded. If, in addition, we consider the \textit{homogeneous \footnote{Inhomogeneous modes (with non-zero wavenumbers) are called \textit{magnetostatic}: for them you can still neglect the wave propagation in the sample and apply the magnetostatic equations, but the exchange forces are still small (see Lecture 4 on spin waves).} mode} of magnetization oscillations (in which $ \mathbf{M} $ does not depend on coordinates), then the effect of boundary conditions reduces just to the uniform internal \textit{demagnetizing field}, which is linear in magnetization.

Following Kittel \cite {Kittel 1948}, we consider a magnetically isotropic ferromagnet of ellipsoidal form with principal axes $x,\,y,\,z$. Let us denote the demagnetizing factors of the sample (see, e.g., \cite{LL8}) $N_x,\,N_y,\,N_z$ respectively. Let the external magnetic field $\mathbf{H}(t)$ have components $(H_x(t),\,0,\,H_z)$, where $H_x(t)$ is a high-frequency field, $H_z$ is the constant field. We will consider the latter to be sufficient for (at least approximately) reaching saturation and ensuring that the domain structure is weakly expressed. In this case we can consider the sample to be always uniformly magnetized. The field $ \mathbf{H}^{int}(t) $ inside the sample then has components:
\begin{equation}
\label{H_int_comp}
H^{int}_x=H_x-N_x M_x, \quad H^{int}_y=-N_y M_y, \quad H^{int}_z=H_z-N_z M_z.
\end{equation}

Let us now return to the equation of magnetization dynamics \eqref{M-time-evol}, in which by $\mathbf{H}(t)$ we mean the effective field $\mathbf{H}^*=-\delta E / \delta \mathbf{M}$. For a ferromagnetic sample, the difference between $\mathbf{H}^*$ and the internal magnetic field $\mathbf{H}^{int}$ is the effective magnetic field $\mathbf{H}_{eff}$ from the Weiss theory \footnote{Further refinements (apart from taking dissipation into account) are associated with \textit{magnetic anisotropy energy} and \textit{influence of domain structure}.}.
However, since $\mathbf{H}_{eff} \parallel \mathbf{M}$ for a uniformly magnetized sample (see \eqref{eff_field}), we can omit this term in the dynamical equation \eqref{M-time-evol}. The Lorentz field $(4\pi/3)\mathbf{M}$ (which, like the demagnetizing field, stems from dipole-dipole interaction) also does not participate in magnetization dynamics. Thus, the dynamical equation takes the form\footnote{Recall that for ferromagnetic systems in which both the mechanical and magnetic moments of the atom are due only to the total spin of electrons participating in the exchange interaction, the factor $ \gamma < 0$ (see \emph{Conventions}).} 
:
\begin{equation}
\label{DN_Ferro}
\frac{\partial\mathbf{M}(t)}{\partial t}=\gamma[\mathbf{M}\times\mathbf{H}^{int}(t)],
\end{equation}
or in components:
\begin{eqnarray}
\label{DN_Ferro_comp}
\frac{\partial M_x}{\partial t}&=&\gamma\left(M_y H^{int}_z-M_z H^{int}_y\right)=\gamma M_y \left(H_z+(N_y-N_z)M_z\right), \nonumber \\
\frac{\partial M_y}{\partial t}&=&\gamma\left(M_z H^{int}_x-M_x H^{int}_z\right)=\gamma\left(M_z H_x - (N_x-N_z)M_x M_z - M_x H_z\right), \nonumber \\
\frac{\partial M_z}{\partial t}&=&\gamma\left(M_x H^{int}_y-M_y H^{int}_x\right)=\gamma \left(M_x M_y (N_x-N_y)-M_y H_x\right). \nonumber
\end{eqnarray}
The dynamical equation of the type \eqref{DN_Ferro} is called the \emph{Landau-Lifshitz equation for magnetization dynamics}, and was first introduced by Landau and Lifshitz in their seminal work~\cite{LandauWorks}.

Since the longitudinal component of magnetization $M_z$ is assumed to be saturated, and the two transverse components $ M_x, M_y \ll M_z $, then in the third equation we can set approximately $d M_z/dt\simeq 0$, whence $M_z\simeq M$. Seeking the dependences of longitudinal magnetization components in time-harmonic form $M_x(t)=M^0_x e^{i \omega t}$, $M_y(t)=M^0_y e^{i \omega t}$, and the high-frequency field in form $H_x(t)=H^0_x e^{i \omega t}$, from the first two equations we obtain the following condition for the existence of nontrivial solutions for the amplitudes $M^0_x,\, M^0_y$:
\begin{equation}
\label{DisEq_FMR}
\omega^2_0=\gamma^2
\left[
(N_x-N_z) M + H_z
\right]
\left[
(N_y-N_z) M + H_z
\right],
\end{equation}
which is essentially the FMR resonant frequency value.
\begin{figure}
	\vspace{-0pt}
	\centering
	\begin{subfigure}[b]{0.18\textwidth}
		\includegraphics[width=\textwidth]{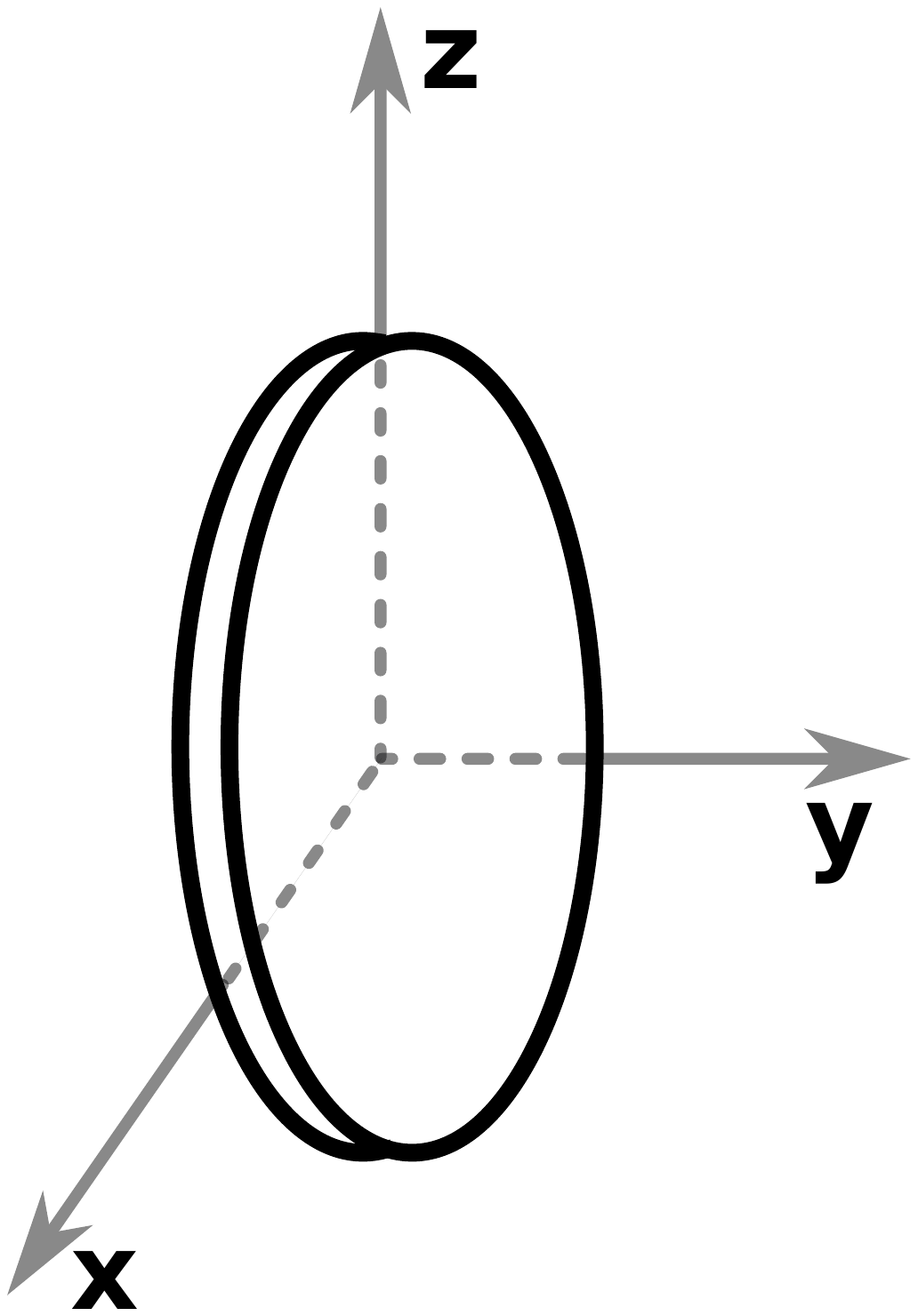}
		\caption{}
		\label{xz_disc}
	\end{subfigure}
	~ 
	\begin{subfigure}[b]{0.18\textwidth}
		\includegraphics[width=\textwidth]{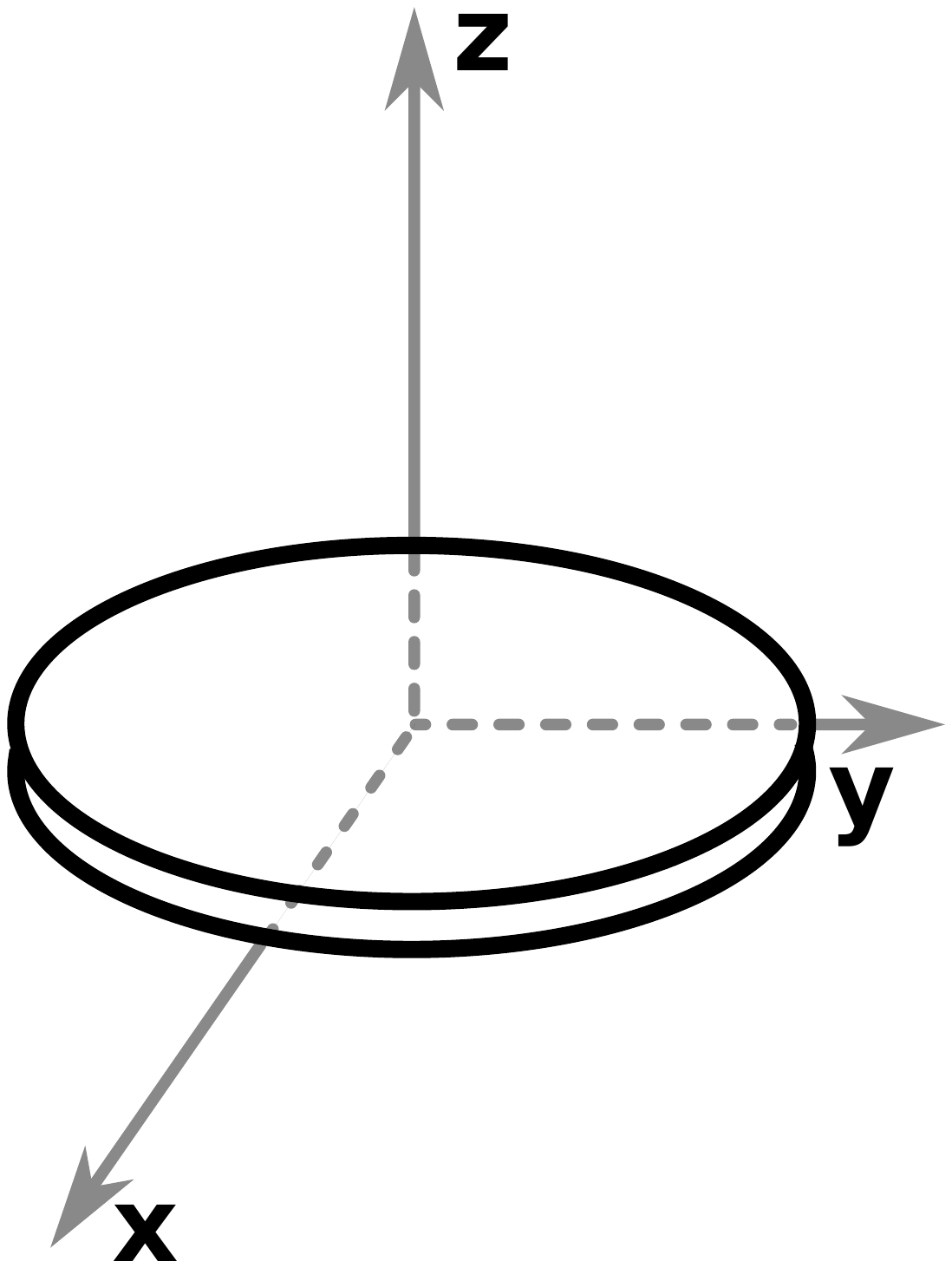}
		\caption{}
		\label{xy_disc}
	\end{subfigure}
	~ 
	\begin{subfigure}[b]{0.18\textwidth}
		\includegraphics[width=\textwidth]{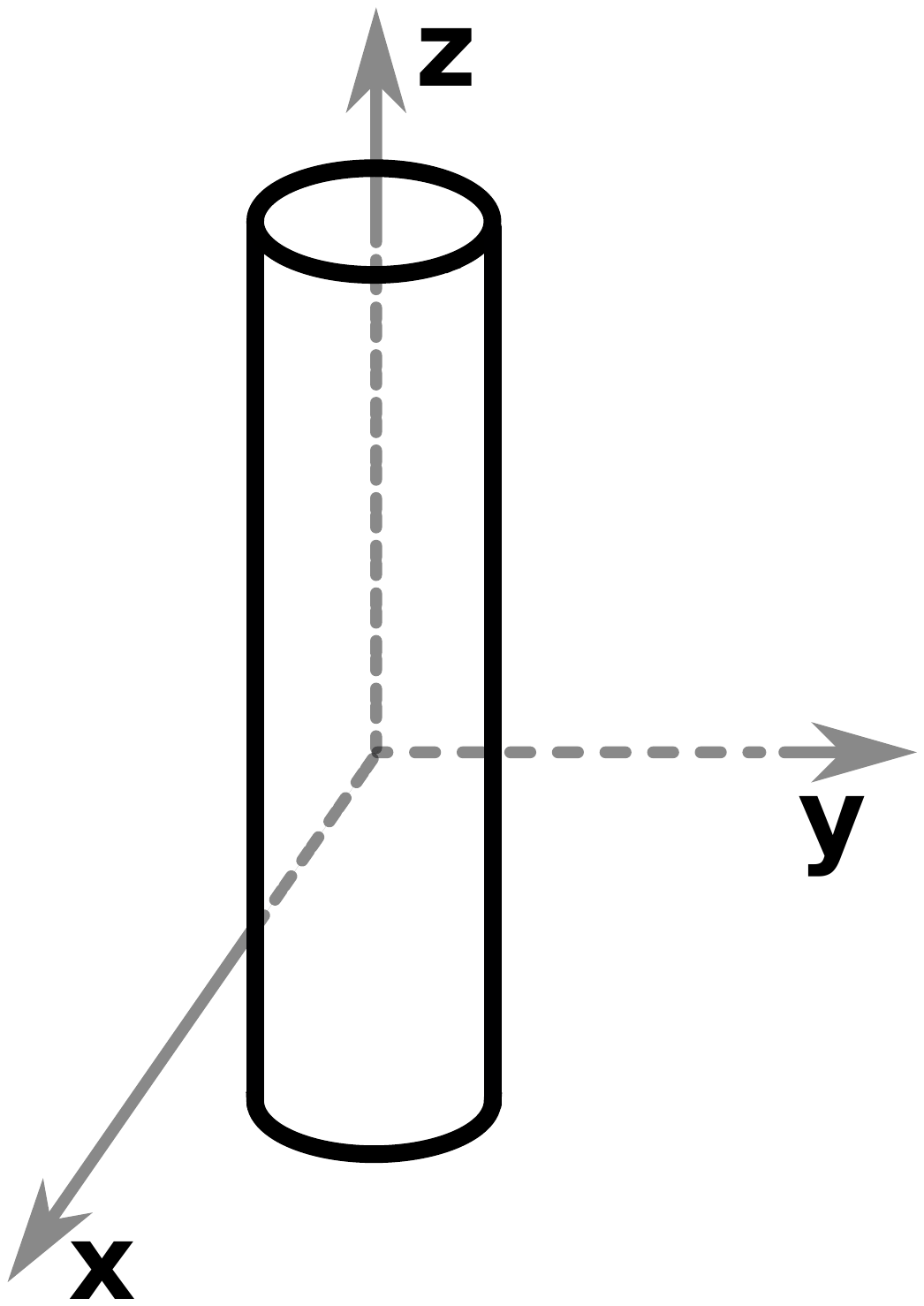}
		\caption{}
		\label{z_cylinder}
	\end{subfigure}
	~ 
	\begin{subfigure}[b]{0.18\textwidth}
		\includegraphics[width=\textwidth]{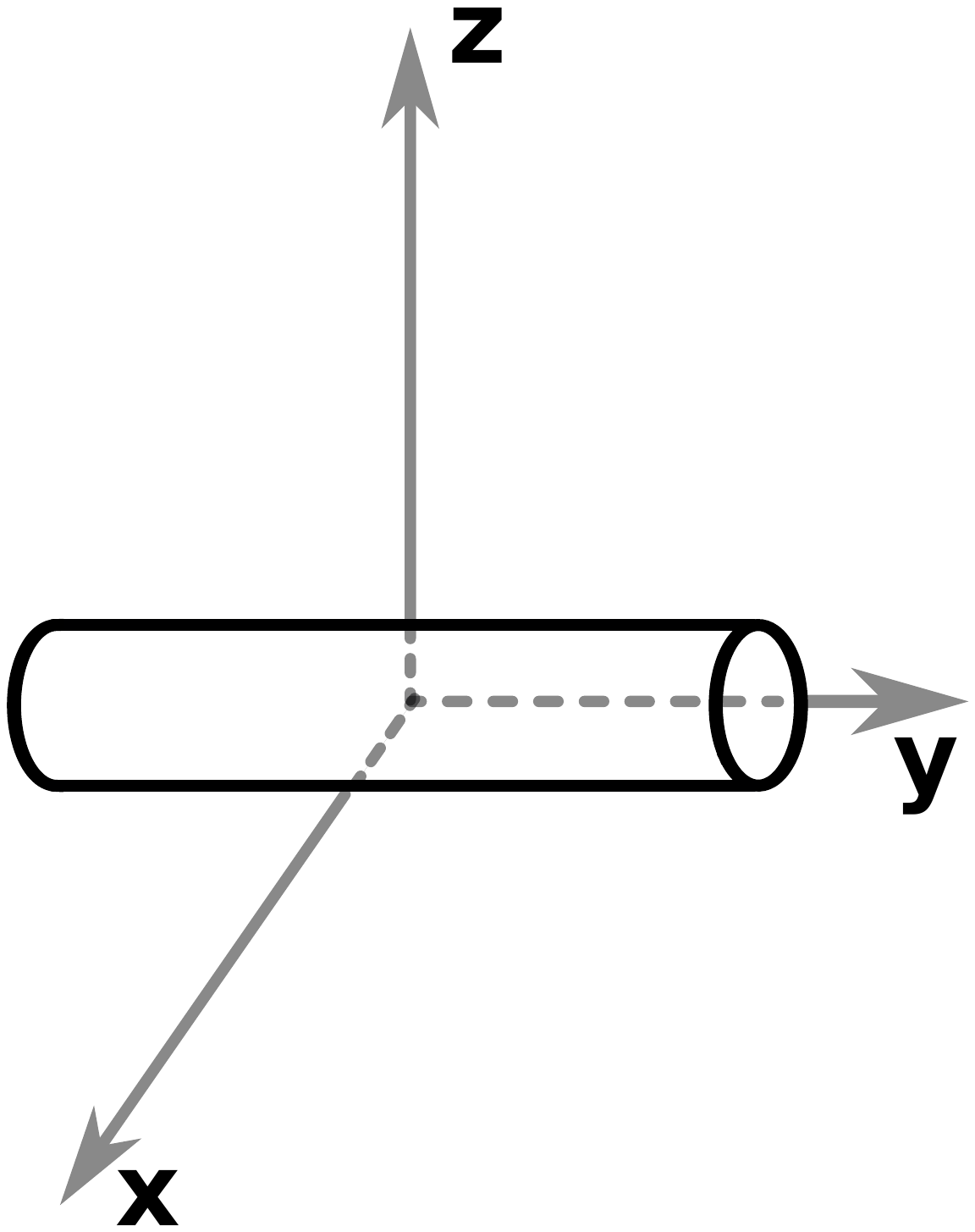}
		\caption{}
		\label{y_cylinder}
	\end{subfigure}
	~ 
	\begin{subfigure}[b]{0.18\textwidth}
		\includegraphics[width=\textwidth]{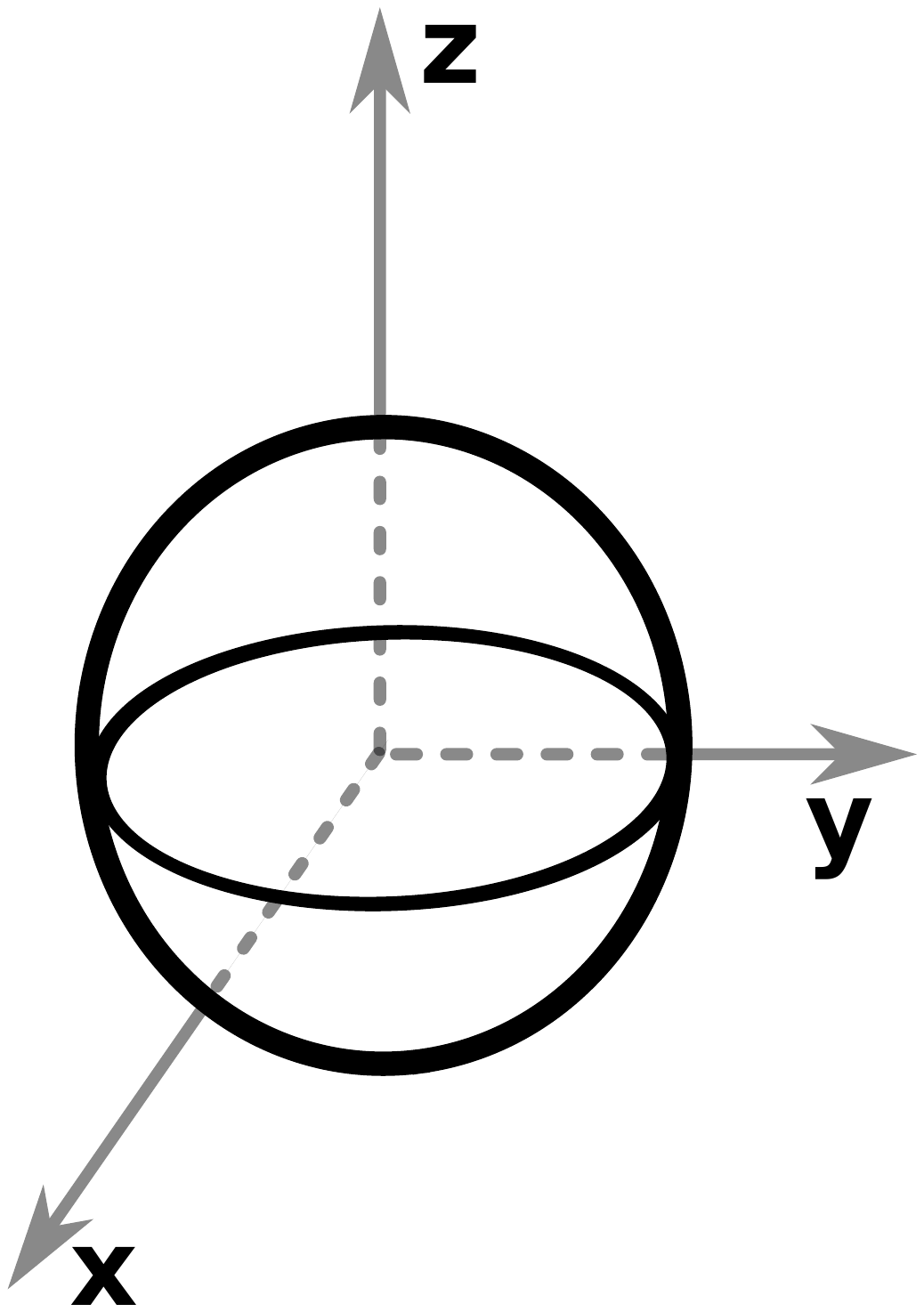}
		\caption{}
		\label{a sphere}
	\end{subfigure}
	
	\caption{Limiting forms of an ellipsoid (\cite{Gurevich}, p. 54). An external constant magnetic field $ \mathbf{H}(t) $ has components $(H_x(t),\,0,\,H_z)$, where $H_x(t)$ is the high-frequency field, $ H_z $ is the constant field.}\label{animals}
	\vspace{-10pt}
\end{figure}
Note that the same expression for the resonance frequency can be obtained by finding the component $\chi_{xx}\equiv\chi_x=M_x/H_x$ of the susceptibility tensor from the first two dynamical equations:
\begin{equation}
\label{chi_x}
\chi_x=\frac{\chi_0}{1-(\omega/\omega_0)^2},
\end{equation}
where the susceptibility $ \chi_0 $ in a static external field $H_x=const$ reads
\begin{equation}
\label{chi_0}
\chi_0=\frac{M}{H_z+(N_x-N_z) M}.
\end{equation}
The expression for $\chi_0$ can also be derived directly, assuming that a small constant additional field along the $ x $ axis only rotates the magnetization vector by a small angle (see \cite{VV 1951}).

The resulting resonant precession frequency $\omega_0$ is called the \emph{uniform mode} frequency of magnetization oscillations, since we assumed that all elementary \footnotemark\footnotetext{In terms of macrodifferentials in electrodynamics of continuous media.} magnetic moments of the sample precess in phase and with the same amplitude.
Apart from homogeneous resonance, ferromagnets host resonance phenomena for which the inhomogeneity of magnetization oscillations is essential (see e.g.~\cite{ABP}, §11).

Specific cases relevant for the experiment are schematically shown in Fig. \ref{animals} (all of these shapes are the limiting cases of an ellipsoid), with corresponding formulas for $\omega_0$ listed in the table \ref{FMR_table}. Note that in the case of a spherical sample, the result coincides with the result obtained when considering non-interacting magnetic dipoles (see section 3.1 in this course). For other cases, the contributions of demagnetizing fields to the dispersion law are present. In this context, the demagnetizing fields effect is also called the \textit{shape anisotropy}.







\begin{table}[h!]
	\vspace{0pt}
	\begin{flushleft}
		\begin{center}
			\begin{tabular}{|c|m{3cm}|m{1cm}|m{1.2cm}|m{1.2cm}|m{1.2cm}|c|}
				\hline
				\multicolumn{ 1}{|c|}{Sample form} & \multicolumn{ 1}{m{3cm}|}{Magnetization \quad direction} & \multicolumn{ 1}{m{2cm}|}{Panel in Fig. \ref{animals}} & \multicolumn{ 3}{b{3.6cm}|}{Demagnetizing factors} & \multicolumn{ 1}{c|}{$\omega_0/\gamma$} \\ \cline{ 4- 6}
				\multicolumn{ 1}{|l|}{} & \multicolumn{ 1}{l|}{} & \multicolumn{ 1}{l|}{} & $N_x$ & $N_y$ & $N_z$ & \multicolumn{ 1}{l|}{} \\ \hline
				\multicolumn{ 1}{|c| }{Plate} & tangent & \multicolumn{1}{c|}{(a)} & \multicolumn{1}{l|}{$ 0 $} & $4\pi$ & \multicolumn{1}{l|}{$ 0 $} & $\sqrt{H_z (H_z+4\pi M)}$ \\ \cline{ 2- 7}
				\multicolumn{ 1}{|l|}{} & normal & \multicolumn{1}{c|}{(b)} & \multicolumn{1}{l|}{$ 0 $} & \multicolumn{1}{l|}{$ 0 $} & $4\pi$ & $H_z-4\pi M$ \\ \hline
				\multicolumn{ 1}{|l|}{Cylinder} & longitudinal & \multicolumn{1}{c|}{(c)} & $2\pi$ & $2\pi$ & \multicolumn{1}{l|}{$ 0 $} & $H_z+2\pi M$ \\ \cline{ 2- 7}
				\multicolumn{ 1}{|l|}{} & transverse & \multicolumn{1}{c|}{(d)} & $2\pi$ & \multicolumn{1}{l|}{$ 0 $} & $2\pi$ & $\sqrt{H_z (H_z-2\pi M)}$ \\ \hline
				Sphere &  & \multicolumn{1}{c|}{(e)} & $4\pi/3$ & $4\pi/3$ & $4\pi/3$ & $H_z$ \\ \hline
			\end{tabular}
		\end{center}
	\end{flushleft}
	\caption{Limiting cases of Kittel formula (\cite{Gurevich}, p. 54) \label{FMR_table}}
	\label{}
\end{table}

The next approximation is taking into account the \emph{magnetic anisotropy energy} $ E_a $. 
The basic origin of magnetic anisotropy is crystallographic anisotropy, which induces the preferential magnetization directions determined by the orientation of crystallographic axes. Anisotropy of magnetic properties can also originate from external elastic stresses~\cite{Gurevich}. 
The anisotropy energy could be accounted for in the magnetic part of free energy by the terms~\cite{Gurevich,KosKov}: 
\begin{equation}
\label{aniz_energy_full}
E_a
=
-\frac{1}{2} K_{1} m_{x}^{2}-\frac{1}{2} K_{3} m_{z}^{2}
,
\end{equation}
where $ K_{1} $ and $ K_{2} $ are the effective anisotropy constants, and $ m_x,\, m_z $ are the projections of the magnetization unit vector\footnote{Note that $ m_y = \sqrt{1- m_x^2 + m_z^2} $ is not included here, as the contribution associated with it will be not independent.}. 
When $ K_1=0 $, the ferromagnet possesses \textit{uniaxial anisotropy}. If, moreover, $ K_3>0 $, then the anisotropy is of \textit{easy-axis} type: in the ground state, the magnetization vector $ \mathbf{M} $ is directed along the ``easy axis'' $ z $. 
If $ K_1=0 $ and $ K_3<0 $, then the anisotropy is of \textit{easy-plane} type: in the ground state, $ \mathbf{M} $ lies in a plane $ (xy) $ perpendicular to the $ z $-axis. 
If $ K_1\neq0 $, by appropriate rotation of the coordinate frame it is always possible to set $ K_1=0 $.

\begin{figure}[h!]
	\vspace{-50pt}
	\begin{center}
		\includegraphics[width=0.45\textwidth]{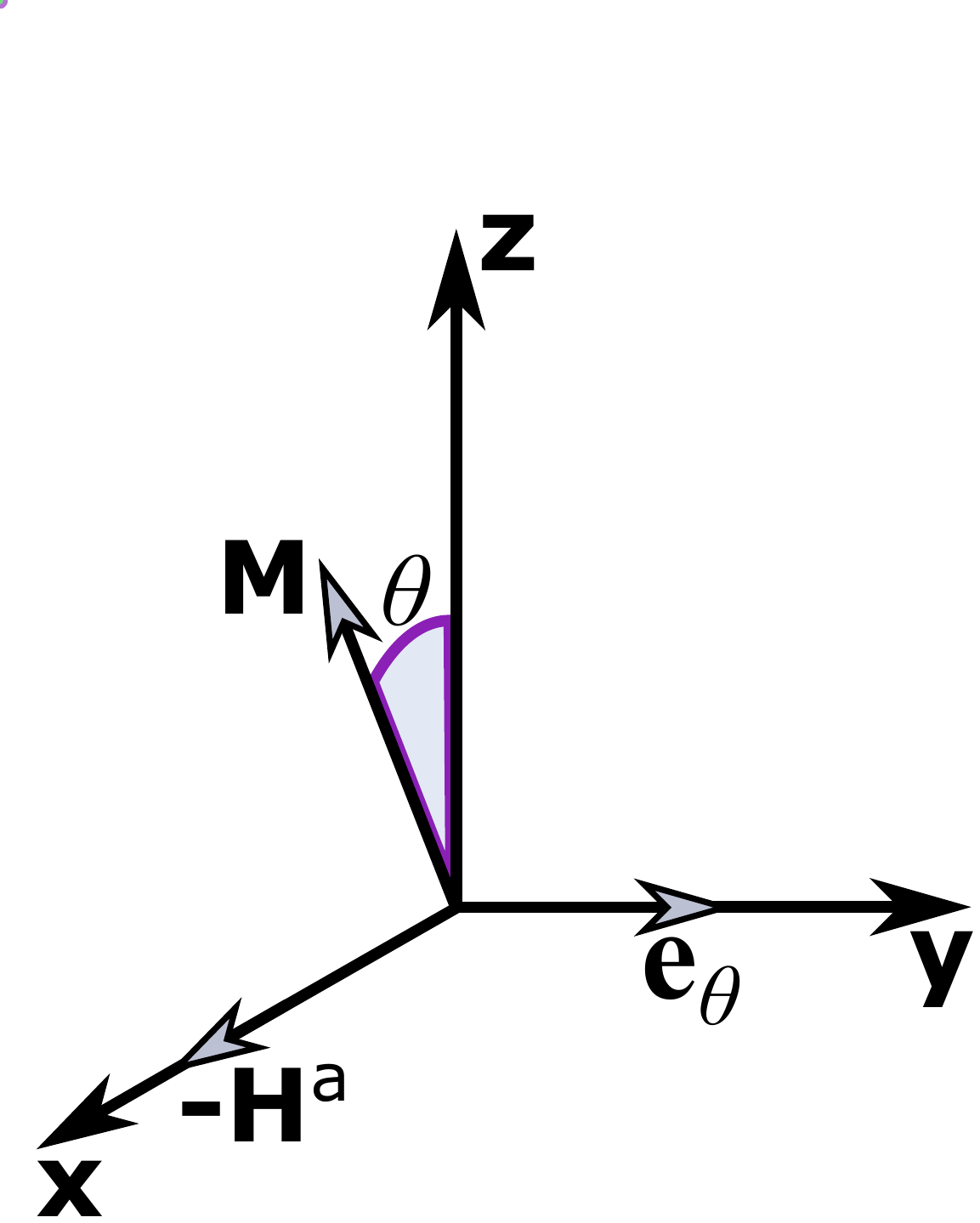}
	\end{center}
	\vspace{-75pt}
	\caption{To the calculation of effective anisotropy field $\mathbf{H}^{a}$.}\label{aniz}
	\vspace{5pt}
\end{figure}

In his phenomenological approach, Kittel \cite{Kittel,Kittel 1948} proposed to determine the \textit{effective anisotropy field} $ \mathbf{H}^{a} $ corresponding to anisotropy energy through the generalized moment associated with it:
\begin{equation}
-\frac{\partial E_a}{\partial \theta} \mathbf{e}_\theta = [\mathbf{M}\times\mathbf{H}^{a}],
\end{equation}
where $\theta$ is the angle between the magnetization vector $ \mathbf{M} $ and an easy axis $ \mathbf{z} $, and $ \mathbf{e}_\theta $ is the vector in $ (x,z) $ plane with polar angle $ \theta $. Although there is a certain degree of freedom in the choice of the magnitude and direction of $ \mathbf{H}^{a} $, we will choose the most natural configuration (see Fig. \ref{aniz}).

Consider a crystal with one direction of easy magnetization (a uniaxial ferromagnet). For small deviations from equilibrium $ \theta \ll 1 $, $ m_z \simeq 1 - \theta^2 / 2 $. The uniaxial symmetry then dictates the form of anisotropy energy:
\begin{equation}
\label{aniz_ene}
E_a \simeq K \theta^2,
\end{equation}
where $ K \equiv K_3/2 $ in Eq.\eqref{aniz_energy_full} (note that constant term has been excluded). 
Hence, 
\begin{equation}
\label{a_field}
\frac{\partial E_a}{\partial \theta} \mathbf{e}_\theta 
\simeq 
2K \theta \mathbf{e}_\theta \equiv -M H^a \mathbf{e}_\theta.
\end{equation}
Let us take into account the effective magnetic anisotropy field $\mathbf{H}^{a} \uparrow\downarrow \mathbf{x} \uparrow\uparrow M_x \mathbf{e}_x$ in the form of an effective additional demagnetizing field $H^{a}=-N^a_x M_x$. Besides, $M_x\simeq M \theta$. Then from \eqref{a_field} we have
\begin{equation}
\label{a_field_2}
2K \theta \simeq M^2 N^a_x \theta, 
\end{equation}
whence we obtain 
\begin{equation}
\label{N^a_x}
N^a_x\simeq 2K / M^2
.
\end{equation}
From the uniaxial symmetry then follows $N^a_y\simeq 2K / M^2$. 
Thus, taking into account the anisotropy energy is reduced to replacing the demagnetizing factors $N_x,\,N_y$ in formula \eqref{DisEq_FMR} for the resonant frequency by $N_x+N^a_x,\,N_y+N^a_y$. 

Finally, the \textit{dissipation} in the sample should be included in the model. The expression for the dissipation in ferromagnets should capture the main feature of ferromagnetic state: the magnetization is saturated throughout the sample (at least in the case of uniform oscillations). The simplest assumption is the following: the action of dissipative forces reduces to an additional effective field proportional in magnitude and inverse in sign to the rate of change of the magnetic moment $ \mathbf{M} $ \cite{Gurevich}:
\begin{equation}
\label{Hilb_eqn}
\frac{\partial\mathbf{M}(\mathbf{r},t)}{\partial t}=\gamma[\mathbf{M}\times\mathbf{H}^*(t)]+\frac{\alpha}{M} [\mathbf{M}\times\frac{\partial\mathbf{M}(t)}{\partial t}].
\end{equation}
This equation is known as the \emph{Landau-Lifshitz-Gilbert equation}.
The dimensionless quantity $ \alpha $ in adiabatic approximation is independent of the magnetization vector $ \mathbf{M} $, its spatial and temporal derivatives, and the parameters of the external field. One can also prove~\cite{One more time LLeq}, using $(\mathbf{M}\cdot d\mathbf{M}/dt)=0$, that the equation \eqref{Hilb_eqn} is equivalent to
\begin{equation}
\label{LL_eqn}
\frac{\partial\mathbf{M}(\mathbf{r},t)}{\partial t}=\tilde{\gamma}[\mathbf{M}\times\mathbf{H}^*(t)]-\frac{\alpha \tilde{\gamma}}{M} [\mathbf{M}\times[\mathbf{M}\times\mathbf{H}^*(t)]],
\end{equation}
where $\tilde{\gamma}=\gamma / (1+\alpha^2)$. 
In this form, the equation for the dynamics of magnetization was first proposed by Landau and Lifshitz in~\cite{LandauWorks}, therefore it is usually called (like the corresponding equation without damping) the \emph{Landau-Lifshitz equation}.

Using any of the equations \eqref{Hilb_eqn}, \eqref{LL_eqn}, one can show \cite{Gurevich} that dissipation leads to damping of free magnetization oscillations and keeps the amplitude of forced vibrations finite at resonance, and determines the resonance width. 


\bigskip
\bigskip
\begin{center}
	\uppercase{Control questions}
\end{center}
\noindent 1. Complementing the equation \eqref{mperp} with the relaxation term $-\mathbf{m}_\perp/\tau$ ($\tau\equiv T_2$), show that the linearized solution \eqref{mperp} again has the form \eqref{chi_pm} where now
\begin{equation}
\chi_\pm=\chi_0\omega_0/\left(\omega_0\mp(\omega+i/\tau)\right)
.
\nonumber
\end{equation}

\noindent 2. What terms are added to the magnetic energy in ferromagnets compared to paramagnets?

\noindent 3. Check that, with account for dissipation in the form \eqref{Hilb_eqn} or \eqref{LL_eqn}, the absolute value of magnetization vector $ M=|\mathbf{M}| $ is still conserved.

\noindent 4. Using any of the equations \eqref{Hilb_eqn}, \eqref{LL_eqn}, show that taking dissipation into account leads to damping of free magnetization oscillations (when the external alternating field is switched off).

\setchapterpreamble{%
	\dictum{%
		
		Spin waves (magnons) form the foundation for understanding the dynamics of magnetic texture and thermodynamic properties of magnets, similar to elastic waves (phonons) in a lattice.
		The exchange interaction between atoms spontaneously breaks the lattice symmetry, setting the order parameter in the ferromagnet: the magnetization vector. Thermal fluctuations tend to restore the broken symmetry and activate spin waves responsible for the reduction of sample magnetization with increasing temperature~\cite{LL9,ABP}.
		Coherent spin waves can carry an exchange spin current in ferromagnetic insulators. Long-wavelength spin waves describe the dynamics of homogeneous and weakly inhomogeneous magnetization.
		
	}%
	\vspace{24pt}%
}

\chapter{Spin waves in ferromagnets}
\section{Macroscopic derivation of the dispersion law for spin waves (long-wavelength limit)\footnotemark}\
\footnotetext{In this section, we follow the derivation in §11.3 of the book \cite{Turov}.}

By definition, the normal (or natural) electromagnetic waves in a medium are the waves that propagate in a medium even after the sources are ``turned off''.
If the wavelength $\lambda\ll L$, where $ L $ is the characteristic size of the sample (quasistationary approximation \cite{LL8}), but at the same time is not too small for averaging over micro-volumes $\Delta V$, $\lambda^3\gg \Delta V$, then the fields $\mathbf{E}$ and $\mathbf{H}$ can be found as solutions of Maxwell macroscopic equations for quasistationary fields in the medium. Let us now consider, by analogy, \emph{normal (free) spin waves}, in which the spin dynamics is driven by the exchange interaction. Since such a wave is the excitation of a ``discrete field'' of spins, in the macroscopic approach, waves of magnetization $ \mathbf{M}(\mathbf{r}) $ will be excited. Therefore, it is necessary to take into account that the field $ \mathbf{H}_{eff}(\mathbf{r}) $ is also determined by magnetization texture in some vicinity of $ \mathbf{r} $; the concept of a molecular field is thus generalized in the case of inhomogeneous magnetization. The general linear relationship between $ \mathbf{H}_{eff}(\mathbf{r}) $ and $ \mathbf{M}(\mathbf{r}) $ follows from the microscopic relation \eqref{Heff_micro} and can be written as
\begin{equation}
\label{int_H}
\mathbf{H}_{eff}(\mathbf{r})=\int \lambda(\mathbf{r}-\mathbf{r'})\,\mathbf{M}(\mathbf{r'})\,dV',
\end{equation}
where the kernel $\lambda(\mathbf{r}-\mathbf{r'})$ generalizes the molecular field constant.Indeed, if the magnetization $\mathbf{M}(\mathbf{r})$ is homogeneous, then after placing $\mathbf{M}(\mathbf{r'})$ outside the integral sign, we obtain the previous relation for $\mathbf{H}_{eff}(\mathbf{r})$ with $\lambda_w=\int \lambda(\mathbf{r}-\mathbf{r'})\,dV'$.

In ferromagnets, due to the strong exchange interaction, a local \emph{quasi-equilibrium} distribution of magnetic moment is very quickly established~\cite{ABP}, which justifies further use of \emph{equilibrium magnetization} and \emph{static magnetic permeability}. In view of the short-range nature of the exchange interaction, the $ \lambda(\mathbf{r} - \mathbf{r'}) $ kernel is nonzero at distances comparable to the lattice constant. This allows, by expanding the magnetization $ \mathbf{M}(\mathbf{r}) $ in a Taylor series
\begin{equation}
\mathbf{M}(\mathbf{r'})=\mathbf{M}(\mathbf{r}) + 
(x_i'-x_i)\frac{\partial\,\mathbf{M}}{\partial x_i} + 
\frac{1}{2}(x_i'-x_i)(x_j'-x_j)\frac{\partial^2\, \mathbf{M}}{\partial x_i\,\partial x_j} +...,
\end{equation}
to apply the theorem for the integral mean in \eqref{int_H}:
\begin{equation}
\mathbf{H}_{eff}(\mathbf{r})=
\lambda_w \mathbf{M}(\mathbf{r}) + 
\eta_i a \frac{\partial\,\mathbf{M}}{\partial x_i} + 
\alpha_{ij} a^2 \frac{\partial^2\, \mathbf{M}}{\partial x_i\,\partial x_j} +...,
\end{equation}
where $\eta_i$, $\alpha_{ij}$ are a constant vector and a constant tensor of the second rank, respectively, having the same dimension as $\lambda_w$, $a$ is a lattice constant defining the size of the region with nonzero kernel $\lambda(\mathbf{r}-\mathbf{r'})$. The number of nonzero components of the constants $\eta_i$, $\alpha_{ij}$ is determined by the crystal symmetry. For crystals with the center of symmetry $\eta_i=0$; in the simplest case of a cubic crystal (as well as in isotropic case) $\alpha_{ij}=\alpha \delta_{ij}$. \emph{Nonuniform exchange constant} (or \textit{magnetic stiffness}) $\alpha$ is of the order of exchange interaction energy, $ \alpha \sim T_c / (a M_s^2) $ (see \cite{LL9}, p.367). The terms with the third derivative  $\mathbf{M}(\mathbf{r})$ can be neglected due to weak inhomogeneity of magnetization field. Thus, in the considered approximation
\begin{equation}
\label{H_eff}
\mathbf{H}_{eff}(\mathbf{r})=
\lambda_w \mathbf{M}(\mathbf{r}) + 
\alpha a^2 \left(\frac{\partial^2\, \mathbf{M}}{\partial x^2} + \frac{\partial^2\, \mathbf{M}}{\partial y^2} + \frac{\partial^2\, \mathbf{M}}{\partial z^2} \right)=
\lambda_w \mathbf{M}(\mathbf{r}) + 
\alpha a^2 \Delta \mathbf{M},
\end{equation}
where $\Delta$ is the Laplace operator in 3D.

\begin{figure}[h]
	\includegraphics [width=0.95\textwidth]{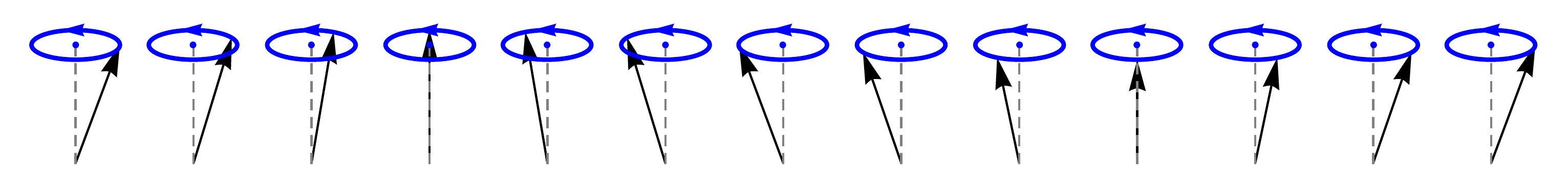}
	\includegraphics [width=0.95\textwidth]{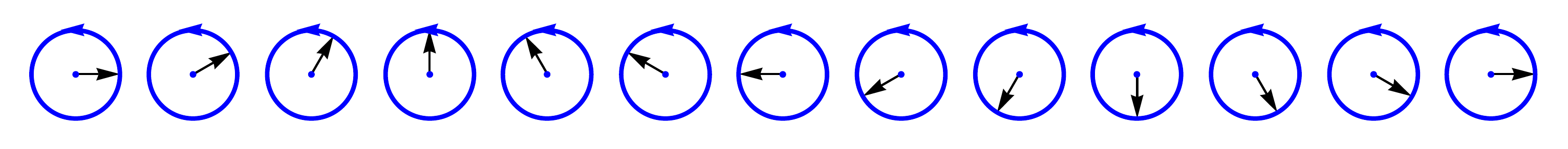}
	\caption{Spin wave in a ferromagnet. Shown is a section on which one spin wavelength fits.}\label{spwa}
\end{figure}

As mentioned in Section \ref{FMR}, the equation of magnetization dynamics remains applicable for a ferromagnetic sample, and the difference between the effective field $\mathbf{H}^*=-\delta E / \delta \mathbf{M}$ and the internal magnetic field $\mathbf{H}^{int}$ in the first approximation is the effective magnetic field $\mathbf{H}_{eff}$ from the Weiss theory. In the long-wavelength limit for spin waves, it is also possible to abstract from the demagnetizing factors: their influence only renormalizes the internal molecular field $ \mathbf{H}_{eff} $.
Note that the first term in \eqref{H_eff} can be omitted since it is parallel to $ \mathbf{M} $. Thus, the dynamic equation takes the form
\begin{equation}
\label{M-time-evol_+_Veiss}
\frac{\partial\mathbf{M}(\mathbf{r},t)}{\partial t}=\gamma[\mathbf{M}\times\mathbf{H}(\mathbf{r},t)]=
\gamma[\mathbf{M}\times\left(\mathbf{H}_0 + \mathbf{h}(t) + \alpha a^2 \Delta \mathbf{M}\right)],
\end{equation}
Let $\mathbf{h}(t) \perp \mathbf{H}_0$\footnotemark. Suppose that weak oscillations of elementary magnetic moments occur only by rotating them practically without changing the length of the local magnetic moment vector (see Fig. \ref{spwa}). 
The constancy of the modulus of the magnetization vector $ | \mathbf{M} | $ in time and space follows directly from the equation~\ref{M-time-evol_+_Veiss} after scalar multiplication of both sides by $ \mathbf{M}(\mathbf{r},t) $. 
In this case, a weak field $\mathbf{h}(t)$ causes small deviations $\mathbf{M}$ from the equilibrium orientation $\mathbf{M}_0 \parallel \mathbf{H}_0$. Therefore, as a first approximation, $M_z\simeq M_0$, $\mathbf{M}=\mathbf{M}_0+\mathbf{m}$, where $\mathbf{m} \perp \mathbf{M}_0,\,\, m\ll M_0$. Thus, the field on the right-hand side of the magnetization dynamics equation is $\mathbf{H}=\mathbf{H}_0 + \mathbf{h}(t) + \alpha a^2 \Delta \mathbf{m}$. We linearize the equation \eqref{M-time-evol_+_Veiss} with respect to small values $\mathbf{m},\,\mathbf{h}$:
\begin{equation}
\label{M-time-evol-lin_+Veiss}
\frac{\partial\mathbf{m}(\mathbf{r},t)}{\partial t}=\gamma[\mathbf{m}\times\mathbf{H}_0]+
\gamma \alpha a^2 [\mathbf{M}_0\times\Delta\mathbf{m}]+
\gamma[\mathbf{M}_0\times\mathbf{h}].
\end{equation}
\footnotetext{Recall that the component $\mathbf{h}_{\parallel}(t)$ has no effect on the dynamics of magnetization, cf. \eqref{chi_matrix}.}
In this equation, in comparison with \eqref{M-time-evol-lin}, the term $\gamma \alpha a^2 [\mathbf{M}_0\times\Delta\mathbf{m}]$ appears, containing the spatial derivatives of the vector $ \mathbf{m} $, which now allows us to consider the oscillations of the magnetization not only taking into account the \textit{time dispersion} $ \mathbf{m}(t) $, but also \emph{spatial dispersion} $ \mathbf{m}(\mathbf{r}) $\footnotemark.
\footnotetext{The characteristic parameter associated with the spatial dispersion of spin waves is the atomic scale $\sim a$ on which the exchange forces appear.}
In accordance with this, we choose the external variable field $\mathbf{h}(\mathbf{r},t)$ in the form of a plane wave,
\begin{equation}
\mathbf{h}(\mathbf{r},t)=\mathbf{h}_0 e^{i\mathbf{k}\mathbf{r} - i\omega t},
\end{equation}
and seek the solution $\mathbf{m}(\mathbf{r},t)$ in the same form,
\begin{equation}
\mathbf{m}(\mathbf{r},t)=\mathbf{m}_0 e^{i\mathbf{k}\mathbf{r} - i\omega t},
\end{equation}
The equation \eqref{M-time-evol-lin_+Veiss} with this substitution becomes:
\begin{equation}
\label{mp+h_rec_+Veiss}
-i\omega\mathbf{m}(t)
=\gamma[\mathbf{m}\times\mathbf{H}_0]\left( 1+\chi_0 \alpha a^2 k^2 \right)
-\chi_0\gamma[\mathbf{h}\times\mathbf{H}_{0}],
\end{equation}
where $\chi_0=M_0/H_0$ is the static 
susceptibility of a ferromagnet. 
The resulting equation differs from \eqref{mp+h_rec} only by the factor $\left( 1+\chi_0 \alpha a^2 k^2 \right)$ for the first term on the right-hand side. Therefore, one could immediately use the results for paramagnetic resonance (see Lecture 3.1). However, we will solve this equation independently by writing it in $x$- and $y$-components (recall that one can choose $m_z(t)=0$):
\begin{equation}
\label{m_sys}
i\omega m_x + \omega_k m_y = \chi_0 \omega_0 h_y, \qquad
\omega_k m_x - i\omega m_y = \chi_0 \omega_0 h_x.
\end{equation}
Here, the notation
\begin{equation}
\label{omega_k}
\omega_k = |\gamma| H_0 \left( 1+\chi_0 \alpha a^2 k^2 \right) =
|\gamma| \left[ H_0 + H_e (a k)^2 \right],
\end{equation}
is introduced, where $H_e=\alpha M_0$ is the exchange field (of the order of the strength of Weiss molecular field, since $\alpha\simeq\lambda_w$), and $\omega_0\equiv\left.\omega_k\right|_{k=0}=|\gamma| H_0$. 
Part of magnon energy is the Zeeman energy in an external magnetic field, $ \hbar \omega_k^{(H)} = \hbar |\gamma| H_0 = 2 \mu_B H_0 $; 
therefore, the \emph{excitation of each magnon in the magnet decreases its total magnetic moment by $ 2 \mu_B $} \cite{LL9}. \emph{One magnon, therefore, carries the spin $ \hbar $}, which confirms that it is a \textit{boson}.

Having solved the inhomogeneous linear system \eqref{m_sys}, we get:
\begin{equation}
\label{m_sys_sol}
m_x=\chi_0 \frac{\omega_k \omega_0}{\omega^2_k - \omega^2} h_x + i\chi_0 \frac{\omega \omega_0}{\omega^2_k - \omega^2} h_y, \quad
m_y=\chi_0 \frac{\omega_k \omega_0}{\omega^2_k - \omega^2} h_y - i\chi_0 \frac{\omega \omega_0}{\omega^2_k - \omega^2} h_x,
\end{equation}
So, the susceptibility $\chi_{ij}$ of a ferromagnet in an external alternating magnetic field $\mathbf{h}(\mathbf{k},\omega)$ is a \textit{tensor}: $m_i=\chi_{ij} (\mathbf{k},\omega) h_j$, with components
\begin{equation}
\label{chi_ij_+Veiss}
\chi_{xx}=\chi_{yy}=\chi_0 \frac{\omega_k \omega_0}{\omega^2_k - \omega^2}, \quad
\chi_{xy}=-\chi_{yx}=i\chi_0 \frac{\omega \omega_0}{\omega^2_k - \omega^2}, \quad
\chi_{xz}=\chi_{yz}=\chi_{zz}=0.
\end{equation}
Note that $\chi_{ij}(\mathbf{H}_0)=\chi_{ji}(-\mathbf{H}_0)$, i.e. again magnetic gyrotropy of the medium is present. We also emphasize that the susceptibility tensor $ \chi_{ij} $ is a function of $ \mathbf{k} $ and $ \omega $, i.e. when the inhomogeneity of the exchange interaction field is taken into account, \textit{along with the frequency dispersion, spatial dispersion also appears}. The latter is small to the extent that $ a k $ is small in the long-wavelength limit $\lambda=2\pi/k\gg a$.

Let us now turn to the definition of normal spin waves propagating in a ferromagnet. To this end, it is necessary to substitute the obtained expressions for the susceptibility tensor $ \chi_{ij} $ into Maxwell's equations, assuming that the \emph{electromagnetic field is itself created by oscillations of magnetization}.
As follows from quantitative estimates, the frequency of normal waves in the considered long-wave approximation turns out to be low enough to to safely use, instead of the full Maxwell equations which include retardation effects, their abbreviated magnetostatic version \cite{LL8}:
\begin{equation}
\operatorname{rot} \mathbf{H}=0, \quad \operatorname{div} \mathbf{B}=0.
\end{equation}
Considering that $\mathbf{H}=\mathbf{H}_0 + \mathbf{h}$, $\mathbf{B}=4\pi\left(\mathbf{M}_0+\mathbf{m}\right)$, we have
\begin{equation}
\label{30}
\operatorname{rot} \mathbf{h}=0, \quad \operatorname{div} \mathbf{h} + 4\pi\operatorname{div} \mathbf{m}=0.
\end{equation}
Next we write the last of these relations in components, using the constitutive equation $m_i=\chi_{ij}h_j$:
\begin{equation}
\frac{\partial h_i}{\partial x_i} + 4\pi\chi_{ij}\frac{\partial h_j}{\partial x_i}=0.
\end{equation}
The first relation \eqref{30} implies the potentiality of the field
$\mathbf{h}(\mathbf{r})$, i.e.  $\mathbf{h}(\mathbf{r})=-\operatorname{grad}\varphi$ for some scalar function $\varphi(\mathbf{r})$, or $h_i=-\partial \varphi / \partial x_i$. Then for the potential $ \varphi $ we have the equation
\begin{equation}
\frac{\partial^2 \varphi}{\partial x_i \partial x_i} + 4\pi\chi_{ij}\frac{\partial^2 \varphi}{\partial x_i \partial x_j}=0,
\end{equation}
whose solution is sought in the form of a plane wave $\varphi(\mathbf{r},t)=\varphi_0 e^{i\mathbf{k}\mathbf{r} - i\omega t}$. Then
\begin{equation}
\left[k^2 + 4\pi k_i k_j \chi_{ij}(\mathbf{k},\omega)\right]\varphi(\mathbf{r},t)=0\,.
\end{equation}
Nontrivial $(\varphi\neq 0)$ solutions of this equation exist under the condition
\begin{equation}
\label{drel}
k^2 + 4\pi k_i k_j \chi_{ij}(\mathbf{k},\omega)=0.
\end{equation}
The equation \eqref{drel} is the \textit{dispersion law of spin waves in the long-wave approximation}. Substituting into it the components of the susceptibility tensor \eqref{chi_ij_+Veiss}, we get
\begin{equation}
\label{drel_finale}
k^2 + 4\pi \chi_0 \frac{\omega_k \omega_0}{\omega^2_k - \omega^2} \left(k^2_x + k^2_y\right)=0,
\end{equation}
Let us consider the obtained dispersion law in the case when the wave vector $ \mathbf{k} $ is directed at an angle $ \theta_k $ relative to the magnetization vector $\mathbf{M}_0\parallel\mathbf{z}$. Then $k^2_x + k^2_y=k^2\sin^2{\theta_k}$, and \eqref{drel_finale} takes the form
\begin{equation}
\label{drel_vybor_koord}
1=\chi_0 \frac{\omega_k \omega_0}{\omega^2 - \omega^2_k} 4\pi \sin^2{\theta_k},
\end{equation}
from where, using \eqref{omega_k}, we find the explicit form of the dispersion law:
\begin{equation}
\label{drel_yv}
\omega(k,\theta_k)=\gamma \left[ \left(H_0 + H_e (a k)^2\right) \left(H_0 + H_e (a k)^2 + 4\pi M_0 \sin^2{\theta_k}\right) \right]^{1/2}.
\end{equation}
In the limit of a strong external field $ H_e \gg H_0 $, we return to the dispersion law for spin waves \eqref{omega_k}. $ \omega $ depends not only on the modulus of the wave vector, but also on its direction with respect to the magnetization vector $ \mathbf{M}_0 $ due to the \textit{magnetostatic term} $4\pi M_0 \sin^2{\theta_k}$: the direction of the magnetic moment breaks the isotropy of the dispersion law of spin waves. Consider two limiting cases:

1) $\theta_k=0$. Then $\omega=\gamma \left(H_0 + H_e (a k)^2\right)$;

\begin{figure}[h!]
	\vspace{-0pt}
	\begin{center}
		\includegraphics[width=0.55\textwidth]{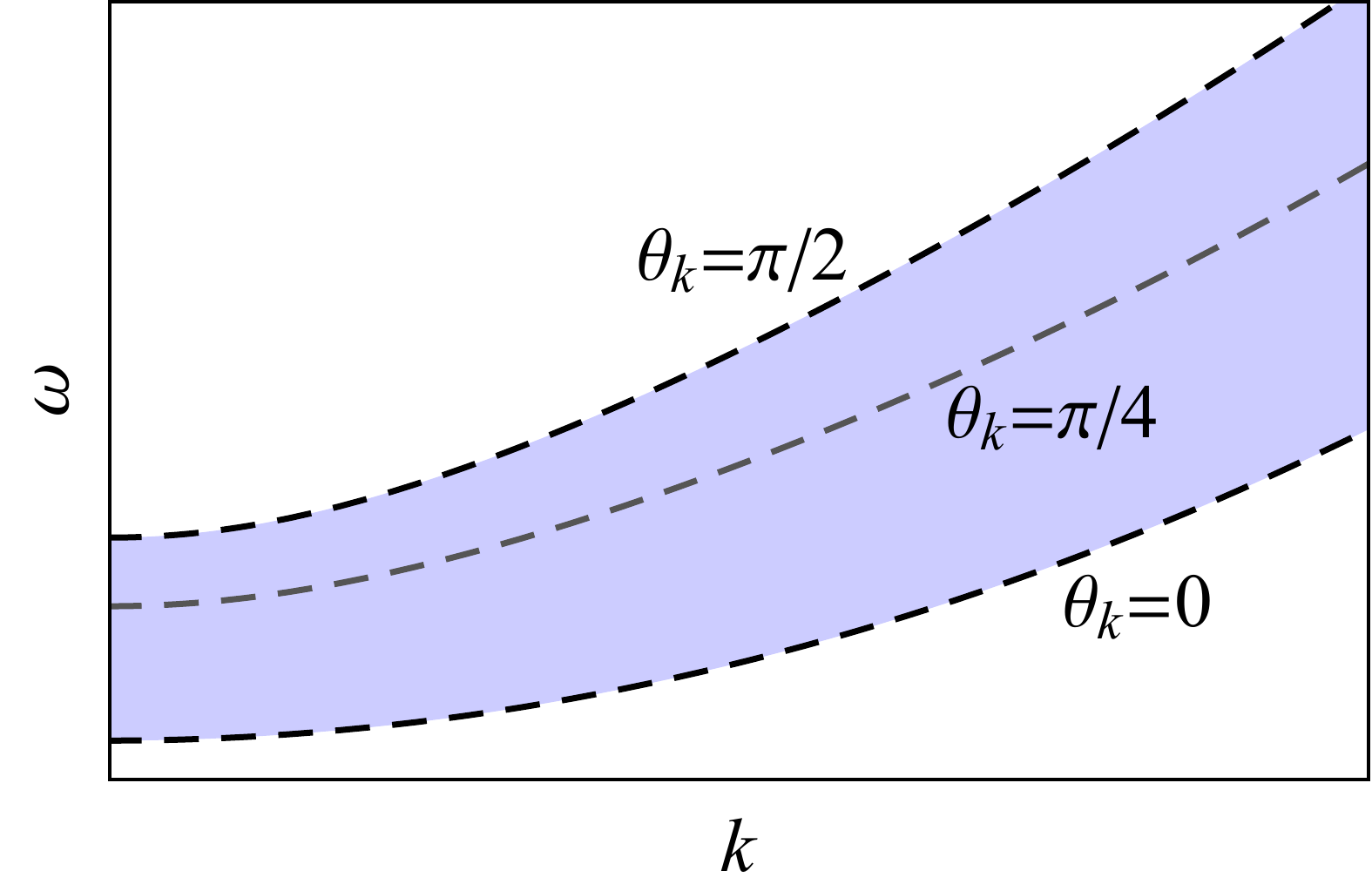}
	\end{center}
	\vspace{-25pt}
	\caption{The dispersion law for spin waves \eqref{drel_yv}. Dispersion curves $\omega(k,\theta_k)$ fill the area between the curves $\omega(k,0)$ and $\omega(k,\pi/2)$.}\label{drell}
	\vspace{+0pt}
\end{figure}

2) $\theta_k=\pi/2$. Then $\omega=\gamma \left[ \left(H_0 + H_e (a k)^2\right) \left(H_0 + H_e (a k)^2 + 4\pi M_0\right) \right]^{1/2}$.\\
The dispersion laws $\omega=\omega(k)$ corresponding to these two limiting cases are shown schematically in Fig. \ref{drell} with black dashed lines.

Spin waves in ferromagnets can be excited by external fields (for example, an alternating microwave magnetic field) and propagate in a ferromagnet similarly to sound waves. They are also excited by the thermal motion of atoms, which is the reason for the special temperature dependence of the magnetic and other properties of ferromagnets: the thermal activation of spin waves leads to a decrease in the magnetization of the sample (see, for example, \cite{Kittel}, p. 559), as well as to additional scattering conduction electrons on them and, consequently, to a decrease in the conductivity of the sample.

In quantum language, spin waves correspond to quasiparticles~-- magnons with energy $\varepsilon_k=\hbar\omega(\mathbf{k})$, which are elementary excitations of the spin system. The dispersion law of magnons can be alternatively obtained by considering a specific Hamiltonian of a ferromagnet (see e.g. \cite{ABP}; also \cite{Maekawa}, section 3.3.1).

\section{Exchange spin current\footnotemark \label{exchange spc}}\

\footnotetext{This section builds on the presentation of Part 3 of the book \cite{Maekawa}; see also the article \cite{Kajiwara}.}


The spin current $ \mathbf{J}^s $ in the nondissipative case can be determined from the continuity equation for the magnetic moment (or spin) \eqref{js_eqn}, in which $ \boldsymbol{\mathcal{S}} $ is the density of the spin moment of any nature, including the spin of lattice ions.

The dynamical equation for the magnetization in ferromagnets, \eqref{Hilb_eqn} or \eqref{LL_eqn}, can be rewritten as a continuity equation for the magnetic moment (see equation (69.11) in the book \cite{LL9}, p. 368), in which the magnetic flux tensor moment $ \mathbf{J}^s $ is the so-called \textit{exchange spin current} (associated, as we will show, with inhomogeneous exchange interaction, and hence with spin waves).

Consider the equation \eqref{Hilb_eqn} without damping.
In the absence of an external field, the effective field is, according to \eqref{H_eff}, $ \mathbf{H}_{eff}(\mathbf{r})=\lambda_w \mathbf{M}(\mathbf{r}) + \alpha a^2 \Delta \mathbf{M} $, and the first term can be omitted, as before, when writing the equation of motion. The external field $ \mathbf{H}_i $ directed along the $ \textbf{z} $ axis can be taken into account by adding it to $ \mathbf{H}_{eff} $.
Thus, we have the equation:
\begin{equation}
\label{eqn_obm_sp_tok}
\frac{\partial\mathbf{M}(\mathbf{r},t)}{\partial t}=\gamma [\mathbf{M}\times \mathbf{H}_i]-A [\mathbf{M}\times\Delta \mathbf{M}(\mathbf{r})],
\end{equation}
where the so-called \textit{exchange stiffness} $ A = -\gamma \alpha a^2 > 0 $ is introduced (here $ \alpha $ is the inhomogeneous exchange constant defined in the previous section).
Let us show that the second term on the right-hand side of \eqref{eqn_obm_sp_tok} is reduced to the form 
\begin{equation}
\label{rhs_Js obmen}
-\nabla \cdot \mathbf{J}^s,
\end{equation}
where $ \mathbf{J}^s $ is the exchange spin current (second rank) tensor:
\begin{equation}
\label{Js obmen}
J^s_{\alpha \beta}=A[\mathbf{M}\times \nabla_\beta \mathbf{M}]_\alpha=A\varepsilon_{\alpha\mu\nu}M_{\mu}\nabla_\beta M_\nu.
\end{equation}
Here $ \nabla_\beta $ denotes a partial derivative $ \partial/\partial x_\beta $, and the result of the action of $ \nabla \cdot $ operator in \eqref{rhs_Js obmen} is the vector $ \nabla_\beta J^s_{\alpha \beta} $. We have:
\begin{eqnarray}
\label{}
-\left(\nabla \cdot \mathbf{J}^s\right)_\alpha=-\nabla_\beta J^s_{\alpha \beta}
&=&
-A \nabla_\beta (\varepsilon_{\alpha\mu\nu}M_{\mu}\nabla_\beta M_\nu) \nonumber \\
&=&
-A\varepsilon_{\alpha\mu\nu} (\nabla_\beta M_{\mu}\nabla_\beta M_\nu + M_{\mu}\nabla_\beta \nabla_\beta M_\nu).
\end{eqnarray}
The term $ -A \varepsilon_{\alpha\mu\nu} \nabla_\beta M_{\mu}\nabla_\beta M_\nu = -A [\nabla_\beta \mathbf{M} \times \nabla_\beta \mathbf{M}]=0$, and finally we obtain
\begin{equation}
\label{}
-\left(\nabla \cdot \mathbf{J}^s\right)_\alpha=-A\varepsilon_{\alpha\mu\nu} M_{\mu}\nabla_\beta \nabla_\beta M_\nu,
\end{equation}
which coincides with the coordinate notation of the second term on the right-hand side of~\eqref{eqn_obm_sp_tok}. So, we have the equation
\begin{equation}
\label{eqn_obm_sp_tok_good}
\frac{\partial\mathbf{M}(\mathbf{r},t)}{\partial t}=\gamma [\mathbf{M}\times \mathbf{H}_i]-\nabla \cdot \mathbf{J}^s,
\end{equation}
which, in the absence of an external field $ \mathbf{H}_i $, is a spin conservation law in the form \eqref{js_eqn}. Note that the $ \mathbf{z}$-component of this equation (for external field $ \mathbf{H}_i\,||\,\mathbf{z}$), $ \partial M_z/\partial t + \nabla_\beta \mathbf{J}^s_{z \beta} = 0 $, is the conservation law of the spin projection onto the $ \mathbf{z} $ axis along which the external magnetic field is directed. 

Let us emphasize that the nature of the exchange spin current is \textit{fundamentally different} from the nature of the electron spin current \eqref{spin current} associated with the physical transfer of magnetic moments of individual electrons. \emph{Exchange spin current can exist in ferromagnetic insulators}, while the spin current of conduction electrons can only flow in metallic magnets.

Finally, considering the dynamic equation \eqref{Hilb_eqn} with the relaxation term on the right-hand side, we arrive at an equation of the form \eqref{js_eqn+T} with
\begin{equation}
\mathbf{T}=\frac{\alpha}{M} [\mathbf{M}\times\frac{\partial \mathbf{M}(t)}{\partial t}].
\end{equation}

In the key experiment \cite{Kajiwara}, the transfer of spin current through the ferromagnetic insulator $ Y_3 Fe_5 O_{12} $ by means of an exchange spin current of spin waves was demonstrated. In the three-layer structure $ Pt^{(1)}/Y_3 Fe_5 O_{12}/Pt^{(2)} $, the electric current in the $ Pt^{(1)} $ layer caused a spin current in the perpendicular direction due to spin Hall effect. This spin current, reflected from the metal($Pt^{(1)}$)-magnetic insulator($Y_3 Fe_5 O_{12}$) boundary, exerted a spin moment \eqref{passive diffusion} on the magnetic dielectric in which magnons with a finite wavelength were excited\footnotemark\footnotetext{The finite length of the excited spin waves (in comparison with magnons with $ \lambda \rightarrow \infty $ in ``ideal'' spin pumping) is explained by the nonlocality of the moment acting from the external (spin) current on the magnetic structure of the insulator.}.
The insulator thickness was chosen to be shorter than the characteristic damping length of spin waves. At the boundary between a magnetic dielectric($Y_3 Fe_5 O_{12}$)-metal($Pt^{(2)}$), spin waves were converted back into a spin current in $ Pt^{(2)} $ by the mechanism of spin pumping, and due to the inverse spin Hall effect, a voltage has been detected at the boundaries of the $ Pt^{(2)} $ sample.


\bigskip
\bigskip
\bigskip
\begin{center}
	\uppercase{Control questions}
\end{center}
\noindent 1. What are the \textit{free} spin waves?

\noindent 2. What is the physical nature of the magnetostatic term $ 4\pi M_0 \sin^2{\theta_k} $ in the dispersion law \eqref{drel_yv}?

\noindent 3. Show that for the ``non-uniform exchange'' term $ \alpha a^2 \sum_i \left(\partial^2 \mathbf{M} / \partial x_i^2\right) $ in \eqref{H_eff} (which plays a key role for spin waves), there corresponds a part of the magnetic energy density of the form $ W_\alpha = C \alpha \cdot \sum_i \left( \partial \mathbf{M} / \partial x_i \right)^2 $, where $ C > 0 $.

\noindent 4. What is the physical nature of the exchange spin current? Why is it necessarily a tensor? Explain what each of its tensor indexes stands for.

\setchapterpreamble{%
	\dictum{%
		
		In spintronics, a number of effects of spin-dependent interaction of spin-polarized electrons and magnetization texture are known.
		The variety of such interactions is reflected in different types of additional \textit{spin torques} on the right side of the Hilbert \eqref{Hilb_eqn} or Landau-Lifshitz equation \eqref{LL_eqn}. In this lecture, we will discuss the main types of spin torques and give an example of the derivation of spin torques in a conducting ferromagnet.
		
	}%
	\vspace{24pt}%
}

\chapter{Spin torques}\

\section{Types of spin torques}\ 

The Landau-Lifshitz equation without damping~\eqref{Hilb_eqn} is similar to the equation of motion of a top under the action of an external torques (in the case of a top~-- the torque of gravity)\footnotemark: magnetization $ \mathbf{M} $ is similar to the angular momentum of the top $ \mathbf{L} = I \boldsymbol{\omega} $, the term $ \gamma\mathbf{M}\times\mathbf{H}^* $ is similar to the torque of external forces $ m\mathbf{r}_c\times \mathbf{g}$. This similarity is a consequence of the similarity of the potential energies $ \mathcal{U} $ of two systems: in the case of a top, $ \mathcal{U}=-m (\mathbf{g}\cdot\textbf{r}_c )$, and in the case of a magnet, the free energy in the effective field is $ \mathcal{U} \equiv \mathcal{F}=-(\mathbf{M}\cdot\textbf{H}^* )$. Thus, only the generalized forces differ. This analogy allows us to call the term $ \gamma\mathbf{M}\times\mathbf{H}^* $ the \emph{spin torque}. Continuing the analogy, we will call by the spin torque any term $ \mathbf{T}_i $ on the right-hand side of the magnetization dynamic equation $ d\mathbf{M}(\mathbf{r},t)/dt=\sum_i \mathbf{T}_i(\mathbf{r},t) $.

\footnotetext{However, with a difference in the character of damping: in the case of a top, dissipation leads to an increase in the angle between the axis of the top and the vertical, and in the case of a magnet, vice versa.}

Many types of spin torques are known in spintronics. First, we can conditionally divide the moments into the \emph{active}, the spin physics of which is due to either \emph{dynamics} or \emph{spatial inhomogeneity} of magnetization, and the \emph{passive}, arising when a magnet with static homogeneous magnetization is either irradiated by a \textit{spin current}, or surrounded by a \textit{near-surface spin density}. Of course, there often could be torques of mixed type, but these could be usually described by the sum of corresponding elementary torques. 

Let us consider the active torques first.
There are two types of active spin torques: \emph{dynamical} and \emph{induced by the current $ \mathbf{j}_e $ if the texture is inhomogeneous}. These can be both surface and volume torques. Their phenomenological expressions read:
\begin{eqnarray}
\label{dynamical ST}
\mathbf{T}_{dynamical}
&=&
-
\left(
\mathcal{A} 
-
\tilde{\alpha} \mathbf{m} \times 
\right)
\partial_t \mathbf{m}
,
\\
\label{current-induced ST}
\mathbf{T}_{current-induced}
&=&
\frac{\mu_B P}{e M_s} (1 - \xi \mathbf{m} \times)(\mathbf{j}_e \cdot \nabla) \mathbf{m}
,
\end{eqnarray}
where $ \mathbf{m} = \mathbf{M} / M_s $ is the unit magnetization vector, and $ \mathcal{A} $, $ \tilde{\alpha} $, $ \xi $ are usually positive (model-dependent) constants. 
The torque \eqref{current-induced ST} describes the effect of current on an inhomogeneous magnetic structure (Neel and Bloch domain walls, vortex domain walls)~\cite{Theory of current-driven magnetization dynamics in inhomogeneous ferromagnets}, and therefore applies only to conducting ferromagnets (where $ P $is an equilibrium polarization of conduction electrons). Torque \eqref{dynamical ST} can arise both as a volume effect in conducting ferromagnets~\cite{Zhang Li, Saslow 07} (see section \ref{Zhang Li}), and as a surface effect in ferromagnetic insulators if they are adjacent to the normal metal~\cite{TseBra 2005} (see the lecture on spin pumping \ref{SP}).

\begin{figure}[h!]
	\centering
	\includegraphics [width=0.8\textwidth]{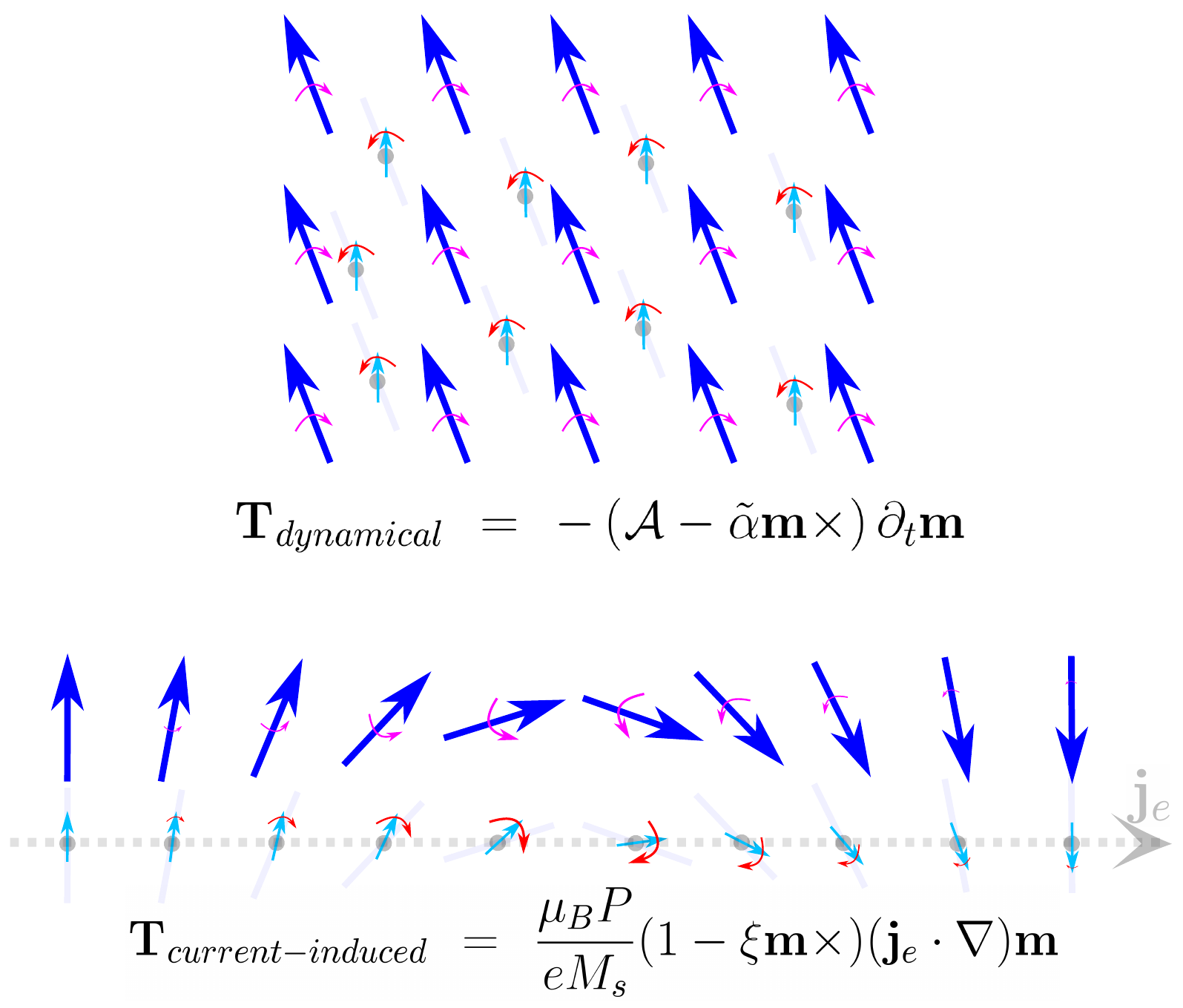}
	\caption{Two types of active spin torques in ferromagnets: dynamic (top panel; schematically shows a variant of the volume effect in a ferromagnetic metal, $ \mathbf{M} $ precesses counterclockwise), and current-induced (bottom panel). Blue arrows represent the spins of lattice ions, blue arrows~-- the spins of conduction electrons, curved arrows~-- mutual torques acting between them.}
	\label{2_STs}
	\vspace{-0 pt}
\end{figure}

The two torques, \eqref{dynamical ST} and \eqref{current-induced ST}, are surprisingly similar in structure, since they are united by one physical mechanism~-- \emph{delay of conduction electrons with respect to the magnetization of lattice spins}. In the case of dynamic torque \eqref{dynamical ST}, this is the ``\emph{time lag}'' 
when magnetization is uniform (Fig. \ref{2_STs}, top panel). In the case of current-induced torque \eqref{current-induced ST}, this is the ``\emph{delay in space}'':
at the same moment in time, the spins of conduction electrons turn parallel to the magnetization at the point $ \mathbf{r} $ only ``further downstream'', at the point $ \mathbf{r} + \delta r (\textbf{j}_e / j_e) $ (Fig. \ref{2_STs}, bottom panel).
The various terms of the torques \eqref{dynamical ST}, \eqref{current-induced ST} describe two delay effects: \emph{adiabatic, or conservative} (without $ \mathbf{m} \times $) and \emph{non-adiabatic, or dissipative} (with $ \mathbf{m} \times $). The former is due exclusively to the exchange interaction; the latter is associated with the dissipation of energy into other degrees of freedom.

\begin{figure}[h!]
	\centering
	\includegraphics [width=0.6\textwidth]{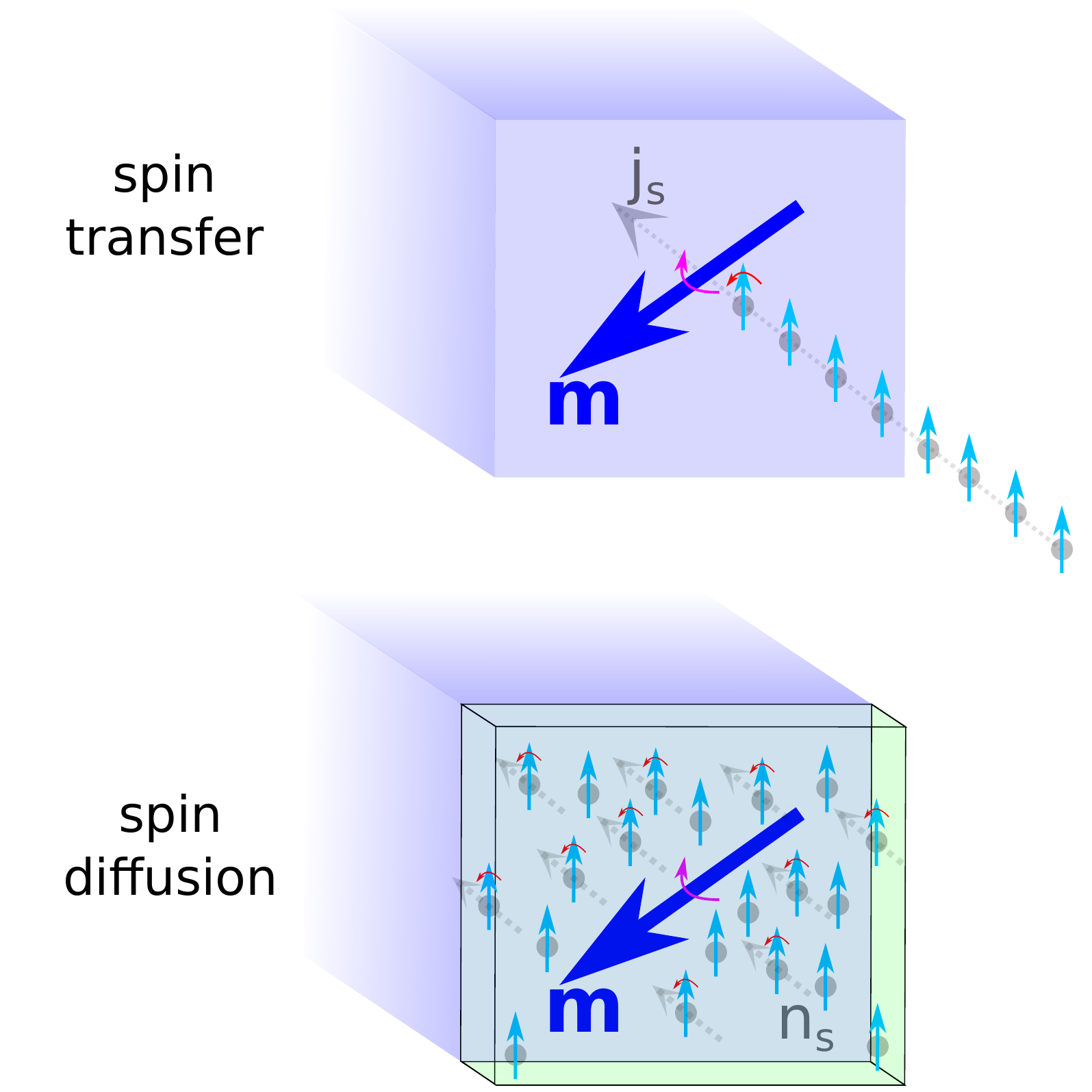}
	\caption{Passive spin torques in ferromagnets: spin current-induced torque \eqref{passive transfer} (also called the \textit{Spin Transfer Torque}, \textbf{STT}) (top panel), and induced by diffusion of the near-surface spin density \eqref{passive diffusion} (bottom panel).}
	\label{SOTs}
	\vspace{-0 pt}
\end{figure}

Let us go back to the current-induced \eqref{current-induced ST} torque. In the Stoner band model, with taking into account the (s-d) interaction of conduction electrons with the lattice, the contribution to Hilbert damping from conduction electrons is $ \tilde{\alpha}_{Stoner\,(s-d)} =\xi n_0 / M_s \lesssim \xi $\footnotemark\footnotetext{See section \ref{Zhang Li}.}, where $ n_0 $ is the local equilibrium spin density of conduction electrons (aligned with $ \mathbf{M} $). The constant $ \alpha $ also includes other contributions: magnon-phonon interaction, spin-orbit interaction of conduction electrons with impurities, and other mechanisms. Therefore, in moderately pure metallic ferromagnets it may occur that $ \alpha \sim \xi $. As it turns out, this regime has a distinguished, ``exactly integrated'' dynamics of inhomogeneous magnetization~\cite{Theory of current-driven magnetization dynamics in inhomogeneous ferromagnets}, in which the domain wall moves without compression, and the critical current at Walker breakdown, at which the magnetization begins to oscillate and the average velocity of the domain wall drops rapidly, diverges as $ \propto 1 \Big/ | 1 - \xi / \alpha | $.


Let us now consider the passive torques caused by an external spin current (Fig. \ref{SOTs}, top panel) or near-surface spin density (Fig. \ref{SOTs}, bottom panel):
\begin{eqnarray}
\label{passive transfer}
\mathbf{T}_{transfer}
&=&
-
\frac{\gamma}{M_s V}
\mathbf{m} \times 
\mathbf{I}_s \times \mathbf{m}
,
\\
\label{passive diffusion}
\mathbf{T}_{diffusion}
&=&
-
\frac{\gamma}{M_s V}
\frac{1}{4\pi}
\left(
g^{\uparrow\downarrow}_i
+
g^{\uparrow\downarrow}_r \mathbf{m} \times 
\right)
\boldsymbol{\mu}_s \times \mathbf{m}
.
\end{eqnarray}
The expression for the transport spin moment \eqref{passive transfer} describes the effect of the spin current $ \mathbf{I}_s $~\cite{Slonczewski 96} precessing around $ \textbf{m} $. For a given spin current density $ \mathbf{j}_s = \mathbf{I}_s / S $, this moment is inversely proportional to the sample thickness $ d $: $ T_{transfer} \propto S / V \propto 1 / d $.
The spin torque \eqref{passive diffusion} describes the contribution from the diffusion of the nonequilibrium spin density with a \textit{spin chemical potential} (\textit{spin accumulation}) $ \boldsymbol{\mu}_s $ at the boundary of the ferromagnet; complex coefficient $ g^{\uparrow \downarrow} = g^{\uparrow \downarrow}_r + i g^{\uparrow \downarrow}_i $, called \textit{spin-mixing conductance}) \eqref{spin-mixing g}, characterizes the spin-dependent reflection of electrons from a surface during diffusion. Such a spin torque can be induced by a current in the spin Hall effect or the inverse spin-galvanic effect (\textbf{ISGE})~\cite{SOT}, when electrons become spin-polarized near the boundary of a heavy metal and a ferromagnet due to the enhanced spin-orbital interaction~\cite{SOT}. The effect inverse to the diffusion moment, the \textit{spin pumping}, will be discussed in detail in the next section.

The \eqref{passive diffusion} expression and the \eqref{dynamical ST}, \eqref{current-induced ST} torques have a similar structure.
The term in \eqref{passive diffusion} $ \propto g^{\uparrow \downarrow}_i $ is called \emph{field-like torque}, because it describes the interaction of the ``macrospin'' of the ferromagnet and the incoming / reflected spins through exchange field. The term $ \propto g^{\uparrow \downarrow}_r $ is called \emph{dissipative-like, or (anti)damping torque}, since it describes the additional damping / gain of the dynamics of the ``macrospin'' of a ferromagnet due to ``absorption'' the spin component of the incoming / reflected spins which is transverse to the magnetization.
The term $ \propto g^{\uparrow\downarrow}_r $ usually turns out to be an order of magnitude larger than $ \propto g^{\uparrow\downarrow}_i $ due to the averaging of the field component from incoming / reflected electrons with different wave vectors~\cite{TseBra 2005, SOT}. The dissipative analogue of the moment \eqref{passive transfer} is also rarely taken into account, since for typical transition metal ferromagnets the spin coherence time is much longer than the spin precession period.

In fact, experiments are often explained by the contributions from both types of moments \eqref{passive transfer} and \eqref {passive diffusion}, since the spin current leads to the accumulation of a dynamically equilibrium spin density near the boundary, which depends on the surface properties and the ratio of dynamic frequency $ \omega \propto \partial\mathbf{m} / \partial t $ to the spin relaxation frequency $ 1/\tau_{sc} $.
Finally, we note that the transport~\eqref{passive transfer} and diffusive~\eqref{passive diffusion} spin torques can cause a reorientation of the magnetization of a ferromagnet, and at a current density/surface spin density exceeding the threshold set by the Gilbert damping constant, they can cause \textit{auto-oscillations} of the sample magnetization vector. The phenomenon of auto-oscillations of magnetization induced by an external spin current is used in the so-called \textit{spin-torque nano-oscillators}~\cite{Demidov17}.

\section{Role of nonequilibrium conduction electrons response in the dynamics of magnetization of ferromagnets: 4 spin torques
\label{Role of nonequilibrium elns in FM} \label{Zhang Li}}\

In this section, we will consider the 4 volume spin torques with which nonequilibrium electrons in a conducting ferromagnet act on the magnetization, as was obtained in a seminal paper by Zhang and Li~\cite{Zhang Li}. The derivation of the equations is semiclassical, and is based on the simplest Hamiltonian of the (s-d) interaction, the relaxation time approximation, and perturbation theory.

Non-equilibrium electrons can appear in a metallic magnet, for example, when it is placed in an external (constant) electric or alternating magnetic field. The electric field directly induces a charge current carrying spin due to the splitting of the electronic levels by the exchange field; an alternating magnetic field causes the dynamics of magnetization, which, in turn, creates a nonequilibrium spin density due to the exchange field.

Thus, \emph{although the reasons for the emergence of a nonequilibrium spin density may be different, the physics of its interaction with magnetization has universal features of an exchange interaction}.

Following \cite{Zhang Li}, we take the simplest Hamiltonian of the (s-d) interaction of localized spins $ \hat{\mathbf{S}} $ (magnetization) and spins $ \hat{\mathbf{s}} $ of conduction electrons:
\[
\hat{H}_{sd} = 
-J_{ex} \hat{\mathbf{s}} \cdot \hat{\mathbf{S}}
,
\]
where $ J_{ex} $ parameterizes the exchange (ferromagnetic) interaction. Since the ``macrospin'' $ \hat{\mathbf{S}} $ has large eigenvalues, we can consider it as a classical vector $ \mathbf{S} = -S \cdot \mathbf{M}(r,t) / M_s $ ($ M_s $ is the equilibrium magnetization of localized spins).

Next we write the Heisenberg equation for the spin operator of one conduction electron, assuming the presence of two contributions to the spin Hamiltonian: the Hermitian one, which corresponds to the (s-d) interaction, and the non-Hermitian one, which describes the relaxation of the spin (during its collisions with impurities, defects, other electrons, and lattice ions):
\begin{equation}
\label{initial_s_eqn}
\frac{\partial \mathbf{s}}{\partial t}
+
\nabla \cdot \hat{\mathbf{J}}
=
\frac{1}{i \hbar} \left[ \mathbf{s}, \hat{H}_{sd} \right]
-
\Gamma_{relaxation} (\mathbf{s}, t)
,
\end{equation}
where $ \hat{\mathbf{J}} $ is the spin current operator.
Since we need to know the law of relaxation of the spin density, let us average this equation~-- both quantum mechanically and over the states of conduction electrons. Let us define the spin current density of conduction electrons as the average $ \mathbf{m}(\mathbf{r},t) = \langle\mathbf{s}\rangle $, and the spin current density~-- as the average $ \mathcal{J}(\mathbf{r},t) = \langle\hat{\mathbf{J}}\rangle $. Averaging \eqref{initial_s_eqn} and expanding the commutator gives
\begin{equation}
\label{m_dynamic_eqn}
\frac{\partial \mathbf{m}(\mathbf{r},t)}{\partial t}
+
\nabla \cdot \mathcal{J}(\mathbf{r},t)
=
-
\frac{1}{\tau_{ex} M_s} \left[ \mathbf{m}(r,t) \times \mathbf{M}(r,t) \right]
-
\langle \Gamma_{relaxation} (\mathbf{s}, t) \rangle
,
\end{equation}
where $ \tau_{ex} = \hbar / S J_{ex} $.

Let us linearize the equation \eqref{m_dynamic_eqn}, taking as the first approximation the quantities adiabatically following $ \mathbf{M}(\mathbf{r},t) $: 
\[
\mathbf{m}(\mathbf{r},t) = n_0 \mathbf{M}(\mathbf{r},t) / M_s + \delta\mathbf{m}(\mathbf{r},t)
,
\]
where $ n_0 $ is the equilibrium spin density of conduction electrons,
\[
\mathcal{J}(\mathbf{r},t) = (-\mu_B P / e) \mathbf{j}_e \otimes \mathbf{M}(\mathbf{r},t) / M_s + \delta\mathcal{J}(\mathbf{r},t) 
,
\]
where $ \mathbf{j}_e$ is the charge current density, $ P$~-- is the degree of polarization of the equilibrium spin current in a ferromagnet (here $ \otimes $ stands for the tensor product sign). The relaxation term $ \langle \Gamma_{relaxation} (\mathbf{s}, t) \rangle $ is linearized in the kinetic tau-approximation with respect to nonequilibrium magnetization,
\[
\langle \Gamma_{relaxation} (\mathbf{s}, t) \rangle
=
\delta\mathbf{m}(\mathbf{r},t) / \tau_{sf}
,
\]
where $ \tau_{sf} $ is the spin relaxation time. 
We will consider the linear response $ \delta\mathbf{m}(\mathbf{r},t) $ for the charge current $ \mathbf{j}_e $ and the frequency of magnetization precession $ \partial \mathbf{M} / \partial t \propto \omega $; then the derivative $ \partial (\delta\mathbf{m}) / \partial t $ can be neglected. 
We also assume that nonequilibrium spin current arises as a result of diffusion of nonequilibrium spin density, $ \delta\mathcal{J} = -D_0 \nabla \delta\mathbf{m}(\mathbf{r},t) $, where $ D_0 $ is the diffusion coefficient. Substituting the linearized expressions into the equation \eqref{m_dynamic_eqn}, we get the equation for $ \delta\mathbf{m}(\mathbf{r},t) $ \cite{Zhang Li}. 
Assuming further that the characteristic dimensions of the magnetization inhomogeneity $ \mathbf{M(\mathbf{r},t)} $ (for example, the width of the domain wall $ w $) is much larger than the transport coherence length for the spin $ \lambda \sim \sqrt{D_0 \tau_{sf}} $, we exclude the term $ \propto D_0 \nabla^2 \delta\mathbf{m} $; then the equation for $ \delta\mathbf{m} $ becomes algebraic. 
Next, we calculate the generalized (spin) torque acting on the magnetization $ \mathbf{M(\mathbf{r},t)} $, $ \mathbf{T}(\mathbf{r},t) = + \frac{1}{\tau_{ex} M_s} \left[ \mathbf{m}(r,t) \times \mathbf{M}(r,t) \right] = + \frac{1}{\tau_{ex} M_s} \left[ \delta\mathbf{m}(r,t) \times \mathbf{M}(r,t) \right] $ (torque in \eqref{m_dynamic_eqn} with opposite sign), and finally get \cite{Zhang Li}:
\begin{eqnarray}
\label{bulk_ST}
\mathbf{T}
&=&
\frac{1}{1+\xi^2}
\Big(
-
\frac{n_0}{M_s}
\frac{\partial \mathbf{M}}{\partial t}
+
\frac{\xi n_0}{M_s^2}
\mathbf{M} 
\times 
\frac{\partial \mathbf{M}}{\partial t}
- 
\nonumber \\
&&
-\frac{\mu_B P}{e M_s^3}
\mathbf{M} \times 
\left[
\mathbf{M} 
\times 
(\mathbf{j}_e \cdot \nabla)\mathbf{M}
\right]
-
\frac{\mu_B P \xi}{e M_s^2}
\left[
\mathbf{M} 
\times 
(\mathbf{j}_e \cdot \nabla)\mathbf{M}
\right]
\Big),
\end{eqnarray}
where $ \mathbf{T} $, $ \mathbf{M} $, $ \mathbf{j}_e $, as before, are functions of $ (\mathbf{r},t) $, and the notation $ \xi = \tau_{ex} / \tau_{sf} $ is introduced.
The first two terms in \eqref{bulk_ST} renormalize the gyromagnetic constant $ \gamma $ and the damping constant $ \alpha $ in the Hilbert equation \eqref{Hilb_eqn}. These two terms describe the \emph{breathing Fermi surface effect} proposed by Kambersky \cite{Kambersky_BFS} \footnotemark\footnotetext{Recall that in this section we have obtained the result taking into account the usual spin relaxation on impurities, defects, other electrons, phonons, without taking into account spin-dependent (spin-flip) scattering. The latter also contributes to the Hilbert damping, as, for example, in the case of spin pumping (see lecture~\ref{SP}).}:
when the magnetization changes in time, the spins of conduction electrons follow its direction, with a lag of the order of spin relaxation time.
The last two terms represent the \textit{current-driven effect} and are proportional to the magnetization gradient. Moreover, the third term could be simplified by expanding the double cross product and noticing\footnotemark\footnotetext{Note that those components of the torques obtained in perturbation theory which are aligned with the local magnetization $ \mathbf{M}(\mathbf{r}, t) $, and therefore should change its value $ M_s = |\mathbf{M}(\mathbf{r},t)| $, are a priori \textit{unphysical}, since magnetization is determined only by exchange forces at a given temperature. Such perturbations should not lead to non-smooth variations which could invalidate macroscopic approach, and can only \textit{change the direction of the local magnetization}. For this reason, those components of torques could be discarded immediately.} that $ \mathbf{M} \left( \mathbf{M} \cdot (\mathbf{j}_e \cdot \nabla)\mathbf{M} \right) = \mathbf{M} \sum_i j_{e\, i} \left( \mathbf{M} \cdot d \mathbf{M} / d x_i \right) = \mathbf{M} \sum_i (j_{e\, i}/2) d \mathbf{M}^2 / d x_i = 0 $ to the form:
\[
+\frac{\mu_B P}{e M_s}
(\mathbf{j}_e \cdot \nabla)\mathbf{M}
.
\]
The phenomenology of all four terms is then fully consistent with expressions \eqref{dynamical ST} (first two terms in \eqref{bulk_ST}) and \eqref{current-induced ST} (last two terms in \eqref{bulk_ST}). 

The fourth non-adiabatic term in \eqref{bulk_ST} has a simple qualitative explanation: the spin-polarized current $ \mathbf{e}_x \cdot (\mu_B P/e) j_{e\,x} (\mathbf{r}) $, flowing along the $ x $ axis from the point $ \mathbf{r} $ with local magnetization $ \mathbf{M}(\mathbf{r}) $ to the point $ \mathbf{r}+\mathbf{e}_x \delta x $ with local magnetization $ \mathbf{M}(\mathbf{r}+\mathbf{e}_x \delta x) $, transfers the component torque $ \propto (\mu_B P/e) j_{e\,x} \cdot \left( \mathbf{M}(\mathbf{r}+\mathbf{e}_x \delta x) - \mathbf{M}(\mathbf{r}) \right) \propto (\mu_B P/e) j_{e\,x} \left( \mathbf{e}_x \cdot \nabla \right) \mathbf{M} \equiv (\mu_B P/e) \left( \mathbf{j}_{e} \cdot \nabla \right) \mathbf{M} $ to the magnetization. An additional factor $ \xi = \tau_{ex} / \tau_{sf} $ describes the efficiency of torque transfer. A similar explanation is applicable to the second non-adiabatic term in \eqref{bulk_ST}, if we replace in the above reasoning $ \delta x $ by $ \delta t $ and the equilibrium current density $ j_{e\,x} $ by the local equilibrium spin density $ n_0 $.
The adiabatic terms (first and third in~\eqref{bulk_ST}) describe conservative moments of exchange forces.


\begin{center}
	\uppercase{Control questions}
\end{center}
\noindent 1. What is the mechanism underlying the spin torques \eqref{dynamical ST}, \eqref{current-induced ST}? What are the first (adiabatic) and second (non-adiabatic) terms related to?

\noindent 2. Which component of the spin current $ \mathbf{I}_s $, according to \eqref{passive transfer}, usually acts on the magnetization with a torque? Why? 

\noindent 3. Briefly explain the physical nature of the four terms in the bulk spin torque \eqref{bulk_ST} with which conduction electrons act on the magnetization $ \mathbf{M} $ of the lattice.

\setchapterpreamble{%
	\dictum{%
		
		In this lecture, we will consider the effect reciprocal to the spin transfer torque~-- the \textit{spin pumping}~\cite{TseBra 2005}, in which the dynamics of magnetization in a magnet is damped, while a spin current is released into adjacent layers of normal metal.

	}%
	\vspace{24pt}%
}

\chapter{Spin pumping \label{SP}}\

The classical dynamics of magnetization $ \mathbf{M} = \mathbf{m} M $ in the case of homogeneous FMR is described by the Landau-Lifshitz \eqref{LL_eqn} or Hilbert \eqref{Hilb_eqn} equation: $ \mathbf{m} $ precesses with the frequency $ \omega_0 $ \eqref{DisEq_FMR}, which generally depends on the external field $ H_0 $, demagnetizing factors $ N_i $ (sample shape), sample magnetization $ M $ and anisotropy constant $ K $. Damping coefficient $ \alpha $ in the Hilbert equation \eqref{Hilb_eqn}
is determined by the nature of the dissipative processes \emph{in the bulk} of the sample, i.e. damping is represented by a volume effect (moreover, $ \alpha \propto \omega $, where $ \omega $ is the precession frequency).

Nevertheless, in a number of FMR experiments with \emph{thin two-layer films} composed of ferromagnet and normal paramagnetic metal \textbf{FM/NPM} (\emph{Cu-Co} and \emph{Pt-Co}, with thickness $ \sim10\text{\AA}\div 10\mu m $), a significant \textit{increase in the Hilbert damping} coefficient was observed with a decrease in the film thickness, compared to its value in bulk \emph{Co} ferromagnets \cite{Tserkovnyak PRL, Tserkovnyak PRB}.
The observed correction to damping was found to be inversely proportional to the sample thickness, $ \tilde\alpha\propto 1/d $, indicating a \emph{surface effect}. An explanation of the experimental dependence of the correction to damping on the type of the normal metal was also required: the observed additional attenuation was much greater for two-layer \emph{\textbf{Pt}-Co} films than for \emph{\textbf{Cu}-Co}.

\begin{figure}[h!]
	\centering
	\includegraphics [width=0.5\textwidth]{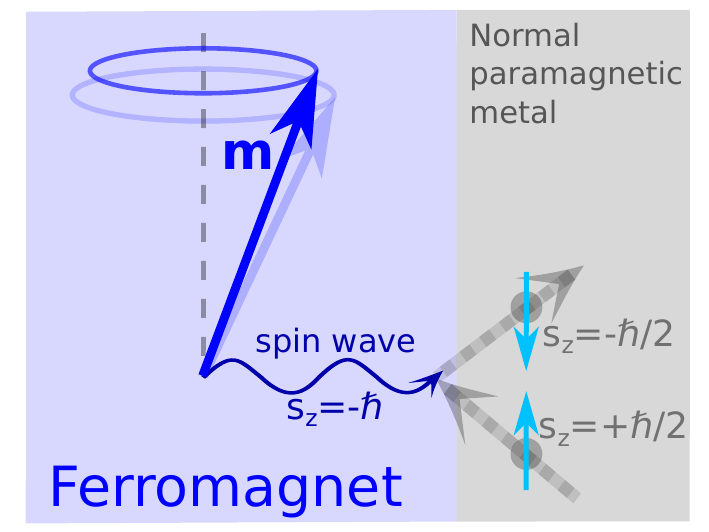}
	\caption{The elementary process underlying spin pumping: spin-flip of an electron ($ s_z = +\hbar/2 \rightarrow s_z = -\hbar/2 $) with the annihilation of a magnon of uniform magnetization dynamics with spin $ s_z = -\hbar $.
		In this process, only spin is transferred from the electron to the magnetic structure (the linear momentum is transferred to the lattice).
	}
	\label{spin_pumping_by_spin_flip}
	\vspace{-0 pt}
\end{figure}

A simple explanation \cite{Berger,Urban 01} was soon found: electrons flowing into/out of a ferromagnet in a thin surface layer are polarized parallel to layer magnetic moment, transferring the transverse component of spin to the magnetic structure as a whole (nonlocally), in the form of spin waves. In quantum language this corresponds to \emph{spin-flip of electron spin with emission/absorption of magnon} (see Fig. \ref{spin_pumping_by_spin_flip}). Thus, the surface torque $ \tau \propto S $ acts on a magnetic structure with magnetic moment $ \mathcal{M} = M V $, therefore the additional Hilbert damping reads $ \tilde\alpha\propto \tau / \mathcal{M} \propto S/V \propto 1/d $.

Let us consider in more detail the inverse effect of spin pumping~-- the \emph{spin-transfer torque} (\textbf{STT}) exerted by the spin current on magnet.
In the case of a ferromagnetic insulator, the current is completely reflected from the boundary, so that the spin-flip of the electrons occurs during scattering. 
In the case of a metallic ferromagnet, some of the electrons undergo scattering+spin-flip as well, while some of them enter the ferromagnet and begin to precess in the exchange field, due to which the transverse component of the spin density decays (algebraically) at a distance of the order of so-called \textit{spin coherence length},
$
\label{lambda fc}
\lambda_{fc}=\pi / (k^\uparrow_f-k^\downarrow_f)
$, 
where $ k^\uparrow_f \text{ and } k^\downarrow_f $ are the spin-dependent Fermi wave vectors \cite{Anatomy of spin-transfer torque, Penetration Depth of Transverse Spin Current}. 
In both cases, \emph{spin moment} appears on the right-hand side of the Hilbert equation, equal to the transverse component of the spin current transferred to the magnet as a whole \cite{Anatomy of spin-transfer torque}: $ \mathbf{I}_{s\,\perp} = \mathbf{I}_s - \mathbf{m}(\mathbf{I}_s \cdot \mathbf{m}) $. 

\emph{Spin pumping} is in fact the time-reverse of \textbf{STT}, and is responsible for additional damping: magnetization, losing the intensity of precession, injects the \textit{pure spin current} into the neighboring normal metal \footnotemark\footnotetext{In this case, the charge current, as we we will show, is absent on average.} The increased constant $ \tilde\alpha $ in the Hilbert equation can therefore be explained by the leakage of the spin-polarized current component $ \mathbf{I}_{s\,\perp} $ into the metal, if the expression for the spin current induced by the dynamics of $ \mathbf{M} $ will have a component of the form $ \frac{const}{M} [\mathbf{M}\times\frac{d\mathbf{M}(t)}{dt}] $. In the next section, we will discuss the derivation of the expression for the spin current $ \mathbf{I}_s $ with such a phenomenology, first performed in the works \cite{Tserkovnyak PRL, Tserkovnyak PRB}.

Let us return to the fact that the effects for normal metals $ Pl $ and $ Cu $ are quite different. Platinum (Pt), as a heavy metal (atomic number $ Z_{Pl}=78 $), has a much shorter \emph{spin-flip relaxation time $ \tau_{sf} $} than lighter copper (atomic number $ Z_{Cu}=29 $), since, as it is known, $ \tau_{sf}\propto Z^{-4} $.
The outflowing spin current in the case of copper creates a nonequilibrium concentration of spin-polarized electrons near the ferromagnet-normal metal interface, so that a significant spin countercurrent appears $ |\mathbf{I}^{back}_s| \sim |\mathbf{I}_s| $. Platinum, on the other hand, acts as an almost \textit{ideal spin sink}, and the spin countercurrent is almost negligible $ |\mathbf{I}^{back}_s| \ll |\mathbf{I}_s| $~-- in this case, the effect of increased damping is noticeable for sufficiently thin films.

\section{Scattering matrix method}\ 

The expression for the outflowing spin current~\cite{Tserkovnyak PRL,Tserkovnyak PRB} is naturally obtained in the scattering matrix formalism, known in mesoscopic physics as the \textit{Landauer approach}. In this section, we start with a short introduction to this formalism.
For a more detailed introduction, we refer te reader to the books~\cite{Nazarov book,Moskalets book}, and a review~\cite{UFN review}. 

Imagine an arbitrary (conducting) one-dimensional nanostructure in which electrons are scattered in quantized channels. For an adequate theoretical description of electron transport in a structure, it is necessary to construct its model~-- i.e. break it down into elements that are parameterized by certain characteristics.
The simplest case is a \emph{two-terminal conductor}~\cite{Nazarov book}. Let us break the nanostructure into a \emph{scattering region} connecting two \emph{large conductive reservoirs (leads)} through \emph{ideal waveguides (ballistic contacts)}. The distributions of electrons in the reservoirs will be assumed to be the equilibrium Fermi distributions at a given temperature with certain chemical potentials, $ \mu_L $ and $ \mu_R $. 
It turns out that all the necessary information (both about semiclassical transport properties and about quantum effects) can be obtained from a single matrix that depends on the electron energy~-- the \emph{nanostructure scattering matrix}, $ \hat{\mathbf{S}}(\varepsilon) $. 

In the simplest case of two reservoirs, the \emph{scattering matrix expresses a linear relationship between the asymptotics of the amplitudes of the incoming scattering states} \emph{far in the left reservoir ($ L $) and far in the right reservoir ($ R $)}. The scattering states are quantized in the transverse direction, therefore they are also called the \emph{transverse channels/modes}. After that, an analogy with the \emph{transfer-matrix method} in the problem of scattering by a one-dimensional potential in quantum mechanics is obvious.

The Fig. \ref{Tse2005-structure} shows an example of splitting a three-layer ``sandwich'' structure (\textbf{N/F/N}) into elements, as well as reflection ($ r,r' $) and transmission ($ t,t' $) amplitudes for one of the quantum channels. The scattering matrix has the following block structure (each block is a square matrix $ Ch(\varepsilon)\times Ch(\varepsilon) $, where $ Ch(\varepsilon) $ is the number of channels at an electron energy $ \varepsilon $) \cite{Nazarov book}:
\begin{equation}
\hat{S}(\varepsilon)
=
\begin{pmatrix}
\hat{S}_{LL} & \hat{S}_{LR} \\
\hat{S}_{RL} & \hat{S}_{RR}
\end{pmatrix}
\equiv
\begin{pmatrix}
\hat{r} & \hat{t}' \\
\hat{t} & \hat{r}'
\end{pmatrix},
\end{equation}
where all matrix elements depend on the electron energy $ \varepsilon $. The $ \hat{r} $ matrix describes the leftward reflection of electrons incident on the structure. Its element $ r_{\alpha \alpha'} $ is the amplitude of the following process: an electron incident from the left lead in the $ \alpha' $ channel is reflected to the left lead into the $ \alpha $ channel. $ |r_{\alpha \alpha'}| $ is then the probability of this process \cite{Nazarov book}. Three other blocks are defined similarly~-- for example, the matrix $ \hat{t}' $ describes the transmission of electrons incident on the structure from the right lead into the the left lead.

\section{Parametric charge pumping}\

The most famous formula in quantum transport~-- \textit{the Landauer formula}~-- gives the relationship between the current flowing through the nanostructure and the voltage between the reservoirs, through the matrix elements of $ \hat{\mathbf{S}} $ (the current flows into the reservoir with less chemical potential, all channels are assumed to be occupied at any given time)~\cite{Brower}:
\begin{equation}
\label{landauer}
\delta I_L = 
G \delta V 
=
\frac{2_s e^2}{h} \delta V
\sum_{\substack{\alpha\in L \\ \beta \in R}}
|S_{\alpha\beta}|^2
=
\frac{2_s e^2}{h} \delta V
\sum_{\alpha\in L}
\mathbf{Tr}
\left(
\mathbf{S}_{\alpha \beta}
\mathbf{S}^\dag_{\alpha \beta}
\right),
\end{equation}
where the sum is taken by all channels $ \alpha $ in the left reservoir and $ \beta $ in the right reservoir. All matrix elements (at low voltages $ \delta V = \mu_R - \mu_L $) are taken at the Fermi level. The coefficient $ 2_s $ originates from spin degeneracy: while we are studying the charge (spin-independent) transport, there will always be spin degeneracy.

Consider now \emph{parametric charge pumping}, when the reservoirs have the same chemical potential $ \mu_R = \mu_L $, and the scattering matrix is parametrized by several(at least by two) parameters $ X_i(t) $, which are \emph{out of phase} and depend on time \cite{Brower} (the importance of out-of-phase will be shown below using the example of two parameters $ X_1, X_2 $). Parameters $ X_i(t) $ can be values that parameterize the gate voltage profile, external magnetic field, Fermi level, and other quantities that change (quantum mechanical) properties of the scatterer.

Let us take the following approximations:

1) the adiabatic effect is studied, i.e. the first correction to the current, in terms of the characteristic frequency of parameter $ X_i $ variation, $ \omega\sim \dot{X_i}/X_i $, is found;

2) the electrons are scattered elastically, which corresponds to the unitary scattering matrix $ \mathbf{S} $, which immediately (adiabatically) follows the parameters $ X_i(t) $;

3) the transfer matrix $ \mathbf{S} $ weakly depends on the energy $ E $ of the scattering states, and the temperature difference and voltage applied to the system are small, which corresponds to taking the derivatives of the matrix $ \mathbf{S} $ at Fermi level, $ \varepsilon = \varepsilon_F $ ;

4) electron-electron interaction in the scatterer can be neglected. \\
In this approximation, P.W. Brower (1998) \cite{Brower}, building on the work of \cite{Buttiker 1994}, derived an expression for the current flowing through the reservoir $ m\,\,(m=Left/Right) $ (the contribution from all channels $ \alpha\in m $ entering the $ m $ reservoir is counted):
\begin{equation}
\label{current}
I(m,t)=\frac{\delta Q(m,t)}{\delta t} = e \frac{d n(m)}{d t} = e \sum_i \frac{\partial n(m)}{\partial X_i} \frac{d X_i (t)}{d t},
\end{equation}
where the so-called \textit{emissivities} $ \partial n(m)/\partial X_i $ read
\begin{eqnarray}
\label{charge emissivity}
\frac{\partial n(m)}{\partial X_i}
&=&
\frac{1}{4\pi i} \sum_\beta \mathbf{Tr}
\left(
\frac{\partial \mathbf{S}_{\alpha \beta}}{\partial X_i} 
\mathbf{S}^\dag_{\alpha \beta}
-
\mathbf{S}_{\alpha \beta} \frac{\partial \mathbf{S}^\dag_{\alpha \beta}}{\partial X_i}
\right)
=\nonumber \\
&=&
\frac{1}{2\pi} \sum_{\substack{\beta \in \{m,m'\} \\ \alpha\in m}} \textbf{Im} \left( \frac{\partial S_{\alpha \beta,m m'}}{\partial X_i} S^*_{\alpha \beta,m m'} \right)
.
\end{eqnarray}
Keeping in mind the further consideration of spin pumping, it is important to show where the Hermitian conjugation $ \dag $ comes from in the first expression in \eqref{charge emissivity}. In the secondary quantization representation, $ \mathbf{S} $ linearly connects the operators $ \mathbf{b}_{\alpha,m}(E) $ of annihilation of particles with energy $ E $ entering the reservoir $ m $ through the channel $ \alpha $, and the operators $ \mathbf{a}_{\beta,m'}(E) $ of annihilation of particles with energy $ E $ leaving the reservoir $ m' $ through the channel $ \beta $:
\begin{equation}
\label{abS}
\mathbf{b}_{\alpha,m}(E)=\sum_{\beta,m'} S_{\alpha \beta,m m'} \mathbf{a}_{\beta,m'}(E).
\end{equation}
Hermitian conjugation provides a connection between corresponding creation operators:
\begin{equation}
\label{abS dag}
\mathbf{b}^\dag_{\alpha,m}(E)=\sum_{\beta,m'} S^*_{\alpha \beta,m m'} \mathbf{a}^\dag_{\beta,m'}(E).
\end{equation}
When calculating the statistical average for the charging current in the reservoir $ m $ \cite{Buttiker 1994},
\begin{equation}
\label{current in a b}
I_m (t)=\frac{e}{h}\sum_{\alpha \in m} \int dE dE' e^{i(E-E')t/\hbar} 
(
\mathbf{a}^\dag_{\alpha,m}(E)\mathbf{a}_{\alpha,m'}(E')-
\mathbf{b}^\dag_{\alpha,m}(E)\mathbf{b}_{\alpha,m'}(E')
),
\end{equation}
it is necessary to find statistical averages of the form $ \langle \mathbf{b}^\dag_{\alpha,m}(E) \mathbf{b}_{\alpha,m}(E') \rangle $. But since we know the averages of this kind only for the operators $ \mathbf{a}_{\beta,m'}(E) $ of particles leaving the reservoir $ l $ (which are the equilibrium Fermi functions of the reservoir $ l $, multiplied by the corresponding Kronecker symbols $ \delta $-functions of energy):
\begin{equation}
\label{stat for a-charge}
\langle \mathbf{a}^\dag_{\alpha,m}(E) \mathbf{a}_{\beta,m'}(E') \rangle
=
f_m(E) \delta_{\alpha \beta} \delta_{m m'} \delta(E-E')
,
\end{equation}
in averages of the form $ \langle \mathbf{b}^\dag_{\alpha,m}(E) \mathbf{b}_{\alpha,m}(E') \rangle $ we should use the expressions \eqref{abS}, \eqref{abS dag}. Thus, $ S^*_{\alpha \beta,m m'} $ appears in the expression for the current, and after simplification, the Hermitian conjugate matrices $ \mathbf{S}^\dag $ appear.


Let us give the simplest example~-- when two parameters $ X_1, X_2 $ \cite{Brower} change in time. Then, according to \eqref{current}, the charge passing through the contact $ m $ with a small change in $ \delta X_1,\,\delta X_2 $ parameters is
\begin{equation}
\label{charge 2 params}
\delta Q(m,t)
=
e \frac{\partial n(m)}{\partial X_1}\delta X_1
+
e \frac{\partial n(m)}{\partial X_2}\delta X_2.
\end{equation}
For a change in $ X_1, X_2 $, in which $ X_1, X_2 $ perform exactly one cycle, one needs to take the integral
\begin{eqnarray}
\label{charge 2 params_int-green}
Q(m,\tau)
&=&
e
\int^\tau_0
d t
\left(
\frac{\partial n(m)}{\partial X_1}\frac{\partial X_1}{\partial t}
+
\frac{\partial n(m)}{\partial X_2}\frac{\partial X_2}{\partial t}
\right)
= \nonumber \\
&=&
\int_\mathcal{L}
\left(
\frac{\partial n(m)}{\partial X_1}d X_1
+
\frac{\partial n(m)}{\partial X_2}d X_2
\right)= \nonumber \\
&=&
\int_\mathcal{A}
d X_1\,d X_2
\left(
\frac{\partial}{\partial X_1}\frac{\partial n(m)}{\partial X_2}
-
\frac{\partial}{\partial X_2}\frac{\partial n(m)}{\partial X_1}
\right),
\end{eqnarray}
where $ \tau $ is the period, $ \mathcal{L} $ is the contour in two-dimensional parameter space $ (X_1, X_2) $, $ \mathcal{A} $ is the area of this contour (in the last equality we have used Green's theorem). 
From Eq.\eqref{charge 2 params_int-green} it follows that a nonzero charge per cycle will be only for a nonzero area $ \mathcal{A}\neq 0 $. In the case of (harmonic) in-phase change of parameters, the circuit $ \mathcal{L} $ is a straight line, $ \mathcal{A} = 0 $, and the average charge over the period is $ 0 $. In the case of an antiphase change of parameters, $ \mathcal{L} $ is an ellipse, $ \mathcal {A} \neq 0 $, and the average charge over the period is nonzero, which means that charge pumping occurs.

\section{Spin pumping\footnotemark}\

\footnotetext{Recommended literature: a classic review \cite{TseBra 2005}, and possibly two original articles \cite{Tserkovnyak PRB}, \cite{Tserkovnyak PRL}.}

\begin{figure}[h]
	\vspace{-0pt}
	\centering
	\includegraphics[width=0.6\textwidth]{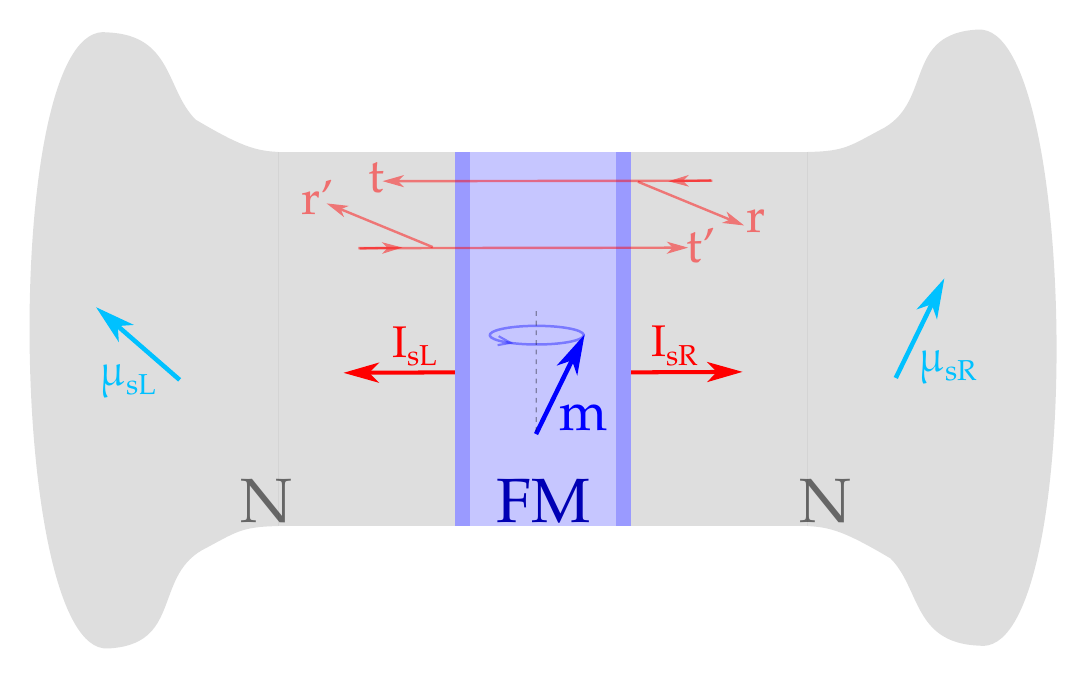}
	\vspace{-0pt}
	\caption{The spin pumping three-layer system as viewed in mesoscopic scattering theory. \textit{Ferromagnetic scatterer} (blue) with dynamically varying magnetization $ \mathbf{m} $ pumps pure spin currents $ \textbf{I}_{sL},\,\textbf{I}_{sR} $ into adjacent \emph{diffuse normal metal reservoirs} \textbf{N} (gray) through \emph{ballistic contacts}. The scatterer contains the ferromagnet FM (light blue area in the center), as well as areas near the interfaces (dark blue area) of the order of the spin coherence length $ \lambda_{fc} $ in the ferromagnet. Each of the two \textbf{N}-regions is divided into a \emph{reservoir} with chemical potential \eqref{mu_c} and \emph{spin accumulation vector} \eqref{mu_s}, and a \emph{ballistic contact (lead)} with a quantized number of transverse modes at the Fermi level. 
	The scatterer is described by the spin-dependent transfer matrix \eqref{abS Spin}, which consists of the reflection ($ r, r '$) and transmission ($ t, t' $) amplitudes for each of the quantized channels in the ballistic region.
	}\label{Tse2005-structure}
	\vspace{0pt}
\end{figure}

Now we turn to the case of spin-dependent transport, when the elements of the scattering matrix $ \hat{\mathbf{S}} $ are complex $ 2\times2 $ matrices in spin indices, which are not reducible to unit matrices. The distribution function $ \hat{f}(\varepsilon) $ is the $ 2\times2 $ \textit{density matrix} in spin indices, and is still assumed to be isotropic (the isotropy approximation is valid if the counterflow $ \mathbf{I}^{back}_s $ does not lead to significant drift in the distribution function in the counterflow direction). In thermal equilibrium and in the absence of spin accumulation in reservoirs ($ \tau_{sf} \rightarrow 0 $, normal metal is an \emph{ideal spin sink}), $ \hat{f}(\varepsilon) $ is an isotropic distribution function in spin indices:
\begin{equation}
\hat{f}(\varepsilon)
=
f_{FD}(\varepsilon)
\hat{\sigma}_0.
\end{equation}
Local chemical potential $ \mu_c $ of the reservoir reads
\begin{equation}
\label{mu_c}
\mu_c
=
\int^{\infty}_{\varepsilon_0}
d\varepsilon
\,
\mathbf{Tr} \left[\hat{\sigma}_0 \hat{f}(\varepsilon) \right]
\equiv
\int^{\infty}_{\varepsilon_0}
d\varepsilon
\,
\mathbf{Tr} \left[ \hat{f}(\varepsilon) \right]
,
\end{equation}
where energy $ \varepsilon_0 $ lies below the Fermi level by an amount much greater than the average thermal energy of electrons in the reservoir $ k_B T $ and the potential difference in the reservoirs $ V=\mu^R_c-\mu^L_c $, but otherwise is arbitrary \cite{TseBra 2005}.
In the general case ($ \tau_{sf}> 0  $), a metal has a nonzero locally equilibrium spin density with a \textit{spin chemical potential}, or \textit{spin accumulation} $ \boldsymbol{\mu}_s $:
\begin{equation}
\label{mu_s}
\boldsymbol{\mu}_s
=
2 \int^{\infty}_{\varepsilon_0}
d\varepsilon
\,
\mathbf{Tr} \left[\hat{\boldsymbol{\sigma}} \hat{f}(\varepsilon) \right]
.
\end{equation}
The approximations are basically the same as in charge pumping. The important details of dividing the system into mesoscopic elements are shown in Fig.\ref{Tse2005-structure}.
Instead of spinless scattering states in charge transport, we take pure electronic scattering states polarized either parallel to the $ z $ axis ($ \mathbf{P}=\mathbf{e}_z,\, spin=\uparrow $), or antiparallel to the $ z $ axis ($ \mathbf{P}=-\mathbf{e}_z $, $ spin=\downarrow $). The matrix $ \hat{\mathbf{S}} $ linearly connects the destruction operators $ \mathbf{b}^{spin}_{\alpha,m}(E) $ for particles with energy $ E $ and spin $ spin $, entering the reservoir $ m $ through the channel $ \alpha $, and the destruction operators $ \mathbf{a}^{spin'}_{\beta,m'}(E) $ for particles with energy $ E $ and spin $ spin' $ leaving the $ m'$ reservoir via the channel $ \beta $:
\begin{equation}
\label{abS Spin}
\mathbf{b}^{spin}_{\alpha,m}(E)=\sum_{\beta,m',spin'} \mathbf{S}^{spin\,spin'}_{\alpha \beta,m m'} \mathbf{a}^{spin'}_{\beta,m'}(E).
\end{equation}

Symmetry breaking with respect to scattering of different spin states occurs in the scatterer~-- at the boundary of the ferromagnet \textbf{F}. The scattering matrix of a ferromagnet can be decomposed in the space of spinors into two \emph{unitary}\footnotemark\footnotetext{The unitarity follows from the fact that states with a mean spin $ \mathbf{P} $ parallel/antiparallel to the magnetization $ \mathbf {m} $ must scatter independently.} matrices describing transport parallel ($ S^\uparrow $) and antiparallel ($ S^\downarrow $) to the direction of magnetization $ \mathbf{m} $ \cite{Nazarov book}:
\begin{equation}
\mathbf{S}=S^\uparrow \frac{1+\mathbf{m}\cdot \hat{\boldsymbol{\sigma}}}{2} 
+ 
S^\downarrow \frac{1-\mathbf{m}\cdot \hat{\boldsymbol{\sigma}}}{2},
\end{equation}
where $ \hat{\boldsymbol{\sigma}} $ is a vector of Pauli matrices. In expanded form,
\begin{equation}
\hat{S}_{\alpha \beta,m m'}
=
S^\uparrow_{\alpha \beta,m m'} \frac{1+\mathbf{m}\cdot \hat{\boldsymbol{\sigma}}}{2} 
+ 
S^\downarrow_{\alpha \beta,m m'} \frac{1-\mathbf{m}\cdot \hat{\boldsymbol{\sigma}}}{2}.
\end{equation}
The scattering matrix $ \mathbf{S} $, therefore, depends only on the direction of magnetization $ \mathbf{m} $, which is equivalent to dependence on \textit{two parameters}: for example, angular coordinates $ (\theta, \phi) $ of the vector $ \mathbf{m} $ ($ \mathbf{m} $ has 3 components, but they are related by normalization: $ m^2_x + m^2_y + m^2_z = 1 $). Following Brower argument in charge pumping \cite{Brower}, we can therefore expect that a nonzero average spin current could be pumped when magnetization vector $ \mathbf{m} $ follows a closed path on a unit sphere\footnote{Recall that the typical dynamics of $ \mathbf{m} $ is precession (as in FMR).}.

Now we write out the most natural generalization of the expression \eqref{current} for the spin current (this time more explicitly, using full time derivatives):
\begin{equation}
\label{current-s}
\hat{I}(m,t)=
e \frac{d \hat{n}(m)}{d t},
\end{equation}
where
\begin{equation}
\label{spin emissivity}
\frac{d \hat{n}(m)}{d t}
=
\frac{1}{4\pi i} \sum_\beta \mathbf{Tr}
\left(
\frac{d \hat{\mathbf{S}}_{\alpha \beta}}{d t} 
\hat{\mathbf{S}}^\dag_{\alpha \beta}
-
\hat{\mathbf{S}}_{\alpha \beta} \frac{d \hat{\mathbf{S}}^\dag_{\alpha \beta}}{d t}
\right).
\end{equation}
Let us rewrite the last expression in expanded form, leaving hats over scattering matrix elements in the channel space (however, now these elements are $ 2 \times2 $ matrices in spinor space):
\begin{eqnarray}
\label{spin matrix emissivity}
\frac{d \hat{n}(m)}{d t}
\equiv
\frac{1}{4\pi i} \sum_{\alpha\beta m'}
\left(
\frac{d \hat{S}_{\alpha \beta,m m'}}{d t} 
\hat{S}^\dag_{\alpha \beta,m m'}
-
\hat{S}_{\alpha \beta,m m'} \frac{d \hat{S}^\dag_{\alpha \beta,m m'}}{d t}
\right)
,
\end{eqnarray}
\emph{where the Hermitian conjugation $ \dag $ in the last expression refers only to individual elements $ \hat{S}^\dag_{\alpha \beta, m m'} $ of the scattering matrix as $ 2 \times2 $ matrices in the space spinors}.
In the equation~\eqref{spin matrix emissivity}, $ \hat{n}(m) $ is the $ 2 \times2 $ matrix in spin indices, just like the spin current $ \hat{I}(m,t) $.
Let us recall the new concept of the scattering matrix \eqref{abS Spin} and write the analogue of the expression \eqref{current in a b} that would be matrix in spin indices:
\begin{equation}
\label{matrix current in a b}
I^{spin\,spin'}_m (t)=\frac{e}{h}\sum_{\alpha \in m} \int dE dE' e^{i(E-E')t/\hbar} 
(
\mathbf{a}^{spin'\dag}_{\alpha,m}(E)\mathbf{a}^{spin}_{\alpha,m'}(E')-
\mathbf{b}^{spin'\dag}_{\alpha,m}(E)\mathbf{b}^{spin}_{\alpha,m'}(E')
).
\end{equation}
An analogue of the assumption \eqref{stat for a-charge} about the statistical average in our case with spin, 
\begin{equation}
\label{stat for a-spin}
\langle \mathbf{a}^{spin'\dag}_{\alpha,m}(E) \mathbf{a}^{spin}_{\beta,m'}(E') \rangle
=
f_m(E) \boldsymbol{\delta}_\mathbf{{spin\,spin'}} \delta_{\alpha \beta} \delta_{m m'} \delta(E-E')
\end{equation}
postulates \emph{spin isotropy} of states entering from \textbf{NM} reservoirs into FM. This assumption is equivalent to specifying a completely unpolarized density matrix for the states entering the $ \textbf{m} $ reservoir:
\begin{equation}
\label{}
\hat{\rho}_m=
\begin{pmatrix} 
1/2 & 0\\
0 & 1/2\\
\end{pmatrix}.
\end{equation}
The matrix current is related to the spin density matrix by the conservation law:
\begin{equation}
\label{conserv law I-rho}
\frac{\partial \hat{\rho}_m}{\partial t} + \frac{\hat{I}_m}{e} = 0.
\end{equation}
For two spin states ($ \mathbf{P}=\mathbf{e}_z,\, spin=\uparrow $), and ($ \mathbf{P}=-\mathbf{e}_z $, $ spin=\downarrow $), the same quantum mechanical superposition principle holds as for spinless scattering states, so they are effectively just \emph{two spin channels}. In the Landauer formula \eqref{landauer}, the trace ($ \textbf{Tr} $) is also taken along the spin channels (from where $ 2_s $ comes from). Therefore, equation~\eqref{spin emissivity} is just a partial $ \textbf{Tr} $ across the channels\footnotemark.
\footnotetext{If we do not take $ \textbf{Tr} $ at all in the Landauer formula \eqref{landauer}, then we get the matrix current in the indices of all channels. However, such a detail turns out to be superfluous, since in experiments it is impossible to separate the contributions from different channels. In contrast, \emph{spin-polarized contributions can be detected separately due to the Inverse Spin Hall Effect \cite{SHE_Sinova}}.} 

Let us move on to the calculations. We take the first term in the sum \eqref{spin matrix emissivity} (for convenience, we will now omit the same indices $ _{\alpha \beta,m m'} $ in $ S^\uparrow_{\alpha \beta,m m'} $ and $ S^\downarrow_{\alpha \beta,m m'} $):
\begin{eqnarray}
\label{}
&&\frac{d \hat{S}_{\alpha \beta,m m'}}{d t} 
\hat{S}^\dag_{\alpha \beta,m m'}
=
\frac{d}{d t}
\left[
S^\uparrow \frac{1+\mathbf{m}\cdot \hat{\boldsymbol{\sigma}}^\dag}{2} 
+ 
S^\downarrow \frac{1-\mathbf{m}\cdot \hat{\boldsymbol{\sigma}}^\dag}{2}
\right] \cdot \nonumber \\
&&	\cdot \left(
S^{\uparrow*} \frac{1+\mathbf{m}\cdot \hat{\boldsymbol{\sigma}}^\dag}{2} 
+ 
S^{\downarrow*} \frac{1-\mathbf{m}\cdot \hat{\boldsymbol{\sigma}}^\dag}{2}
\right)
=
\nonumber \\
&& =\frac{1}{4}
\left[
S^\uparrow \hat{\boldsymbol{\sigma}}  \mathbf{\dot{m}}
- 
S^\downarrow \hat{\boldsymbol{\sigma}}  \mathbf{\dot{m}}
\right]
\cdot 
\left(
S^{\uparrow*} (1+\mathbf{m}\cdot \hat{\boldsymbol{\sigma}})
+ 
S^{\downarrow*} (1-\mathbf{m}\cdot \hat{\boldsymbol{\sigma}})
\right)
=\nonumber \\
&& =\frac{1}{4}
(\hat{\boldsymbol{\sigma}}  \mathbf{\dot{m}})
\left(
S^\uparrow
- 
S^\downarrow
\right)
\cdot 
\left(
(\hat{\boldsymbol{\sigma}} \mathbf{m})
(S^{\uparrow*}-S^{\downarrow*})
+ 
(S^{\uparrow*}+S^{\downarrow*})
\right)
=\nonumber \\
&& =\frac{1}{4}
(\hat{\boldsymbol{\sigma}}  \mathbf{\dot{m}})
(\hat{\boldsymbol{\sigma}} \mathbf{m})
\left|
S^\uparrow
- 
S^\downarrow
\right|^2
+
\frac{1}{4}
(\hat{\boldsymbol{\sigma}}  \mathbf{\dot{m}})
\left(
\left|S^{\uparrow}\right|^2 -
\left|S^{\downarrow}\right|^2
+ 	\left(
S^{\uparrow}S^{\downarrow*}-S^{\downarrow}S^{\uparrow*}
\right)
\right)
. \nonumber
\\
\end{eqnarray}
Let us calculate separately the term that appears here,
\begin{eqnarray}
&&(\hat{\boldsymbol{\sigma}}  \mathbf{\dot{m}})
(\hat{\boldsymbol{\sigma}} \mathbf{m})
=
\sum_{i=x,y,z} \hat{\sigma}_i \dot{m}_i
\sum_{j=x,y,z} \hat{\sigma}_j m_j
=
\sum_{i,j} \hat{\sigma}_i \hat{\sigma}_j \dot{m}_i m_j
=\nonumber \\
&&=\{ \text{we use } \hat{\sigma}_i \hat{\sigma}_j
=\sum_k i \varepsilon_{ijk} \hat{\sigma}_k + \delta_{ij} \cdot \hat{1}
\}=
i \sum_{ijk} \varepsilon_{ijk} \hat{\sigma}_k \dot{m}_i m_j 
+
\hat{1} \cdot \sum_{ii} \dot{m}_i m_i
\equiv\nonumber \\
\end{eqnarray}
\begin{eqnarray}
&&\equiv\{ \text{we use $ |\mathbf{m}|=1 $, } \sum_{ii} \dot{m}_i m_i=(1/2) d(\mathbf{m}^2)/dt=0
\}\equiv\nonumber \\
&&\equiv
i (\dot{\mathbf{m}},\mathbf{m},\hat{\boldsymbol{\sigma}})
\equiv
i (\hat{\boldsymbol{\sigma}},\left[\dot{\mathbf{m}}\times\mathbf{m}\right]).
\end{eqnarray}
Then the first term in the sum \eqref{spin matrix emissivity} is, finally,
\begin{eqnarray}
\frac{d \hat{S}_{\alpha \beta,m m'}}{d t} 
\hat{S}^\dag_{\alpha \beta,m m'}
&=&
\frac{1}{4}
i (\hat{\boldsymbol{\sigma}},\left[\dot{\mathbf{m}}\times\mathbf{m}\right])
\left|
S^\uparrow
- 
S^\downarrow
\right|^2
+ \nonumber \\
&& \frac{1}{4}
(\hat{\boldsymbol{\sigma}}  \mathbf{\dot{m}})
\left(
\left|S^{\uparrow}\right|^2 -
\left|S^{\downarrow}\right|^2
+ 	\left(
S^{\uparrow}S^{\downarrow*}-S^{\downarrow}S^{\uparrow*}
\right)
\right)
.
\end{eqnarray}
The second term in the sum \eqref{spin matrix emissivity} is simply the Hermitian conjugate of the first (here, for convenience, we also omit the indices $ _{\alpha \beta,m m'} $):
\begin{eqnarray}
&&\hat{S}_{\alpha \beta,m m'} \frac{d \hat{S}^\dag_{\alpha \beta,m m'}}{d t}
\equiv 
\left(
\frac{d \hat{S}_{\alpha \beta,m m'}}{d t} 
\hat{S}^\dag_{\alpha \beta,m m'}
\right)^\dag = \nonumber \\
&=&
-\frac{1}{4}
i (\hat{\boldsymbol{\sigma}},\left[\dot{\mathbf{m}}\times\mathbf{m}\right])
\left|
S^\uparrow
- 
S^\downarrow
\right|^2
+
\frac{1}{4}
(\hat{\boldsymbol{\sigma}}  \mathbf{\dot{m}})
\left(
\left|S^{\uparrow}\right|^2 -
\left|S^{\downarrow}\right|^2
+ 	\left(
S^{\uparrow*}S^{\downarrow}-S^{\downarrow*}S^{\uparrow}
\right)
\right)
.\nonumber
\\
\end{eqnarray}
As a result,
\begin{eqnarray}
\label{spin matrix emissivity - result}
&&\frac{d \hat{n}(m)}{d t}
\equiv
\frac{1}{4\pi i} \sum_{\alpha\beta m'}
\left(
\frac{d \hat{S}_{\alpha \beta,m m'}}{d t} 
\hat{S}^\dag_{\alpha \beta,m m'}
-
\hat{S}_{\alpha \beta,m m'} \frac{d \hat{S}^\dag_{\alpha \beta,m m'}}{d t}
\right)= \nonumber \\
&=&
\frac{1}{4\pi i} \sum_{\alpha\beta m'}
\Big(
-\frac{1}{2}
i (\hat{\boldsymbol{\sigma}},\left[\mathbf{m}\times\dot{\mathbf{m}}\right])
\left|
S^\uparrow_{\alpha \beta,m m'}
- 
S^\downarrow_{\alpha \beta,m m'}
\right|^2
+\nonumber \\
&&
\frac{1}{2}
(\hat{\boldsymbol{\sigma}}  \mathbf{\dot{m}})
\left[
S^{\uparrow}_{\alpha \beta,m m'} S^{\downarrow*}_{\alpha \beta,m m'}-S^{\downarrow}_{\alpha \beta,m m'} S^{\uparrow*}_{\alpha \beta,m m'}
\right]
\Big).
\end{eqnarray}
The matrix current \eqref{current-s} can now be represented in the canonical form for spin density matrices:
\begin{eqnarray}
\hat{I}(m,t)=
-\frac{e}{\hbar} \left( \hat{\boldsymbol{\sigma}} \cdot \mathbf{I}^{pump}_s \right),
\end{eqnarray}
where \emph{spin current vector} $ \mathbf{I}^{pump}_s $ is the sum of two terms:
\begin{eqnarray}
\label{spin-current}
\mathbf{I}^{pump}_s(m,t)=
\frac{\hbar}{4\pi}
\cdot
\left(
A_r \left[\mathbf{m}\times\dot{\mathbf{m}}\right]
-
A_i \dot{\mathbf{m}}
\right)
,
\end{eqnarray}
where the dimensionless quantities $ A_r,\,A_i $ are introduced, which contain all information about the scattering matrix (we write out explicitly the elements of $ \hat{\mathbf{S}}$-matrix which are the amplitudes of transmission ($ T $) and reflection ($ R $)):
\begin{eqnarray}
A_r 
&=&
\frac{1}{2}
\sum_{\alpha\beta m'}
\left|
S^\uparrow_{\alpha \beta,m m'}
- 
S^\downarrow_{\alpha \beta,m m'}
\right|^2 \equiv \nonumber \\
&\equiv&
\frac{1}{2}
\sum_{\alpha\beta m'}
\left(
\left|
R^\uparrow_{\alpha \beta,m m'}
- 
R^\downarrow_{\alpha \beta,m m'}
\right|^2
+
\left|
T^\uparrow_{\alpha \beta,m m'}
- 
T^\downarrow_{\alpha \beta,m m'}
\right|^2
\right), \\
A_i 
&=&
\sum_{\alpha\beta m'}
\mathbf{Im} \left( S^{\uparrow}_{\alpha \beta,m m'} S^{\downarrow*}_{\alpha \beta,m m'} \right)
=
\mathbf{Im}
\sum_{\alpha\beta m'}
\left( 
R^{\uparrow}_{\alpha \beta,m m'} R^{\downarrow*}_{\alpha \beta,m m'}
+
T'^{\uparrow}_{\alpha \beta,m m'} T'^{\downarrow*}_{\alpha \beta,m m'}
\right)
.\nonumber\\
\end{eqnarray}
Real-valued parameters $ A_r $ and $ A_i $ can be combined into a complex value
\begin{equation}
A_r+iA_i 
=
g^{\uparrow\downarrow}
-
t^{\uparrow\downarrow}
,
\end{equation}
where two quantities are introduced~-- the \emph{spin-mixing conductance},
\begin{equation}
\label{spin-mixing g}
g^{\uparrow\downarrow}
=
\sum_{\alpha\beta}
\left(
\delta_{\alpha\beta}
-
r^\uparrow_{\alpha\beta}
(r^\downarrow_{\alpha\beta})^*
\right),
\end{equation}
and the \emph{spin-mixing transmittance},
\begin{equation}
t^{\uparrow\downarrow}
=
\sum_{\alpha\beta}
t'^\uparrow_{\alpha\beta}
(t'^\downarrow_{\alpha\beta})^*
.
\end{equation}
For not too thin \textbf{F}-films ($ \gtrsim 10 $\AA), $ |t^{\uparrow\downarrow}| \ll g^{\uparrow\downarrow} $ \cite{TseBra 2005}, therefore the spin pumping effect is mainly determined by the \emph{spin-dependent (spin-flip) reflection amplitudes} of a narrow region of the order of spin coherence length $ \lambda_{fc} $ \eqref{lambda fc}; then
\begin{equation}
A_r
\simeq
\Re[g^{\uparrow\downarrow}]
,
\quad
A_i
\simeq
\Im[g^{\uparrow\downarrow}]
.
\end{equation}

Hence, in a simple model (in which the elements of the scattering matrix are not even specified), we got the expression \eqref{spin-current} for the spin current with the expected phenomenology.
Let us assume that the spin current $ \mathbf{I}^{pump}_s $ does not create a spin-polarized region in a normal metal near the surface of the ferromagnet (as in the case of platinum) and the corresponding leakage spin current is negligible. If the ferromagnetic film is surrounded on the left and right by reservoirs of normal metal, then the spin current $ \mathbf{I}^{pump}_s(m,t) $ flows at $ \dot{\mathbf{m}}\neq 0 $ both in the right $ (m=Right) $ and in the left $ (m=Left) $ normal metal reservoirs. The renormalization of coefficients $ \alpha $ and $ \gamma $ can be obtained using either angular momentum conservation, or the so-called Slonchevsky spin torque
\cite{Slonczewski 96}, $ \boldsymbol{\tau} = \mathbf{m}\times(\mathbf{I}^{pump}_s(Left,t) + \mathbf{I}^{pump}_s(Right,t))\times\mathbf{m} $:
\begin{eqnarray}
&&\frac{1}{\gamma}=\frac{1}{\gamma_0} (1+g_L[A^{Left}_i+A^{Right}_i]/{4\pi M_s}) ,\\
&&\alpha'=\frac{\gamma}{\gamma_0} (\alpha_0 + g_L[A^{Left}_r+A^{Right}_r]/{4\pi M_s}))
,
\end{eqnarray}
where $ M_s $ is the magnetization of ferromagnetic film.
Experiment and first-principles calculations show that $ |A_i| \ll |A_r| $. Therefore, the main correction is applied to the damping constant $ \alpha $, while the correction to the precession frequency is small, which is in perfect consistency with experiment. 

\begin{center}
	\uppercase{Control questions}
\end{center}
\noindent 1. Is it possible to consider the spin-transfer torque as localized at the boundary \textbf{FM-NM} if the ferromagnet is insulating? 

\noindent 2. How does the spin pumping affect the Gilbert damping of a thin ferromagnetic film? Why there is a large difference when adjacent normal metal possess either high or low spin-orbit coupling? 

\noindent 3. What does the spin-mixing conductance describe? 

\noindent 4. Referring to the review \cite{TseBra 2005}, p.10, Eq.(26), or to the book \cite {Nazarov book}, p.121-122, Eqs.(188)-(189), follow the derivation of the backflow spin current $ \mathbf{I}^{back}_s $. Calculate the spin torque $ \boldsymbol{\tau}^{back} $ related to the backflow spin current. 

\noindent 5. Referring to the article \cite{Tserkovnyak PRB}, follow the derivation of the equilibrium spin density $ \boldsymbol{\mu}_s(x=0) $ \eqref{mu_s} at the \textbf{FM-NM} boundary which is based on the spin diffusion equation (especially pay attention to eq. (15) for the ``backflow factor'' $ \beta $). Using it, obtain the renormalized (reduced) parameters $ A_r $ and $ A_i $.

\setchapterpreamble{%
	\dictum{%
		
		The spin Hall effect (\textbf{SHE}) \cite{Dyakonov Review, SHE_Sinova, Spin Hall effect devices} consists in spatial separation of electrons with opposite spins in a heavy metal, semiconductor or two-dimensional electron gas (2DEG) due to spin-orbit interaction, first theoretically predicted by Dyakonov and Perel \cite{Dyakonov Perel 71}. 
		In this short lecture, we will discuss the basic phenomenological equations that reproduce \textbf{SHE}, and the associated Spin Hall Magnetoresistance (\textbf{SHMR}) effect.
	}%
	\vspace{24pt}%
}

\chapter{Spin Hall Effect \label{SHE_chapter}}\

Let us write the basic phenomenological equations for electron and spin currents in a conductor without taking into account spin-orbital interaction (we will mark such unperturbed quantities by the subscript $ ^{(0)} $) \cite{Dyakonov Review,Dyakonov 07}:
\begin{eqnarray}
\label{charge0_SH}
\mathbf{q}^{(0)}
&=&
-\mu n \mathbf{E} - D \nabla n
, \\
\label{spin0_SH}
q^{(0)}_{ij}
&=&
-\mu n E_i P_j - D(\partial P_j / \partial x_i)
,
\end{eqnarray}
where $ \mathbf{q} $ is the electron flux density ($ \mathbf{q} =  - \mathbf{j}_e / e $), $ q_{ij} $ is the spin polarization flux density tensor ($ q_{ij} = (2/\hbar) \mathbf{j}_s $), where the first subscript indicates the direction of the current, and the second the direction of the spin.
The charge current $ - e \mathbf{q}^{(0)} $ is caused by the external electric field $ \mathbf{E} $ ($ \mu$ is the electron mobility) and the diffusion of the charge density $ n $ ($ D $ is the diffusion coefficient). The expression \eqref{spin0_SH} for the spin current $ (\hbar/2) q^{(0)}_{ij} $ looks similar, since the spin is transferred by the same electrons that are involved in charge transport \eqref{charge0_SH}. 

\textit{Spin-orbital interaction couples spin and charge currents.} In a conductor with inversion symmetry,
\begin{eqnarray}
\label{SOI_charge}
q_i
&=&
q^{(0)}_i + \gamma \varepsilon_{ijk} q^{(0)}_{jk}
, \\
\label{SOI_spin}
q_{ij}
&=&
q^{(0)}_{ij} - \gamma \varepsilon_{ijk} q^{(0)}_{k}
,
\end{eqnarray}
where the dimensionless coefficient $ \gamma \ll 1 $ parameterizes the spin-orbit interaction. 
There are also other types of spin-orbit interactions arising at the interfaces of two semiconductors or in ultrathin metal films with broken inversion symmetry~\cite{SHE_Sinova, YeremkoLoktev}. Two main types of such interactions, which lead to the splitting of the electronic spectrum in the $ \mathbf{k}$-space, are called the Rashba and Dresselhaus effects. Here and in the following, we will consider only the case of conductors with an inversion center.

\begin{figure}[h!]
	\centering
	\includegraphics [width=0.7\textwidth]{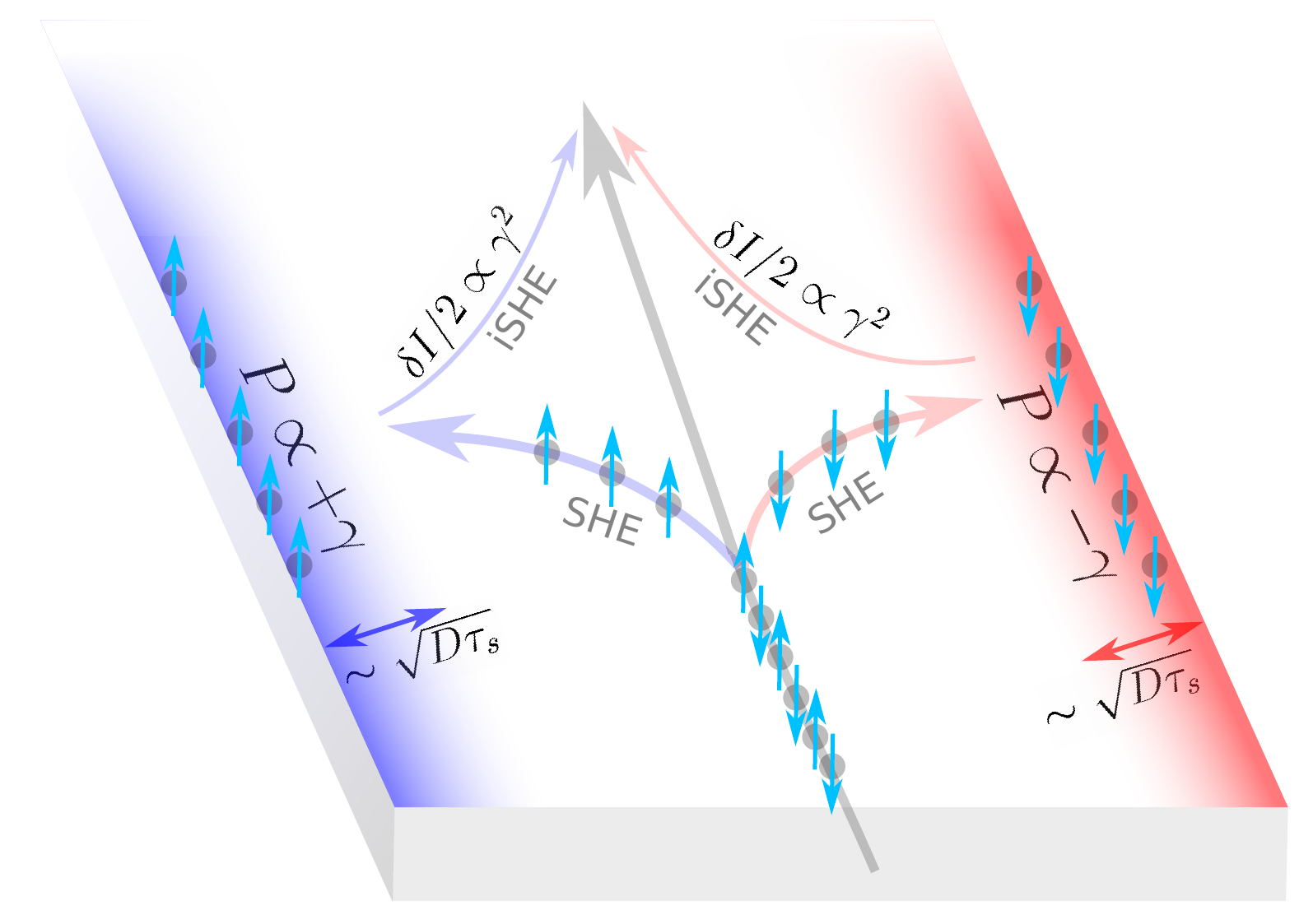}
	\caption{The combination of the direct (SHE) and inverse (ISHE) Spin Hall Effects leads to a decrease in the sample resistance by a value $ \delta R \propto - \gamma^2 $. The spin density $ P \propto \pm \gamma $ is accumulated in stripes of width $ \sqrt{D \tau_s} $ at the opposite edges of the sample.}
	\label{SHE_schamatic}
	\vspace{-0 pt}
\end{figure}

Substituting \eqref{SOI_charge} and \eqref{SOI_spin} in \eqref{spin0_SH} and \eqref{charge0_SH}, we get:
\begin{eqnarray}
\label{SHE_charge_spin_corrected}
\mathbf{j} / e
&=&
\mu n \mathbf{E} - D \nabla n
+ \beta \mathbf{E} \times \mathbf{P}
+ \delta \nabla \times \mathbf{P}
, \\
q_{ij}
&=&
-\mu n E_i P_j - D(\partial P_j / \partial x_i)
+ \varepsilon_{ijk} \left( \beta n E_k + \delta (\partial n / \partial x_k) \right)
,
\end{eqnarray}
where $ \beta = \gamma \mu $, $ \delta = \gamma D $.
Let us discuss the effects for which the terms arising in these equations are responsible.
The term $ \beta \mathbf{E} \times \mathbf{P} $ describes the anomalous Hall effect and is observed only for a nonzero average spin polarization $ \langle\mathbf{P}\rangle \neq 0 $, for example, in ferromagnetic metals or when irradiation with circularly polarized light is present. The term $ \delta \nabla \times \mathbf{P} $ describes the addition to the current due to inhomogeneous spin density, or \emph{the inverse Spin Hall Effect (ISHE)}. The terms $ \varepsilon_{ijk} \left( \beta n E_k + \delta (\partial n / \partial x_k) \right) $ describe the \emph{Spin Hall Effect (SHE)}: an electric current causes a spin current transverse to it. Hence, oppositely polarized spins accumulate at the opposite boundaries of the sample (see Fig. \ref{SHE_schamatic}).
The continuity equation for the spin polarization density $ \mathbf{P} $ which accounts for \textit{spin relaxation} and \emph{spin precession} in an external magnetic field has the form:
\begin{eqnarray}
\label{SHE_continuity}
\partial P_j / \partial t
+
\partial q_{ij} / \partial x_i
+
\varepsilon_{jkl} \Omega_k P_l
+
P_j / \tau_s
=
0
,
\end{eqnarray}
where $ \Omega \parallel \mathbf{H} $ is the spin precession frequency, $ \tau_s $ is the spin relaxation time. 
Boundary conditions to the equations \eqref{SHE_charge_spin_corrected}-\eqref{SHE_continuity} consist in the absence of spin current components perpendicular to the sample boundaries. 

The equations \eqref{SHE_charge_spin_corrected}-\eqref {SHE_continuity} form the basis of the phenomenology of direct and inverse Spin Hall Effects. With their help, a decrease in resistance in films with spin-orbit interaction \cite{Dyakonov 07}, $ \delta R / R_0 \propto -\gamma^2 $, was discovered theoretically.
The qualitative explanation is the following (see Fig. \ref{SHE_schamatic}): due to the Spin Hall Effect, regions with a width of the order of the spin diffusion length $ L_s \sim \sqrt{D \tau_s} $\footnote{For narrow samples, such that $ L < L_s $, the spins will diffuse to the middle of the sample with almost no relaxation, so that the characteristic size of spin-polarized regions in this case is of the order of $ L / 2 $, and the characteristic spin decay time is determined by time needed by the electron to diffuse to the middle of the sample, $ \tau_d = L^2/4D $. The width of the magnetoresistance curve will, therefore, be related precisely to the time $ \tau_d $, and not to the spin relaxation time $ \tau_s $. The ``strong'' magnetic field in this case will be the one in which the spin precesses with the frequency $ \Omega \gtrsim \tau_d^{-1} $.} 
appear at the film boundaries with opposite spin polarizations $ P\propto \pm \gamma $. The diffusive backflow of spin from these regions causes a secondary charge current $ \delta I \propto\gamma^2 $ aligned with the primary current\footnotemark\footnotetext {This can be verified by time-reversing the process of spin-dependent electron scattering on an atom (remember the change of sign of the effective magnetic field acting on the electron with time reversal).} due to the inverse Spin Hall Effect, which corresponds to a \textit{decrease in the resistance} of the sample.


\textit{Positive magnetoresistance} can be achieved by breaking or decreasing the spin density at the sample boundaries by a magnetic field. In a sufficiently strong magnetic field (such that $ \Omega \gtrsim \tau_s^{-1} $, $ H \sim 5\,\text{T} $) \cite {Dyakonov 07}, the electron spin precesses during the diffusion $ t \gg \tau_s $ by the angle $ \gg \Omega \tau_s \gtrsim 1 $, and part of the secondary spin current disappears due to such dephasing, leading to an \textit{increase in the film resistance}. 
This effect was named \emph{Hanle Magnetoresistance} (\textbf{HMR}, by analogy with the Hanle effect in optics), and was first experimentally confirmed in the work \cite{Velez 16}. Another well-known method for decreasing spin density at the sample edges is by attaching a ferromagnet at the sample edge. When the ferromagnet magnetization is not aligned with spin density at the sample boundary, then a part of the spin density goes into the diffusive spin moment \eqref{passive diffusion}, which depends on the orientation of ferromagnet magnetization. In this case, even a ferromagnetic dielectric (e. g. YIG) can control the conductivity of an adjacent film by changing the direction of its own magnetization. Therefore, by measuring the resistance of a metal film, it is possible to \textit{remotely determine the direction of ferromagnet magnetization}.
This effect was named \emph{Spin Hall Magnetoresistance} (\textbf{SMR} or \textbf{SHMR}) and was first discovered experimentally in the works \cite{Nakayama 13, Hahn 13}.

In modern experiments, the inverse Spin Hall Effect is used as the main method for the electric detection\footnotemark\footnotetext{Along with electric method, optical techniques for detecting spin density are popular for semiconductors. It was in this way that the spin Hall effect was first directly experimentally confirmed.} of a nonequilibrium spin density or spin current in a \textit{heavy} metal ($ Pt $, $ Ta $) \cite{SHE_Sinova, Spin Hall effect devices}. Historically, the first electrical method for detecting non-equilibrium spin density in \textit{normal} metals was discovered by Johnson and Silsby \cite{Johnson Silsbee PRL85}: diffusion of spin from a paramagnetic to a ferromagnetic metal induces a voltage on a metallic ferromagnet (a process opposite to the transport of spin current across the boundary of ferromagnetic metal when the voltage is applied to it). Interestingly, in the same work, the Hanle effect was proposed: the diffusion spin current in a paramagnetic \textit{normal} metal decays due to the precession of spins in an external field, and the voltage across the receiver ferromagnet decreases\footnotemark\footnotetext{The latter effect, therefore, can also be interpreted as a kind of positive Hanle magnetoresistance of the whole circuit (ferromagnetic metal-injector $ \rightarrow $ paramagnetic metal $ \rightarrow $ ferromagnetic metal-receiver).}.

\bigskip
\bigskip
\begin{center}
	\uppercase{Control questions}
\end{center}
\noindent 1. Justify qualitatively the direction of the secondary spin current from the edges of the sample (thin gray arrows) in Fig. \eqref{SHE_schamatic}.

\noindent 2. Explain qualitatively the reason for the positive magnetoresistance when the spin-Hall sample is placed in a magnetic field that destroys the spin polarization.

\noindent 3. Explain in brief the essence of the Spin Hall Magnetoresistance effect. How can this effect be utilized in spintronic devices?

\chapter{Antiferromagnets \label{}}

\indent

Antiferromagnets (\textbf{AFM}s) are substances with long-range anti-parallel Neel magnetic order and near-zero magnetic moment. In the simplest case of two magnetic sublattices, such order often amounts to two ferromagnetic sublattices inserted one into another, such that the nearest-neighbour spins are oppositely polarized. The Neel order therefore coined another name: the so-called \textit{staggered field}. The two order parameters in AFMs read
\begin{eqnarray}
\label{LM}
\textbf{M}=\textbf{M}_{1}+\textbf{M}_{2}, 
\quad 
\textbf{N}=\textbf{M}_{1}-\textbf{M}_{2}
,
\end{eqnarray}
where $ \mathbf{N} $ is the so-called \textit{Neel vector}, and $ \mathbf{M} $ is the magnetization of the antiferromagnet, which vanishes in the absence of external magnetic fields and currents, 
and is usually small ($ \mathbf{M} \ll \textbf{N} $) in their presence.

As compared to ferromagnets, AFMs have a number of additional features desirable in spintronics and magnonics. The most important distinctions from FMs prove to be the following~\cite{GomonayLoktev, GomonayNature, Baltz18}: 
\begin{itemize}[leftmargin=35pt,label={}]
	\item
	1. AFMs operate at a much higher frequencies than FMs, which typically fall into \textit{THz range}. This makes them useful e.g. for ultrafast information processing. Prototypical devices developing towards this direction are the so-called \textit{THz AFM nano-oscillators}~\cite{AFM nano-osc}. 
	
	\item
	2. Since AFMs have near-zero magnetization, there are \textit{no stray fields} in them. This makes them more robust against magnetic perturbations, and therefore useful in spintronic data storage technologies. 
	
	\item
	3. Many quantities are \textit{enhanced by exchange}. The already mentioned high frequency of AFM oscillations is the manifestation of this enhancement. 
	
	\item
	4. The spectrum of spin waves
	in AFMs consists of \textit{two branches} which  
	have linear (sound-like) dispersions for high enough wavevectors. 
	This is a manifestation of the fact that AFMs generally possess two (or more) vector degrees of freedom, in contrast to FMs which have only one. Generally, \textit{the spin wave spectrum is richer than in FMs} owing to the plethora of different equilibrium configurations and complexity of modes. 

	\item
	5. The domain wall velocities that can be achieved in certain metallic AFMs via the so-called \textit{field-like Neel spin-orbit torques} are 2 orders of magnitude greater than the ones in ferromagnets, and the notorious Walker breakdown of domain wall motion may be surpassed adiabatically. This is because the limit of domain-wall velocity in AFMs is set by the \textit{magnon velocity, which is much larger than the typical magnon velocity in FMs}. 

	\item
	6. AFMs materials are generally not rare, and a lot of them are insulating, which makes them favorable since in insulators \textit{magnetic losses are generally less} then in metals. 
\end{itemize}

However, to explore these fascinating features of AFMs, one needs to be able to \textit{couple} to the Neel staggered field (e.g. to ``read out'' or ``write'' the magnetic state of an antiferromagnet). For FMs, as we have already learned, this coupling is usually achieved in one of three ways: 
\begin{itemize}[leftmargin=35pt,label={}]
	\item
	1. By external field(s), either static or alternating (or their combination, as it is for \textbf{FMR}).
	
	\item
	2. By external spin or charge currents (nonequilibrium spin density), which flow from (is concentrated in) the adjacent normal metal (remember that all of these effects rely on the spin torques). 
	
	\item
	3. By placing a ferromagnet in contact with another ferromagnet or another magnetic substance. 
\end{itemize}
For AFMs, the external static magnetic field leads to sublattice canting effects only in the second order of perturbation theory, and thus is not efficient. However, AFMs are much more responsive to \textit{alternating magnetic fields}, and this is why the \textit{antiferromagnetic resonance} (\textbf{AFMR}) is experimentally achievable. 
We will also discuss one of the methods from the second group, namely, the \textit{spin current-induced dynamics}. The third group of methods relies, as in FM case, on the state of an interface and its microscopic features (i.e. it is more often a challenge for first-principles calculations than for the macroscopic phenomenological theories).


Next, we will elaborate on some of the topics outlined above by starting from equations for magnetization dynamics. 
First, note that since the magnetizations of the two sublattices are equal in magnitude $ |\textbf{M}_{1}|=|\textbf{M}_{2}| $, the Neel vector is always orthogonal to the AFM magnetization, $ \textbf{N} \perp \textbf{M}$. Thus, we see explicitly that $ \mathbf{N} $ and $ \mathbf{M} $ are two orthogonal order parameters, which motivates us to write the dynamic equations for them, instead of using two equations for $ \mathbf{M}_1 $ and $ \mathbf{M}_2 $. The latter are the Landau-Lifshits-like equations, which, in the absence of anisotropies, read 
\begin{equation}
\label{LLeq_AFMs}
\dot{\textbf{M}}_{i}=\gamma\left[\textbf{M}_{i} \times\left(\Lambda_{i j} \textbf{M}_{j}+\textbf{H}_{0}\right)\right], \quad i=1,2
,
\end{equation}
where $ {H}_{0} $ is the external magnetic field, $ \Lambda_{i j} $ are the exchange constants, and the index $ i=1,2 $ denotes the magnetic sublattice. Assuming that intra-sublattice interactions are next-nearest-neighbour and therefore much weaker than inter-sublattice ones, $ \Lambda_{11},\,\Lambda_{22} \ll \Lambda_{12} = \Lambda_{21} $, the simple calculation using definitions \eqref{LM} leads us to the dynamic equations for $ \textbf{N} $ and $ \mathbf{M} $: 
\begin{eqnarray}
\label{dynamic_eqs_AFMs}
\left\{\begin{array}{l}
\dot{\textbf{N}}
=
\gamma \Lambda_{12}[\textbf{N} \times \textbf{M}]
+
\gamma\left[\textbf{N} \times \textbf{H}_{0}\right]
, \\
\dot{\textbf{M}}
=
\gamma\left[\textbf{M} \times \textbf{H}_{0}\right]
.
\end{array}\right.
\end{eqnarray}
As in the case of FMs, these equations may be supplemented by the terms describing the magnetic anisotropy, the demagnetizing field (present for $ |\mathbf{M}|\neq 0 $), various spin torques, and damping. The number of the corresponding terms will be, however, much greater then in the FM case, since, as we already see from the simplest Eq.\eqref{dynamic_eqs_AFMs}, AFMs possess \textit{two coupled vector degrees of freedom}. Furthermore, even the phenomenologic form of some of the additional terms is the subject of an ongoing study, especially for the spin torques and current-driven torques. We will firther describe only two possible terms: the \textit{magnetic damping}, and the \textit{spin torques induced by external spin current}. 
Finally, we note that dynamic equations \eqref{dynamic_eqs_AFMs}, even when supplemented by additional torques, usually could be rewritten in the closed for just one order parameter, e.g. $ \textbf{N} $. The corresponding differential equation will be second-order, and, as a consequence, will generally describe two modes.

The damping terms may be generalized starting from the Landau-Lifshits-like equations~\eqref{LLeq_AFMs} when adding Gilbert term to them, and read
\begin{eqnarray}
\label{dynamic_eqs_AFMs_damping}
\left\{\begin{array}{l}
(\dot{\textbf{N}})_{damping}
=
\frac{\alpha}{2 M_{0}}
\left(
[
\mathbf{M} \times \dot{\mathbf { N }}
]
+
[
\mathbf{N} \times \dot{\mathbf{M}}
]
\right)
, \\
(\dot{\textbf{M}})_{damping}
=
\frac{\alpha}{2 M_{0}}
\left(
[
\mathbf{M} \times \dot{\mathbf { M }}
]
+
[
\mathbf{N} \times \dot{\mathbf{N}}
]
\right)
,
\end{array}\right.
\end{eqnarray}
where $ \alpha $ is the Gilbert damping constant, and $ M_0 = |\mathbf{M}_1| = |\mathbf{M}_2| $ is the magnetization of each of the sublattices. 

In order to understand the spin torque exerted by the external spin current on the AFM, we should first describe the equilibrium configurations of AFM in the presence of anisotropies and external magnetic field. 
As in FMs, we would need to consider the classes of \textit{light axis} (uniaxial anisotropy) and \textit{light plane} AFMs, and within each of them~-- two more cases, when external magnetic field is parallel or perpendicular to the anisotropy axis or to the easy plane. This produces quite different equilibrium configurations with various regimes. 
Here, we will discuss only the case often chosen in experiments~-- \textit{the easy-axis AFM with external field parallel to easy axis}. 

\begin{figure}[h!]
	\centering
	\includegraphics [width=0.7\textwidth]{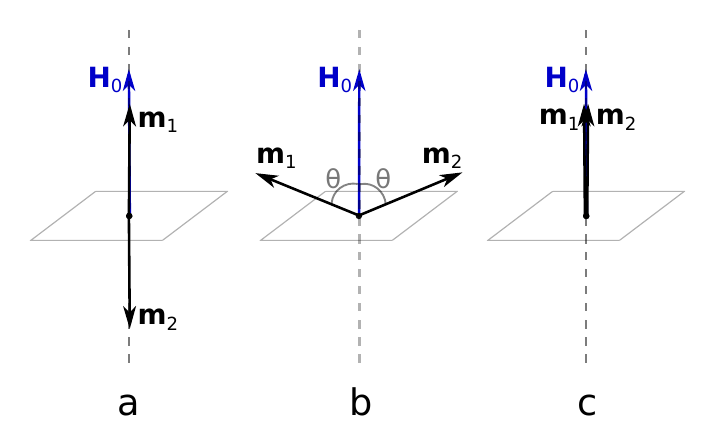}
	\caption{Equilibrium configurations of light-axis AFM. The light axis and the external magnetic field $ \mathbf{H}_0 $ are directed vertically; the field strength increases from left to right. }
	\label{AFMs_equilibr_configs}
	\vspace{-0 pt}
\end{figure}

The corresponding equilibrium configurations could be found rigorously from the energy functional~\cite{Gurevich}, and are depicted in Fig.\ref{AFMs_equilibr_configs}. At low external magnetic fields $ H_0 < \sqrt{2H_E H_A} $, unit sublattice magnetizations vectors $ \mathbf{m}_1,\,\mathbf{m}_2 $ are perfectly anti-parallel in equilibrium (see panel (a)). Here, $ H_E \equiv \Lambda_{12} M_0 $ is the \textit{exchange field}, and $ H_A = K / M_0 $ is the \textit{anisotropy field}. At $ H_0 \simeq \sqrt{2H_E H_A} $, the configuration abruptly changes to one depicted in panel (b), with $ \mathbf{m}_1,\,\mathbf{m}_2 $ forming equal angles $ \theta \simeq \arccos{(H_0 / 2 H_E)} $ with the easy axis; this is the so-called \textit{spin-flop} transition. As the value of field increases further, the vectors $ \mathbf{m}_1,\,\mathbf{m}_2 $ ``collapse'', and finally, at fields $ H_0 \gtrsim 2 H_E $, sublattice magnetizations become parallel, and antiferromagnet becomes (at least in this static picture) a ferromagnet. 

In a typical AFM, the exchange field strength is $ H_E \sim 10^{6} $ [Oe], while the anisotropy field strength is $ H_A \sim 10 \div 10^{4} $ [Oe] $\ll H_E$. 
Hence, we have just mentioned one of the effects of quantily being enhanced by exchange: the \textit{spin-flop field} in AFMs $ H_{sf} = \sqrt{2H_E H_A} $, while in FMs the reorientation field is proportional to anisotropy field $ H_A $; the \textit{exchange enhancement factor} is then $ k_{enh} \sim \sqrt{H_E / H_A} \gg 1 $~\cite{GomonayNature}. This factor will also appear for other quantities enhanced by exchange in AFMs. 


\begin{figure}[h!]
	\centering
	\includegraphics [width=0.7\textwidth]{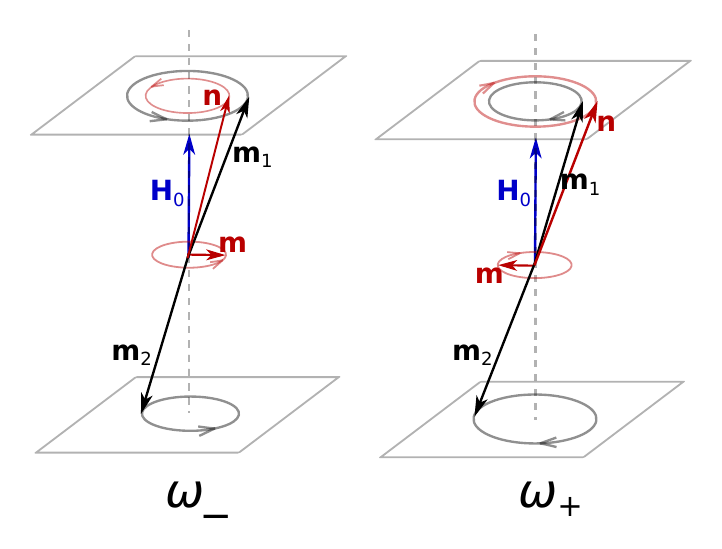}
	\caption{Two oscillation modes of a light-axis AFM in a configuration depicted in the left panel of Fig.\ref{AFMs_equilibr_configs}. All vectors representing the order parameters are assumed to be unitary for clarity: $ \mathbf{m}_1 $ and $ \mathbf{m}_2 $ represent sublattice magnetizations, $ \mathbf{m} $ and $ \mathbf{n} $ represent the magnetization and the Neel vector of AFM, respectively. }
	\label{AFMs_oscillation_modes}
	\vspace{-0 pt}
\end{figure}



As a second step towards understanding the spin torques in AFM, we should briefly consider antiferromagnetic resonance (\textbf{AFMR}) and spin waves in AFM, since these are the type of excitations the spin torques can induce in AFMs, and, conversely, it is these excitations that can produce spin current by spin pumping.

We consider the light-axis parallel-field configuration, with equilibrium as in panel (a) in Fig.\ref{AFMs_equilibr_configs}. By solving the dynamic equations for small deviations of $ \textbf{M}_1 $ and $ \mathbf{M}_2 $, two modes may be obtained, with frequencies:
\begin{equation}
\label{AFMkeq0}
\begin{array}{l}
\omega_{\pm} 
\approx 
\gamma\left(H_{sf} \pm H_{0}\right) 
\end{array}
\end{equation}
These modes are shown in Fig.\ref{AFMs_oscillation_modes}. The two modes represent the (rigid-like) synchronous precession of sublattice magnetization vectors with $ \pi $ phase difference, and differ in handedness. It is very important that the corresponding cone angles are generally different for the sublattice magnetizations: this leads to the \textit{nonzero dynamic magnetization of AFM during oscillations}. Thus, the dynamic susceptibility of AFM is much larger then the static one; i.e., \textit{the AFMs are much more responsive to alternating magnetic fields then to the static fields}. However, the AFM susceptibility at AFMR is still much less than that of the FM at FMR: the ratio of susceptibilities is of the order $ \chi_{AFM} / \chi_{FM} \sim H_A / H_E \ll 1 $. 

The antiferromagnetic resonance is essentially the AFM spin wave mode with $ \textbf{k} = 0 $. The general spin wave spectrum at $ \mathbf{k} > 0 $ in the considered configuration, when neglecting the magnetostatic terms, reads~\cite{Gurevich,ABP}: 
\begin{equation}
\label{AFMk}
\begin{array}{l}
\omega_{\pm} (k)
\approx 
\gamma\left(\sqrt{2 H_E (H_A + H_e (a k)^2)} \pm H_{0}\right) 
,
\end{array}
\end{equation}
where $ H_e $ is (as in FM case) the \textit{imhomogeneous exchange field} (or the \textit{exchange stiffness} constant) parametrizing the additional energy of inhomogeneous exchange, and $ a $ is the lattice constant. For small wavevectors $ ak \ll \sqrt{H_A / H_e} $, the spectrum is quadratic in momentum, $ \omega_{\pm} (k) \approx \gamma\left(k_{enh}\cdot H_e (a k)^2 /\sqrt{2} + H_{sf} \pm H_{0}\right)  $, while for large wavevectors $ ak \gg \sqrt{H_A / H_e} $, the spectrum is \textit{linear} in momentum, $ \omega_{\pm} (k) \approx \gamma\left(\sqrt{2H_E H_e}(a k) \pm H_{0}\right) $. This sound-like spectrum at large $ k $ leads to a \textit{uniform magnon speeds across a broad band}, and is also reflected in unusual magnon contributions to thermodynamical quantities, such as specific heat of AFM. 

Finally, we note that the \textit{circularly polarized spin waves in AFMs carry spin}, as the spin waves in FMs do. In equilibrium, the magnon states with opposite spins are equally populated. However, when a spin current is pumped into AF, the distribution of magnons is modified, and thus a magnon flux with non-zero spin, or \textit{nonzero exchange spin current}, can be generated in AFM. As in FMs, this effect of spin-current to spin-wave-spin-current conversion has been observed in several experiments involving antiferromagnetic insulators~\cite{GomonayNature}. 
However, regarding the mechanisms unlerlying the observed efficient exchange spin transport in AFMs, the agreement has not yet been reached. Some theories claim that \textit{diffusive} magnon transport 
takes place, and some~-- that transport is mediated by the excitation of \textit{coherent} evanescent AFM spin wave modes. 
The counter-intuitive result of \textit{spin current enhancement after passing AFM} for some AFM and excitation parameters is related to the fact that \textit{exchange spin current in AFM is not conserved} due to the interaction between the excited AFM modes and the AFM magnetic texture~\cite{Khymyn16}. 


Now that we are familiar with AFMR in easy-axis parallel-field setting, we can make a heuristic argument about possibility of \textit{spin pumping} in this configuration~\cite{GomonayLoktev, Brataas14}. For small oscillations shown in Fig.\ref{AFMs_oscillation_modes}, we can see that during precession $ \mathbf{m}_1 \simeq -\mathbf{m}_2 $, and $ \mathbf{\dot{m}}_1 \simeq \mathbf{\dot{m}}_2 $. Therefore, at least the fieldlike spin-pumping contributions~\eqref{spin-current} of the two sublattices, 
$ \propto A_r \left(
\left[\mathbf{m_1}\times\dot{\mathbf{m_1}}\right]
+
\left[\mathbf{m_2}\times\dot{\mathbf{m_2}}\right]
\right) $, 
\textit{add up constructively}. 
Hence, the total spin current is approximately proportional to $\mathbf{n} \times \dot{\mathbf{n}}$, where $\mathbf{n}=\left(\mathbf{m}_{1}-\mathbf{m}_{2}\right) / 2$ stands for the unitary Neel vector. Moreover, additional contributions arise due to the finite dynamic magnetization $ \mathbf{m} = \left(\mathbf{m}_{1}+\mathbf{m}_{2}\right) / 2 $ during AFMR. 
Furthermore, cross-terms of the type $ \propto A_r \left(
\left[\mathbf{n}\times\dot{\mathbf{m}}\right]
+
\left[\mathbf{m}\times\dot{\mathbf{n}}\right]
\right) $ 
will be also present due to mixing of scattering channels associated with different sublattices at an interface between AFM and adjacent normal metal.  



Indeed, when the staggered field $\mathbf{n}$ and the magnetization $\mathbf{m}$ are treated as two independent adiabatic parameters, the pumped spin current could be obtained by analogy with the FM case. The two contibutions with nonzero time averages read~\cite{Brataas14}: 
\begin{eqnarray}
&& \mathbf{I}^{pump}_s
=
\frac{\hbar}{e} A_{r}(\mathbf{n} \times \dot{\mathbf{n}}+\mathbf{m} \times \dot{\mathbf{m}})
,\\
&& \mathbf{I}^{pump\, (3)}_{ss}
=
\frac{\hbar}{e} A_{r}(\mathbf{n} \times \dot{\mathbf{m}}+\mathbf{m} \times \dot{\mathbf{n}})
,
\end{eqnarray}
where the \textit{spin current} $\mathbf{I}^{pump}_s$ and the so-called \textit{staggered spin current} $\mathbf{I}^{pump\, (3)}_{ss}$ are measured in units of electrical current. Note however that the value of spin-mixing conductance $A_{r}$  
will be different from that in FM due to the mixing of scattering channels from different sublattices. 
Although $m \ll n$, the contribution of $\left[\mathbf{m} \times \dot{\mathbf{m}}\right]$ to $\mathbf{I}_{s}$ is not completely negligible with respect to $\left[\mathbf{n} \times \dot{\mathbf{n}}\right]$, since $ \mathbf{\dot{n}} $ is proportional to the square of precession angle $ \theta_{pr}^2 \sim m_{i\perp}^2 \ll 1 $. Thus, since $ m \sim m_{i\perp} (H_{sf}/H_E) $, the ratio of two spin pumping contributions  is roughly $ \sim H_{sf}/H_E \sim 1/k_{enh} $, and hence is larger for higher anisotropy fields. 

Finally, the effect reciprocal to spin pumping is also present in AFM~\cite{GomonayLoktev}. Namely, if the external spin current is polarized perpendicularly to $ \mathbf{n} $, 
the spin torques that it exerts on AFM Neel order parameter can, above a certain threshold set by the damping, induce a stable precession of $ \mathbf{n} $ within the plane perpendicular to the spin-current polarization. 
This is an example of a \textit{spin-torque AFM oscillator}, operated just by external spin currents, even in the absence of external magnetic fields.

\bigskip
\bigskip
\begin{center}
	\uppercase{Control questions}
\end{center}
\noindent 1. List several most important distinctions of AFM magnetic dynamics when compared to FM dynamics. 

\noindent 2. What is the typical enhancement factor for the quantities enhanced by exchange in AFM? 

\noindent 3. How many spin-wave modes does a generic AFM have? How does the velocity of a highly excited spin wave in AFM depend on wavevector? 

\noindent 4. Do AFMs support exchange spin current?  

\noindent 5. Is spin pumping possible in AFMs? Give a simple supporting argument using analogy with spin pumping in FMs, treating contributions of two magnetic sublattices as independent.

\selectlanguage{english}

\chapter*{Conclusion \label{}}
\addcontentsline{toc}{chapter}{\protect\numberline{}\noindent Conclusion}%

\ \\
\indent In this short manual, the basic theoretical concepts of magnonics and spintronics were discussed: spin and exchange spin currents, various types of spin torques, the Spin Hall Effect, the Spin Hall and Hanle magnetoresistance. Within the framework of Landauer approach, the theory of Spin Transfer Torque and spin pumping were considered. 
The concepts of magnetism, which play an important role in spintronics, were considered: ferromagnetic and antiferromagnetic resonance, the dynamic susceptibilities of magnets, and the spin waves. 
The most important features distinguishing antiferromagnets from ferromagnets have been outlined. 
Brief information from quantum mechanics, electrodynamics of continuous media, and basic theory of magnetism, given at the beginning of the manual, is sufficient to understand the subsequent material.

\indent We thank A.~S.~Kovalev and M.~Yu.~Kovalevskij for careful reading and insightful comments.

\chapter*{Index \label{}}
\addcontentsline{toc}{chapter}{\protect\numberline{}\noindent Index}%

\setlength{\columnsep}{25pt}
\begin{multicols}{2}

	{
		\begin{itemize}[leftmargin=*,label={}]
			\item
			\NoIndent{Bohr magneton $ \mu_B $, 5, 10, 21, 26}
			
			\item
			\NoIndent{breathing Fermi surface effect, 55}
			
			\item
			\NoIndent{Brillouin function, 24}

			\item
			\NoIndent{Curie-Weiss law, 26}
			
			\item
			\NoIndent{Curie temperature, 26}

			\item
			\NoIndent{gyromagnetic factor $ \gamma $, 5, 28}

			\item
			\NoIndent{damping torque, 37, 57, 75}
			
			\item
			\NoIndent{demagnetizing factors, 34}
			
			\item
			\NoIndent{density matrix (of an electron beam), 12}

			\item
			\NoIndent{diamagnetism, 23}
			
			\item
			\NoIndent{dispersion of permeabilities (temporal and spatial), 20, 32, 36, 43}
						
			\item
			\NoIndent{domain structure, 34, 50, 51}

			\item
			\NoIndent{effective (molecular) field approximation, 25}

			\item
			\NoIndent{electron polarization vector, 11, 13}
			
			\item
			\NoIndent{enhancement by exchange, 73}
			
			\item
			\NoIndent{exchange Hamiltonian, 25}
			
			\item
			\NoIndent{exchange interaction energy, 25}

			\item
			\NoIndent{ferromagnetism, 24}
			
			\item
			\NoIndent{ferromagnetic insulator, 16, 40, 47, 50, 58, 72}

			\item
			\NoIndent{ideal spin sink, 63}
			
			\item
			\NoIndent{Inverse Spin-Galvanic Effect (ISGE), 52}
			
			\item
			\NoIndent{Inverse Spin Hall Effect (ISHE), 16, 71}

			\item
			\NoIndent{Landauer scattering matrix method, 59}
			
			\item
			\NoIndent{Landau-Lifshitz equation, 29, 39}
			
			\item
			\NoIndent{Landau-Lifshitz-Gilbert equation, 39}

			\item
			\NoIndent{Larmor precession frequency, 30}

			\item
			\NoIndent{Maxwell's equations, 18, 20}

			\item
			\NoIndent{natural oscillations of the magnetic moment, 30}
			
			\item
			\NoIndent{Neel spin-orbit torques, 74}
			
			\item
			\NoIndent{Neel vector, 73}
			
			\item
			\NoIndent{nuclear magnetic resonance (NMR), 33}

			\item
			\NoIndent{magnetic anisotropy energy, 36}
			
			\item
			\NoIndent{magnetic dipole-dipole interactions, 34}
			
			\item
			\NoIndent{magnetic gyrotropy, 32}

			\item
			\NoIndent{magnetic stiffness (non-uniform exchange constant) $ \alpha $, 41, 47}
			
			\item
			\NoIndent{magnetic susceptibility (high-frequency), 32, 36, 44}
			
			\item
			\NoIndent{magnetic susceptibility (static), 20, 23, 41}

			\item
			\NoIndent{magnetostatic energy (of spin waves), 44}

			\item
			\NoIndent{normal (free) spin waves, 40}

			\item
			\NoIndent{paramagnetism, 23}
			
			\item
			\NoIndent{Pauli equation, 10, 21}

			\item
			\NoIndent{relaxation times (spin-lattice and transverse), 33}

			\item
			\NoIndent{scattering channels), 60}
			
			\item
			\NoIndent{s-d interaction, 53}
			
			\item
			\NoIndent{spin accumulation, 51}
			
			\item
			\NoIndent{spin coherence time, 14, 16, 53, 55, 59}
			
			\item		
			\NoIndent{spin current, 14, 54, 65}
			leakage spin current, 69

			\item
			\NoIndent{spin density (nonequilibrium), 15, 52, 53}
			
			\item
			\NoIndent{spin-dependent electron reflection, 51}
			
			\item
			\NoIndent{spin-flip (of an electron), 58}
			
			\item
			\NoIndent{spin-flop (of magnetic sublattices), 75}
			
			\item
			\NoIndent{Spin Hall Effect (SHE), 48, 52, 70}
			
			\item
			\NoIndent{spin-mixing conductance, 52, 68}

			
			\item
			\NoIndent{spin pumping, 48, 57}
			
			\item
			\NoIndent{spin torques:}
			
			adiabatic and non-adiabatic, 50, 56
			
			field-like and dissipative-like, 53
			
			transport and diffusive, 51, 58

			\item
			\NoIndent{spinor, 9, 65}
			
			\item
			\NoIndent{spin-orbit interaction, 14, 16, 21, 51, 70}
			
			\item
			\NoIndent{spin-spin interaction, 21}

			\item
			\NoIndent{spin wave dispersion law, 40}

			\item
			\NoIndent{staggered Neel field, 73}

			\item
			\NoIndent{uniform mode (of magnetization oscillations), 34}

			\item
			\NoIndent{Zeeman splitting (of atomic levels), 22}
			
			
		\end{itemize}
	}
	
\end{multicols}


\newpage

\addcontentsline{toc}{chapter}{\protect{Recommended literature}}

\def\bibname{Recommended literature}


\makeatletter
\renewcommand\@bibitem[1]{\item\if@filesw \immediate\write\@auxout
	{\string\bibcite{#1}{S\the\value{\@listctr}}}\fi\ignorespaces}
\def\@biblabel#1{[S#1]}

\def\bibname{Further reading}

\addcontentsline{toc}{chapter}{\protect{Further reading}}

\end{document}